\documentclass[twocolumn]{aastex631}
\usepackage{amsmath}

\def \mkms {\rm ~km~s^{-1}}

\shorttitle{SDSS-IV MaStar: Removal of Interstellar Absorption}
\shortauthors{Rubin et al.}
\graphicspath{{./}{figures/}}

\begin{document}

\title{SDSS-IV MaStar: Quantification and Abatement of Interstellar Absorption in the Largest Empirical Stellar Spectral Library}

\correspondingauthor{Kate H. R. Rubin}
\email{krubin@sdsu.edu}

\author[0000-0001-6248-1864]{Kate H. R. Rubin}
\affiliation{Department of Astronomy, San Diego State University, San Diego, CA 92182 USA}
\affiliation{Department of Astronomy and Astrophysics, University of California, San Diego, La Jolla, CA 92092, USA}
\author[0000-0003-1809-6920]{Kyle B. Westfall}
\affiliation{University of California Observatories, University of California, Santa Cruz, 1156 High St., Santa Cruz, CA 95064, USA}
\author[0000-0001-7711-3677]{Claudia Maraston}
\affiliation{Institute of Cosmology and Gravitation, University of Portsmouth, 1-8 Burnaby Road, Portsmouth PO1 3FX, UK}
\author[0000-0002-6325-5671]{Daniel Thomas}
\affiliation{School of Mathematics and Physics, University of Portsmouth, Lion Gate Building, Portsmouth, PO1 3HF, UK}
\affiliation{Institute of Cosmology and Gravitation, University of Portsmouth, 1-8 Burnaby Road, Portsmouth PO1 3FX, UK}
\author[0000-0003-1025-1711]{Renbin Yan}
\affiliation{Department of Physics, The Chinese University of Hong Kong, Shatin, N.T., Hong Kong S.A.R., China}
\author[0000-0002-2591-3792]{J. Christopher Howk}
\affiliation{Department of Physics and Astronomy, University of Notre Dame, Notre Dame, IN 46556, USA}
\author[0000-0001-9596-845X]{Erick Aguirre}
\affiliation{Department of Astronomy, New Mexico State University, Las Cruces, NM 88003, USA}
\author[0000-0002-8809-4608]{Kaelee S. Parker}
\affiliation{Department of Astronomy, The University of Texas at Austin, 2515 Speedway, Stop C1400, Austin, TX 78712, USA}
\author[0000-0002-9402-186X]{David R. Law}
\affiliation{Space Telescope Science Institute, 3700 San Martin Drive, Baltimore, MD 21218, USA}

\begin{abstract}

We assess the impact of 
\ion{Ca}{2}$~\lambda\lambda3934,3969$ and \ion{Na}{1}$~\lambda\lambda5891,5897$ absorption arising 
in the interstellar medium (ISM) on the SDSS-IV MaNGA Stellar Library (MaStar) and produce corrected  spectroscopy for $80\%$ of the 24,162-star catalog.  We model the absorption strength of these transitions as a function of stellar distance, Galactic latitude, and dust reddening based upon high-spectral resolution studies.  With this model, we identify 6342 MaStar stars that have
negligible ISM absorption ($W^\mathrm{ISM}$(\ion{Ca}{2} K) $<0.07~\rm\AA$ and $W^\mathrm{ISM}$(\ion{Na}{1} 5891) $<0.05~\rm\AA$).  For 12,110 of the remaining stars, we replace their \ion{Na}{1} D profile (and their \ion{Ca}{2} profile for effective temperatures $T_{\rm eff}>9000$ K) with a coadded spectrum of low-ISM stars with similar $T_{\rm eff}$, surface gravity, and metallicity.  For 738 additional stars with $T_{\rm eff}>9000$ K, we replace these spectral regions with a matching ATLAS9-based BOSZ model.  This results in a mean reduction in $W$(\ion{Ca}{2} K) ($W$(\ion{Na}{1} D)) of $0.4-0.7~\rm\AA$ ($0.6-1.1~\rm\AA$) for hot stars ($T_{\rm eff}>7610$ K),
and a mean reduction in $W$(\ion{Na}{1} D) of $0.1-0.2~\rm\AA$ for cooler stars. We show that interstellar absorption in simple stellar population (SSP) model spectra constructed from the original library artificially enhances $W$(\ion{Ca}{2} K) by $\gtrsim20\%$ at young ages ($<400$ Myr); dramatically enhances the strength of stellar \ion{Na}{1} D in starbursting systems (by ${\gtrsim}50\%$); and
enhances stellar \ion{Na}{1} D in older stellar populations (${\gtrsim}10$ Gyr) by ${\gtrsim}10\%$. We provide SSP spectra constructed from the cleaned library, and discuss the implications of these effects for stellar population synthesis analyses constraining stellar age, [Na/Fe] abundance, and the initial mass function.

\end{abstract}


\section{Introduction} \label{sec:intro}

Stellar spectral template libraries are fundamental tools for the analysis of spectroscopy of external galaxies. Catalogs of spectra of individual Milky Way stars were crucial to the earliest efforts to understand the chemical abundances and stellar populations of nearby galaxies \citep{SpinradTaylor1971,Faber1972,OConnell1976,Turnrose1976,Pickles1985}.
Their use continues to serve an exceptionally broad range of science topics, from galactic stellar and gas dynamics \citep[e.g.,][]{Cappellari2011,Cappellari2016,Bloom2017,Westfall2019,Bryant2019,Law2022}, to stellar and gas-phase chemical abundances and evolution \citep[e.g.,][]{Tremonti2004,Ho2015,Belfiore2017,Belfiore2019,Parikh2019,Parikh2021,Neumann2021}, to stellar initial mass function (IMF) studies \citep[e.g.,][]{Treu2010,ThomasJ2011,ConroyvanDokkum2012b,Parikh2018,Bernardi2022}.

Measurement of quantities relevant to each of these topics relies on the modeling of galaxy continua using stellar template spectra. 
These templates may be derived from the coaddition of individual stellar library spectra with similar parameters (e.g., effective temperature, surface gravity, and metallicity;  \citealt{Westfall2019}), or are constructed from selected library stars to represent simple stellar populations (SSPs) having a range of ages and chemical abundances \citep[e.g.,][]{Tinsley1978, Bruzual1983,Guiderdoni1987,Worthey1994,Maraston1998,LeithererHeckman1995,Vazdekis1999,BC2003,Maraston2011,Conroy2013,Maraston2020}.  The stellar continua of distant galaxy populations may then be modeled using linear combinations of these templates.

Theoretical stellar libraries calculated from models of stellar atmospheres \citep[e.g.,][]{Kurucz1993,Coelho2005,Rodriguez-Merino2005,BOSZ2017,Eldridge2017} are advantageous in that they can be produced at arbitrarily high spectral resolution, and can sample any chemical abundance pattern of interest to the user.  However, such calculations have not yet fully accounted for the numerous physical processes that give rise to the profile shapes of stellar absorption lines \citep[e.g.,][]{Kurucz2011}.
For example, the model atmospheres upon which they are built rely on our incomplete knowledge of the relevant atomic and molecular transitions, as well as upon accurate calculation of absorption line opacities.  This becomes challenging at low effective temperatures ($T_{\rm eff}$) due to the profusion of molecular transitions \citep[e.g.,][]{Coelho2014,BOSZ2017}.  

Alternatively, empirical libraries may be constructed through spectroscopic campaigns targeting stars within ${\lesssim} 15$ kpc of the Sun.  Traditionally, such libraries have been limited in their coverage of stellar parameter space, such that they are likely not fully representative of distant galaxy stellar populations \citep{Yan2019}.  However, the recent completion of the Mapping Nearby Galaxies at Apache Point Observatory (MaNGA) survey \citep{Bundy2015,Yan2016} has included the construction and publication of the MaNGA Stellar Library (MaStar), by far the largest empirical stellar library to date \citep{Yan2019,Chen2020,Abdurrouf2022,Hill2022,Imig2022,Lazarz2022}.  MaStar includes over an order of magnitude more stars than any other empirical library in common use (e.g., MILES, STELIB, INDO-US; \citealt{Sanchez-Blazquez2006,LeBorgne2003,Valdes2004}) at a spectral resolution ($\mathcal{R}\sim1800$) and with wavelength coverage ($3622-10354$ \AA) matching that of the MaNGA survey data.  {It is furthermore constructed from high-quality spectra having, e.g., precise and accurate wavelength and flux calibration and telluric correction \citep{Yan2019,Abdurrouf2022}.}
Such a library is required to comprehensively model the broad diversity of spectra among MaNGA's ${\sim}10,000$ galaxy targets \citep{Yan2019}.

Even with the significant advances of MaStar, any empirical library is nevertheless affected by the presence of the Milky Way's interstellar medium (ISM). The broad range of gas temperatures and densities in and around the Galactic plane gives rise to strong absorption in resonant transitions across the ultraviolet and optical \citep[e.g,][]{Hartmann1904,Hobbs1969,Hobbs1974,Sembach1999,Richter2001b,Richter2001a,Wakker2001,Pellerin2002,Robert2003,Howk2003,Yao2009,LehnerHowk2011}.  The profusion of far- and near-UV metal-line transitions tracing warm neutral or ionized material (e.g., \ion{O}{6} $\lambda1031$, \ion{O}{1} $\lambda1302$, \ion{Si}{2} $\lambda1260$, \ion{C}{4} $\lambda1548$, among many others), in combination with the numerous Lyman and Werner transitions of molecular hydrogen at $\rm 980~\AA < \lambda_{rest} < 1120~\AA$ \citep{Tumlinson2002}, introduce significant ``contaminating'' absorption into Galactic stellar UV spectroscopy \citep{Pellerin2002,Robert2003,Crowther2022}.  In the optical, the \ion{Ca}{2} $\lambda\lambda 3934,3969$ and \ion{Na}{1} $\lambda\lambda 5891,5897$ transitions have long been understood to trace the warm (temperature $T<10,000$ K) and cold ($T <1000$ K) phases of the ISM \citep{Hartmann1904,MunchZirin1961,Hobbs1969,Hobbs1974,Crawford1992,Welty1996,Richter2011,Puspitarini2012}.  More recent studies have demonstrated a close association between these transitions and massive \ion{H}{1} cloud complexes observed in 21 cm emission \citep[e.g.,][]{Wakker2001,BenBekhti2008,BenBekhti2012,Bish2019}.  

The absorption strength of \ion{Ca}{2} and \ion{Na}{1} has also been demonstrated to correlate strongly with dust reddening, both within the Milky Way \citep{Phillips1984,Sembach1994,MunariZwitter1997,Welty2006,Poznanski2012,Welty2012,Murga2015} and in extragalactic systems \citep{Wild2005,Zych2009,ChenTremonti2010,Phillips2013,Baron2016,Rupke2021}.
The widely used \citet{Poznanski2012} relation between $E(B-V)$ and the equivalent width of the \ion{Na}{1} D doublet ($W$(\ion{Na}{1})) observed toward extragalactic sources probing the Milky Way ISM and halo predicts that even a modest degree of reddening is associated with significant interstellar \ion{Na}{1}; e.g., it implies an absorption strength of $\approx 0.7$ \AA\ for $E(B-V) = 0.1$.  Given that early-type (O/B/A) stars are preferentially located in dusty star-forming regions, their spectra may suffer an even greater degree of contamination.  Considering that late-type galaxies typically exhibit NaD spectral index strengths of $1.0-3.5$ \AA\ \citep{Parikh2021}, it is plausible that the level of Milky Way \ion{Na}{1} D contamination in empirical stellar library spectra is as strong as that observed in the extragalactic systems these templates are being used to model.  For analyses that draw on the NaD spectral index as an indicator of stellar sodium abundance by linking its observed strength to that predicted in SSP models  \citep[e.g.,][]{Thomas2003,Thomas2011,Martin-Navarro2015,Parikh2019,Parikh2021}, this implies that such [Na/Fe] constraints are systematically underestimated (as we demonstrate in Section~\ref{sec:discussion}).  Analyses making use of the \ion{Na}{1} D doublet to trace the absorption strength and kinematics of cool interstellar gas in extragalactic systems will likewise be affected, as they typically rely on stellar continuum modeling to remove the contribution of stellar atmospheres from the observed line profiles.  Those studies of interstellar gas kinematics that are based upon data sets of modest spectral resolution ($\mathcal{R}\lesssim 3000$; e.g., \citealt{ChenTremonti2010,Cazzoli2014,Concas2019,RobertsBorsani2019,Perna2020,Perna2021,RobertsBorsani2020,Avery2022}) would be most severely impacted by the implied systematic overestimation of the stellar component of \ion{Na}{1} D.

In this work, we address this issue by removing Milky Way interstellar \ion{Ca}{2} $\lambda \lambda 3934,3969$ and \ion{Na}{1} $\lambda\lambda 5891,5897$ absorption from the MaStar empirical stellar library.  We begin by assembling high-resolution spectroscopic observations of interstellar \ion{Ca}{2} and \ion{Na}{1} absorption  toward early-type stars from the literature \citep{Sembach1993,MunariZwitter1997,Welsh2010}.  We use these data to build simple models of the equivalent widths of these transitions ($W^{\rm ISM}$)
as a function of stellar distance, Galactic latitude, and the dust reddening measured along the stellar sightline.  We then use these models to estimate the level of ISM contamination toward each MaStar object, and identify a subset of the library for which the contamination level is minimal ($W^{\rm ISM}$(\ion{Ca}{2} K) $<0.07$ \AA\ and $W^{\rm ISM}$(\ion{Na}{1} 5891) $<0.05$ \AA).  These ``low-ISM'' sightlines comprise $\approx 27\%$ of the spectral sample. For each of the remaining stars, we identify subsets of the low-ISM sample that have similar stellar parameters ($T_{\rm eff}$, surface gravity or $\log g$, and [Fe/H]) where possible, coadd the spectra within these subsets, and replace the portions of the affected star's spectrum immediately surrounding the \ion{Ca}{2} and \ion{Na}{1} transitions with the coadd.  For those stars for which we could not identify a low-ISM replacement subset that was sufficiently close in stellar parameter space, and which have $T_{\rm eff} > 9000$ K, we instead draw on the theoretical stellar library of \citet{BOSZ2017} to replace the affected spectral regions.     

We demonstrate the presence of Milky Way interstellar absorption in a subset of early-type stars from the MaStar library, build our model of interstellar \ion{Ca}{2} and \ion{Na}{1} absorption, and identify our ``low-ISM'' subsample in Section~\ref{sec:Wr-Dist-EBV}. In Section~\ref{sec:correction}, we assess the intrinsic dispersion in the \ion{Ca}{2} and \ion{Na}{1} profiles of low-ISM stars with similar stellar parameter values; describe our spectral coaddition technique; and describe our profile replacement procedures.  Section~\ref{subsec:clean_spectra} presents the final ``cleaned'' MaStar spectroscopy and assesses the change in absorption strength of the stellar \ion{Ca}{2} and \ion{Na}{1} D profiles relative to the original library.  
In Section~\ref{subsec:hc_templates}, we use a hierarchical clustering technique to construct a set of 58 coadded templates from our cleaned spectra that are representative of the library's stars, and that may be used for continuum modeling of galaxy spectra.  In Section~\ref{subsec:MaStar-SSPs}, we present a suite of SSP model spectral templates constructed from our cleaned spectra which may be used for stellar population synthesis studies.
{The cleaned stellar spectra, hierarchically-clustered templates, and SSP templates are all publicly available.}\footnote{The cleaned stellar spectra are available at Zenodo at \url{https://doi.org/10.5281/zenodo.14014915}.  The full set of SSP templates is available at \url{https://doi.org/10.5281/zenodo.14807331}.  The hierarchically-clustered templates, as well as a subset of the SSP templates which have been modified for use with the MaNGA Data Analysis Pipeline are available at \url{https://github.com/sdss/mangadap/tree/4.2.0/mangadap/data/spectral_templates}.}  To our knowledge, this is the only extant stellar empirical library with relatively low spectral resolution ($\mathcal{R}\lesssim 10,000$) for which the effects of absorption lines arising in the Milky Way's ISM have been assessed and corrected (although corrections have been offered for ultraviolet stellar spectroscopy and higher-resolution optical libraries; e.g., \citealt{Taresch1997,Robert2003,Borisov2023}).
Finally, we discuss the implications of our results for studies of stellar population age, sodium abundance, and IMF in extragalactic systems in Section~\ref{sec:discussion}.

\section{A Model of Interstellar \ion{Ca}{2} and \ion{Na}{1} Absorption} \label{sec:Wr-Dist-EBV}

\subsection{Interstellar \ion{Ca}{2} and \ion{Na}{1} in MaStar Stellar Spectroscopy}\label{subsec:MaStar_Wr}

We begin our investigation by searching for direct evidence of the impact of interstellar \ion{Ca}{2} and \ion{Na}{1} absorption on spectra drawn from the MaStar library.  
Spectral types O, B, and A are commonly used as background probes of foreground interstellar absorption, in part due to the weakness of the intrinsic absorption in their atmospheres.  This suggests that an analysis of early-type MaStar stars may provide the most straightforward evidence of contamination from the ISM.  However, first we must quantify the  strengths of the intrinsic absorption that is expected from these stars, since 
the resolution of MaStar spectroscopy is too low to 
differentiate intrinsic stellar profiles from interstellar absorbers.

We note here that high-resolution stellar spectroscopy has also revealed significant telluric absorption in the wavelength range $\rm 5885~\AA < \lambda < 5906~\AA$ \citep[e.g.,][]{Lallement1993,Chen2014,Sandford2023}. Any telluric contamination of the \ion{Na}{1} D spectral region can be modeled and corrected for at high spectral resolution, but has not been removed from the MaStar spectra \citep{Yan2016,Yan2019}.  In Appendix~\ref{sec:appendix-telluric}, we show that the strength of telluric absorption features in this wavelength range is likely to be negligible (${<}0.1$ \AA) given the atmospheric conditions at Apache Point Observatory during the observations of the vast majority of the MaStar sample.  However, we also demonstrate that telluric absorption may be significant (${\sim} 0.1-0.2$ \AA) in a small minority of spectra.  Studies that require precision stellar population modeling of \ion{Na}{1} may therefore benefit from an additional correction for telluric effects.

\begin{figure}[ht]
 \includegraphics[width=\columnwidth,trim={0.5cm 0.5cm 0 0},clip]{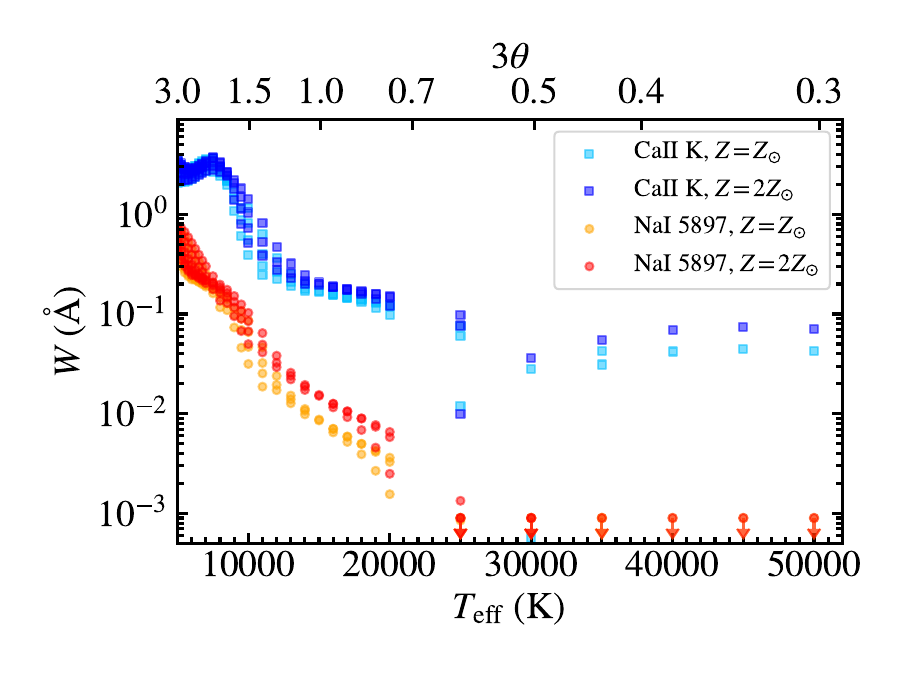}
\caption{Equivalent widths of the \ion{Ca}{2} K (light and dark blue) and \ion{Na}{1} 5897 (orange and red) transitions measured from the \citet{Rodriguez-Merino2005} theoretical spectral templates.  Measurements for solar-metallicity model spectra are shown in light blue and orange, and measurements for models with twice solar-metallicity are shown in dark blue and red. Measurements that fall below the range in $W$ shown are indicated as upper limits.  The scatter in $W$ values at a given $T_{\rm eff}$ and metallicity is due to differences in $\log g$ (which cover the range $0 \le \log g \le 5$).  The top axis shows the quantity $3\theta = 3 \times 5040~\mathrm{K}/T_{\rm eff}$ for reference.   \label{fig:maraston_ews}}
\end{figure}

Returning to our effort to assess the intrinsic strengths of \ion{Ca}{2} and \ion{Na}{1} D in hot stars, we draw on the theoretical stellar spectra computed by \citet{Maraston2011}.  
These templates rely on model stellar atmospheres described in full by 
\citet{Maraston2009}, and are based on the UVBLUE and BLUERED libraries of $\mathcal{R} =$ 10,000 theoretical spectra computed by \citet{Rodriguez-Merino2005}.
We continuum-normalize each template spectrum using a linear fit to feature-free spectral regions on either side of each transition.  We then compute the boxcar $W$ over the ranges $3931.0~\mathrm{\AA} < \lambda_{\rm air} < 3935.5~\mathrm{\AA}$ and $5895.0~\mathrm{\AA} < \lambda_{\rm air} < 5896.5~\mathrm{\AA}$ to assess the strength of \ion{Ca}{2} K and \ion{Na}{1} 5897, respectively. In the case of \ion{Ca}{2} K, the spectral window also includes absorption from neighboring, weak metal-line transitions which dominate the  equivalent width in stars with $T_{\rm eff} \ge $ 40,000 K.
Measurements of these line strengths in all  theoretical spectra having metallicities $Z=Z_{\odot}$ and $2Z_{\odot}$ are shown in Figure~\ref{fig:maraston_ews}.  We find that 
the strength of \ion{Na}{1} 5897 decreases monotonically with temperature in the range 15,000 K $ < T_{\rm eff} < $ 50,000 K. 
The strength of \ion{Ca}{2} K also decreases monotonically over this temperature range, but the equivalent width measurement shown plateaus at $T_{\rm eff} >$ 30,000 K due to the presence of the contaminating transitions mentioned above.  
The maximum measured line strengths within this temperature range are $W$(\ion{Ca}{2} K) $= 0.207$ \AA\ and $W$(\ion{Na}{1} 5897) $ = 0.015$ \AA.  Moreover, above $T_{\rm eff} \ge$ 30,000 K, we measure $W$(\ion{Ca}{2} K) $\le 0.074$ \AA.  We conclude that intrinsic stellar absorption in the \ion{Na}{1} $5897$ transition is minimal for all stars with $T_{\rm eff} \ge $ 15,000 K, while equivalent widths of up to $\approx0.2$ \AA\ may be attributed to stellar absorption in the case of \ion{Ca}{2} K.  

We now proceed with measurement of the absorption strengths of the \ion{Ca}{2} K and \ion{Na}{1} 5891, 5897 transitions in a subset of the MaStar spectra which have been deemed ``high-quality'' (i.e., they have high signal-to-noise ratios and relatively high spectral resolution). We focus here on the 29 stars in the high-quality sample with $T_{\rm eff} > $ 15,000 K, adopting the median $T_{\rm eff}$ value among stellar parameter estimates computed by four groups within the MaStar team  \citep[hereafter $T_{\rm eff,med}$;][Y.\ Chen et al.\ {\it in preparation}]{Imig2022,Hill2022,Lazarz2022}.\footnote{We adopt the median stellar parameter values reported in the v2 catalog available at \url{https://www.sdss4.org/dr17/mastar/mastar-stellar-parameters/}.}  
We determine the continuum level of each star by fitting a spline function to feature-free spectral regions close to the transitions of interest using the \texttt{lt\_continuumfit} GUI, which is part of the \texttt{linetools}\footnote{\url{https://linetools.readthedocs.io/en/latest/}} Python package (version 0.3).  This tool allows for interactive placement of knots and visual inspection of the resulting spline fit.  

\begin{figure*}[ht]
 \includegraphics[width=\textwidth,trim={0.5cm 0.5cm 0.5cm 0.5cm},clip]{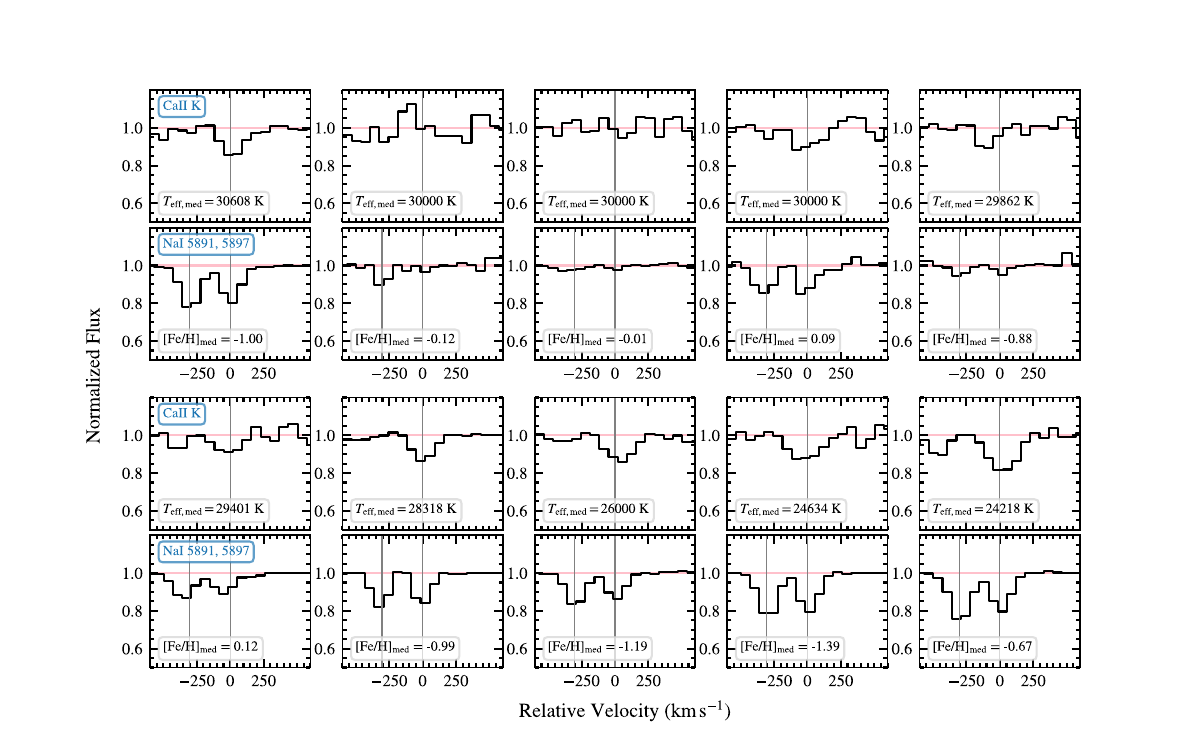}
\caption{Spectroscopy of the ten stars with the highest $T_{\rm eff, med}$ in the high-quality MaStar sample.  Each pair of stacked panels shows the same spectrum in windows surrounding the \ion{Ca}{2} K and \ion{Na}{1} $\lambda \lambda 5891, 5897$ transitions on top and bottom, respectively.  Velocities are computed relative to the \ion{Ca}{2} K $\lambda3934$ and \ion{Na}{1} $\lambda 5897$ rest wavelengths, and the gray vertical lines indicate these transitions along with the relative velocity of $\lambda 5891$. The values of $T_{\rm eff, med}$ and $\rm [Fe/H]_{med}$ for each star are included in each panel pair.  \label{fig:OB_velplots}}
\end{figure*}

We then use the \texttt{XAbsSys} GUI, also available with \texttt{linetools}, to visually inspect the \ion{Ca}{2} K and \ion{Na}{1} transitions in each star.  In all cases, an absorption feature is evident within $\pm300\mkms$ of the rest frame.  We use this GUI to manually select the spectral window to be used to compute the $W$ for each line, and perform this computation using a boxcar sum of the continuum-normalized flux decrement.  Spectral regions surrounding these transitions for the ten stars with the highest values of $T_{\rm eff, med}$ are shown in Figure~\ref{fig:OB_velplots}.  All exhibit clear \ion{Na}{1} absorption, and the vast majority exhibit strong \ion{Ca}{2} K absorption.  Our $W$ measurements for the full sample of high-quality MaStar stars with $T_{\rm eff,med} > $ 15,000 K are included in Figure~\ref{fig:ew_EBV}.

\subsection{Establishing the $W^{\rm ISM}$ $-$ Dust Reddening Relation as a Function of Distance}\label{subsec:wr-distance-model}

To constrain the relationship between the absorption strength of these transitions and the distance and dust reddening of the star toward which they are observed, we draw on a rich literature of Milky Way ISM studies.  \citet{Sembach1993} obtained $\mathcal{R}\sim68,000$ optical spectroscopy of 57 early-type stars at distances beyond 1 kpc, with the goal of probing interarm regions, Galactic center sightlines, and high latitude directions.  They performed Voigt profile fitting of the observed \ion{Ca}{2} and \ion{Na}{1} absorption troughs, and reported the column densities and Doppler parameters of individual velocity components, in addition to the total $W^{\rm ISM}$ of each system.  They determined the color excess of each star ($E(B-V)$) by comparing the intrinsic colors implied by previously published spectral types \citep{Johnson1963} to observed colors, and estimated distances using the spectral type $-$ absolute magnitude relations of \citet{Walborn1972,Walborn1973}.  They assumed a reddening law with $R_V = A_V/E(B-V) = 3.1$.

We also draw on the sample of 32 O and B stars observed by \citet{MunariZwitter1997} at $\mathcal{R}\sim16,500$.  The primary aim of this study was to establish the relation between \ion{Na}{1} D and \ion{K}{1} line strengths and color excess across a wide range of reddening values.  \ion{Ca}{2} K did not fall within their spectroscopic coverage.  $E(B-V)$ estimates for the sample were adopted from \citet{SudziusBobinas1994}, and distances were computed using the \citet{Schmidt-Kaler1982} absolute magnitude scale.  They assumed $R_V = A_V/E(B-V) = 3.2$ when calculating these distances.  The total $W^{\rm ISM}$ for each system was reported, along with the $W^{\rm ISM}$ of individual velocity components in multi-component systems.

Finally, we take advantage of the large sample of sightlines studied by \citet{Welsh2010}.  This work presents $\mathcal{R} > $ 50,000 spectroscopy of \ion{Na}{1} detected along 482 stellar sightlines, and of \ion{Ca}{2} K detected toward 807 sightlines. The authors performed Voigt profile modeling of each of these systems, and then
combined these measurements with previously published data to produce a catalog of absorption toward 1857 early-type stars, all within 800 pc of the Sun.  Their analysis reveals a ``wall" of \ion{Na}{1}-absorbing material at a distance of 80 pc from the Sun, within which only quite weak \ion{Na}{1} ($W^{\rm ISM}$(\ion{Na}{1} 5891) $< 5$ m\AA) is observed.  This finding alone implies that \ion{Na}{1} absorption varies significantly over very short distances.  

\begin{figure*}[ht]
\includegraphics[width=\textwidth]{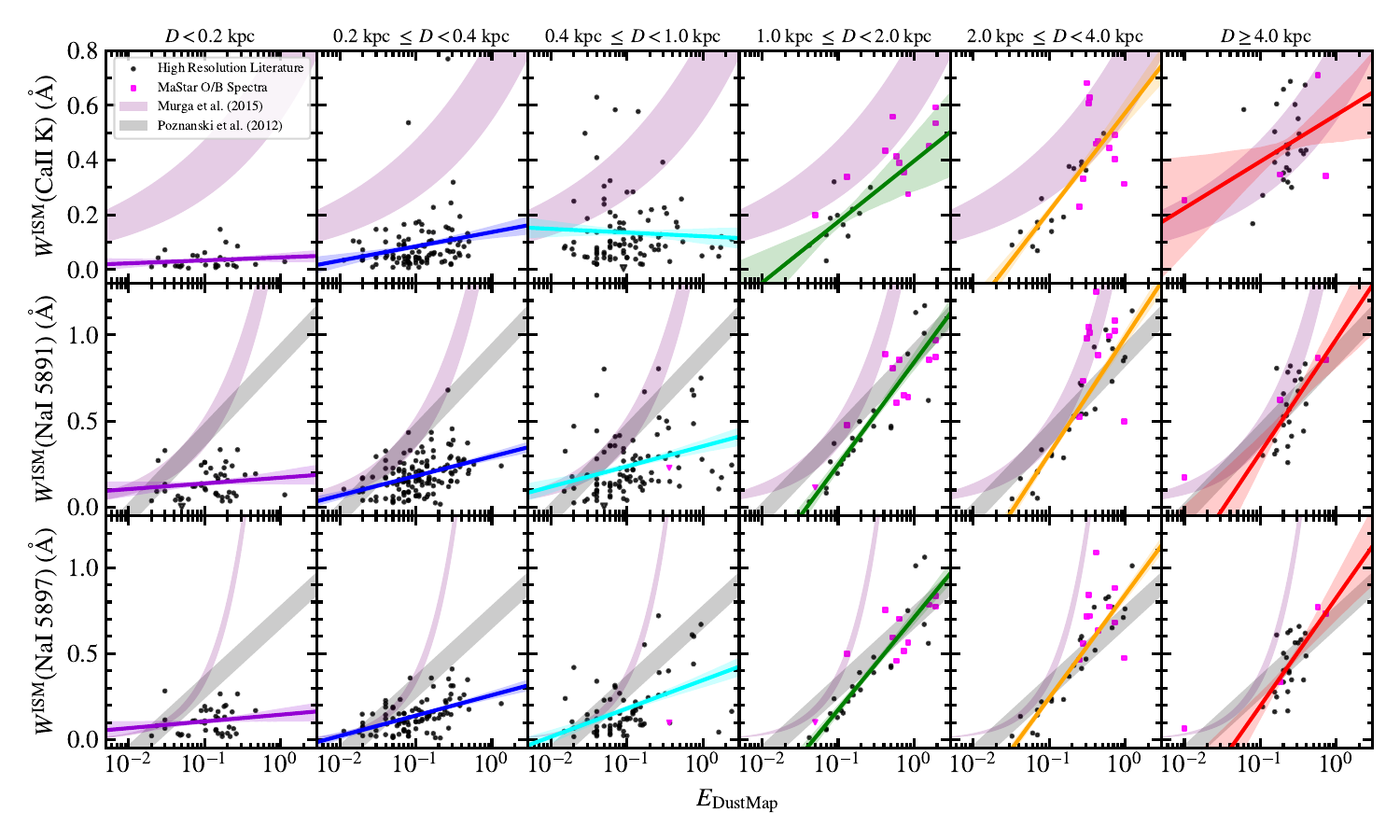}
\caption{Equivalent widths of interstellar \ion{Ca}{2} K (top row), \ion{Na}{1} 5891 (middle row), and \ion{Na}{1} 5897 (bottom row) vs.\ $E_{\rm DustMap}$ reported from analysis of high-spectral-resolution stellar spectroscopy by \citet{Sembach1993}, \citet{MunariZwitter1997}, and \citet{Welsh2010}.  Detections and $2.5\sigma$ upper limits (reported by \citealt{Welsh2010}) are indicated with solid black circles and downward-pointing triangles, respectively.  
Each column shows measurements of stars with distances within the range specified above the top-most panel.  All panels exclude sightlines with $E_{\rm DustMap} < 0.01$.  Magenta squares and triangles indicate detections and upper limits measured from the high-quality MaStar O and B spectral sample.  Colored lines show best-fit linear relations between $W^{\rm ISM}$ and $\log E_{\rm DustMap}$ (see Section~\ref{subsec:wr-distance-model}), and the surrounding transparent contours indicate the inner $\pm34\%$ of the locus of fits drawn randomly from the PPDF of each model.  The $W$ ranges indicated by the gray and magenta contours correspond to the ranges in these quantities that are implied by the \citet{Poznanski2012} and \citet{Murga2015} relations and $\pm1\sigma$ uncertainties, respectively.
\label{fig:ew_EBV}}
\end{figure*}

\begin{figure*}[ht]
 \includegraphics[width=\textwidth,trim={1cm 0cm 1cm 0cm},clip]{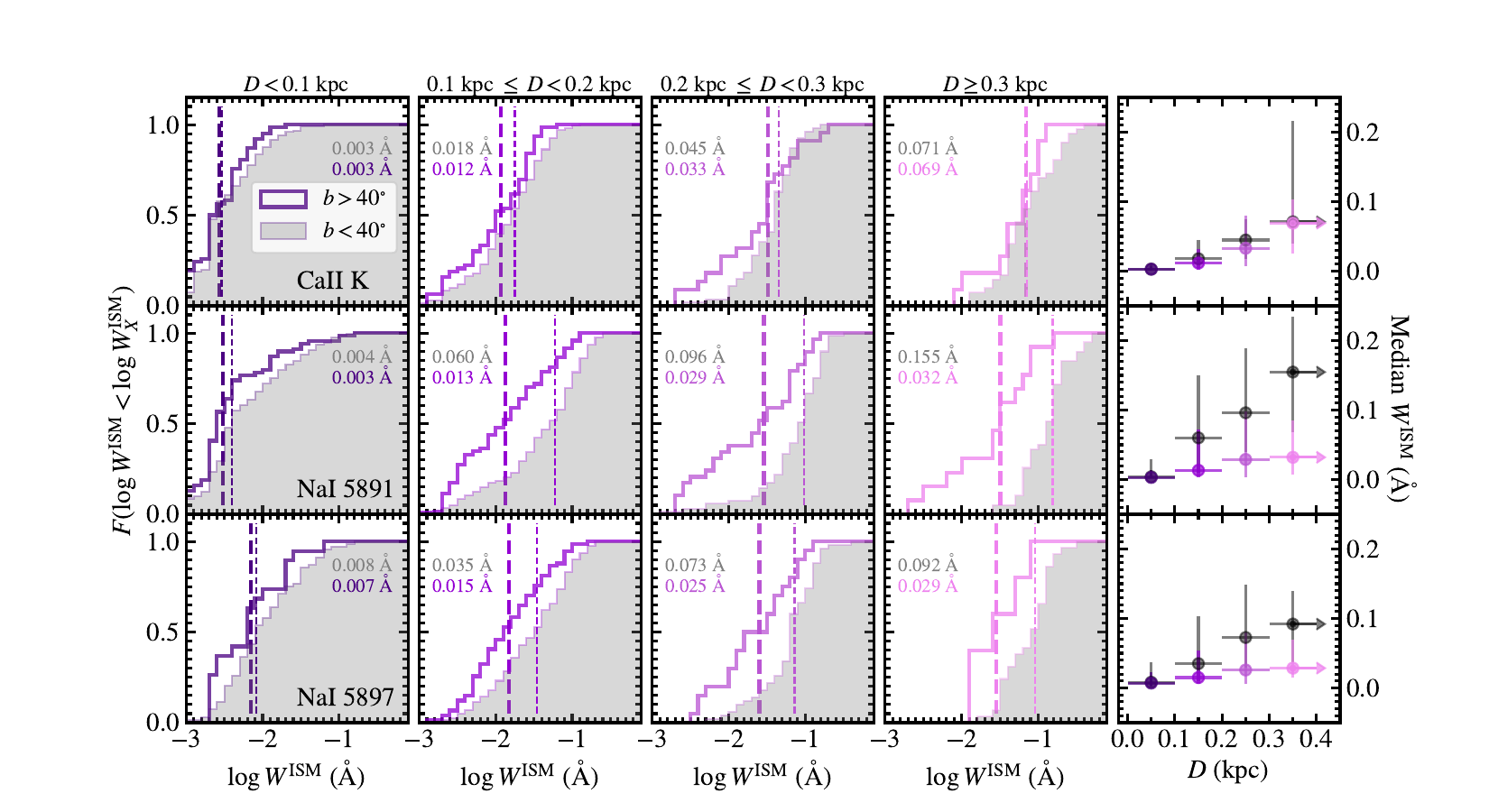}
\caption{Cumulative distributions of $\log W^{\rm ISM}$ for interstellar \ion{Ca}{2} K (top row), \ion{Na}{1} 5891 (middle row), and \ion{Na}{1} 5897 (bottom row) for sightlines having $E_{\rm DustMap} < 0.01$ reported from analysis of stellar spectroscopy by \citet{Sembach1993}, \citet{MunariZwitter1997}, and \citet{Welsh2010}.  Nondetections are included at the value of their $2.5\sigma$ or $3\sigma$ upper limits for \citet{Welsh2010} and \citet{Sembach1993} sightlines, respectively.  Each column includes measurements of stars with distances within the range specified above the top-most panel.  The open histograms with thick outlines include sightlines with high Galactic latitudes ($b > 40^{\circ}$), while the filled gray histograms include sightlines with $b < 40^{\circ}$.  The median values of these distributions are marked  with thick and thin dashed vertical lines, respectively, {and are printed on each panel in colored and gray text}.  The right-most column shows the median $W^{\rm ISM}$ values of each low-latitude (gray) and high-latitude (colored) distribution as a function of distance.  Error bars indicate the 16th- and 84th-percentile values of $W^{\rm ISM}$ for the corresponding subsample.
\label{fig:ew_lowEBV_dist}}
\end{figure*}

To explore the relationship between $W^{\rm ISM}$, stellar distance, and reddening in these samples, we first make use of the three-dimensional (3D) Milky Way dust map of \citet{Green2019} to estimate reddening values for the \citet{Welsh2010} stellar sightlines.  We use the Python package \texttt{dustmaps} \citep{Green2018} to perform our query.  The map returns a reddening $E_{\rm DustMap}$ in arbitrary units designed to be consistent with those used by \citet{SFD98}.   
Here we choose to adjust the $E(B-V)$ values reported in \citet{Sembach1993} and \citet{MunariZwitter1997} to the units of $E_{\rm DustMap}$ so that we may report on relationships between $W^{\rm ISM}$ and $E_{\rm DustMap}$.  

To do so, we refer to \citet{Lazarz2022}, who established the relation between $A_V$ and $E_{\rm DustMap}$ via an analysis comparing the latter to extinction values derived from stellar spectral template modeling of the MaStar sample.  Their relation, 

\begin{equation}\label{eq:av_edustmap}
    A_V = 3.31 E_{\rm DustMap} - 0.076, 
\end{equation}

\noindent is consistent with the relation between $A_V$ and $E(g-r)$ derived by \citet{Green2019}.  We compute the extinction ($A_V$) implied by the $E(B-V)$ values reported by \citet{Sembach1993} and \citet{MunariZwitter1997}, adopting $R_V = 3.1$ and 3.2, respectively.  We then solve Equation~\ref{eq:av_edustmap} for $E_{\rm DustMap}$ and apply the resulting relation to these two samples.

We then divide all of these literature samples (including that of \citealt{Sembach1993}, \citealt{MunariZwitter1997}, and \citealt{Welsh2010}) into several distance bins, and show the relationship between $W^{\rm ISM}$ and $E_{\rm DustMap}$ within each bin in Figure~\ref{fig:ew_EBV} (black filled circles and triangles).  
There are also numerous sightlines with $E_{\rm DustMap} < 0.01$ which we exclude from this figure (many of which have $E_{\rm DustMap}=0$).  Because these sightlines exhibit a broad range of $W^{\rm ISM}$ values that generally fall below $0.1$ \AA, we treat them separately and show their cumulative distributions in two bins of Galactic latitude in Figure~\ref{fig:ew_lowEBV_dist}. 
For completeness, Figure~\ref{fig:ew_EBV} also includes the $W^{\rm ISM}$ values measured from the high-quality MaStar O and B stellar spectra as described in Section~\ref{subsec:MaStar_Wr} (magenta squares).

Comparing the distribution of points between the columns of Figure~\ref{fig:ew_EBV}, we note that these quantities appear correlated, and that the slope of the relation between them steepens with increasing distance.  We also show the relations between $W^{\rm ISM}$ and $E(B-V)$ reported by \citet[][for all transitions]{Murga2015} and \citet[][for the \ion{Na}{1} transitions]{Poznanski2012}  for extragalactic sightlines with transparent magenta and gray contours.\footnote{Both of these works used reddening values obtained from the \citet{SFD98} dust maps ($E(B-V)_{\rm SFD}$), which have similar units to $E_{\rm DustMap}$.  }  Most measurements lie below these relations at stellar distances $D < 1$ kpc, while at $D > 1$ kpc, the \ion{Na}{1} equivalent widths shown here appear broadly consistent with the \citet{Poznanski2012} relations.  The \ion{Ca}{2} measurements shown lie below the fitted \citet{Murga2015} relation at $D<4$ kpc, but come into accord with it at $D>4$ kpc.  
We have also examined these relationships by first dividing the sample into bins in $E_{\rm DustMap}$, and plotting $W^{\rm ISM}$ vs.\ $\log D$.  We find that while $W^{\rm ISM}$ and $\log D$ are indeed correlated at $E_{\rm DustMap} > 0.05$, the $W^{\rm ISM}$ vs.\ $E_{\rm DustMap}$ relations exhibit significantly less scatter.

To understand the relationships shown in Figure~\ref{fig:ew_EBV} quantitatively, we follow \citet{Chen2010a} and \citet{Rubin2018a} to compute the likelihood function for the model 

\begin{equation}
    W^{\rm ISM} = \beta + \alpha \log E_{\rm DustMap}
\end{equation}

\noindent for each transition (\ion{Ca}{2} K, \ion{Na}{1} 5891, and \ion{Na}{1} 5897), with slope $\alpha$ and intercept $\beta$.  All securely-measured $W^{\rm ISM}$ values contribute $\chi^2/2$ to the logarithm of the likelihood.  For non-detections, we integrate the Gaussian contribution to the likelihood from $-\infty$ to the value of the limit (see \citealt{Rubin2018a} for details).  We also assume that the Gaussian variance of each measurement about this model ($s_i^2$) has contributions from both the $W^{\rm ISM}$ measurement error ($\sigma_i$) and intrinsic scatter in the relation ($\sigma_{\rm intr}$), such that $s_i^2 = \sigma_i^2 + \sigma_{\rm intr}^2$.  We sample the posterior probability density function (PPDF) for this model using the affine-invariant ensemble Markov Chain Monte Carlo sampler as implemented in the Python software package \texttt{emcee} \citep{Foreman-Mackey2013}.  We adopt the noninformative priors $-1.4 < \phi = \arctan \alpha < 1.4$, $-10 < \beta_{\perp} = \beta \cos(\phi) < 10$, and $-10 < \ln \sigma_{\rm intr} < 10$ {\citep{Robert2009,VanderPlas2014}}.  We sample this 3D parameter space with 100 ``walkers", each of which takes 5000 steps in total, and the first 1000 of which are discarded.  We adopt the median and $\pm34$th-percentile values of marginalized versions of the resulting PPDF as the best value of each parameter and its uncertainty.

The results of this modeling are shown in Figure~\ref{fig:ew_EBV}.  The solid colored lines indicate the best-fit model for the data shown in the corresponding panel.  The transparent contours around each best-fit relation were obtained by drawing 1000 sets of parameters at random from the PPDF, calculating the $W^{\rm ISM}$ implied by those parameters at each point along the $x$-axis, and then filling in the region containing the inner $\pm34$th-percentile values of that $W^{\rm ISM}$ distribution.  Our modeling indicates that the slope of the $W^{\rm ISM} - \log E_{\rm DustMap}$ relation tends to increase with increasing distance.  This steepening is significant for all transitions shown.  We list the best-fit parameters of each of these models in Table~\ref{tab.ewEBVdistfits}. 

Turning our attention to the equivalent widths measured along very low $E_{\rm DustMap}$ sightlines shown in Figure~\ref{fig:ew_lowEBV_dist}, we note that these cumulative distributions tend to shift to higher $\log W^{\rm ISM}$ values with increasing stellar distance.  We also find that while the $W^{\rm ISM}$ distribution of \ion{Ca}{2} K absorbers does not appear to vary with Galactic latitude, the subsamples of \ion{Na}{1} absorbers close to the disk plane (at $b < 40^{\circ}$) exhibit higher median $\log W^{\rm ISM}$ values than those at higher latitudes by $0.3-0.5$ dex  at $D>0.1$ kpc.  {This is consistent with the survival of \ion{Ca}{2} in the lower-density, higher-temperature gas (${\sim}1000-$10,000 K) above the disk plane \citep[][]{Puspitarini2012}, as well as with
the findings of, e.g., \citet{Sembach1994} and \citet{Welsh2010}, who measure larger exponential scale heights for 
the Milky Way's interstellar \ion{Ca}{2} than for \ion{Na}{1}.}

\begin{deluxetable}{llccc}
\tablewidth{700pt}
\tabletypesize{\footnotesize}
\tablecaption{Best-fit Parameters for $W^{\rm ISM} - \log E_{\rm DustMap}$ Relations\label{tab.ewEBVdistfits}}
\tablehead{
\colhead{Transition} & \colhead{Distances} & \colhead{$\alpha$} & \colhead{$\beta$} & \colhead{$\sigma_\mathrm{intr}$} \\
\colhead{} & \colhead{(kpc)} & \colhead{$(\rm \AA~(\log mag)^{-1})$} & \colhead{($\rm \AA$)} & \colhead{($\rm \AA$)}
}
\startdata
 \ion{Ca}{2} K & [0.0, 0.2] & $0.01\pm0.01$ & $0.04\pm0.02$ & $0.03\pm0.00$ \\
             & [0.2, 0.4] & $0.05\pm0.02$ & $0.14\pm0.02$ & $0.09\pm0.01$ \\
             & [0.4, 1.0] & $-0.01\pm0.03$ & $0.12\pm0.03$ & $0.12\pm0.01$ \\
             & [1.0, 2.0] & $0.22\pm0.11$ & $0.40\pm0.11$ & $0.08\pm0.02$ \\
             & [2.0, 4.0] & $0.36\pm0.05$ & $0.57\pm0.05$ & $0.06\pm0.01$ \\
             & [4.0, 20.0] & $0.17\pm0.13$ & $0.57\pm0.09$ & $0.13\pm0.02$ \\
 \ion{Na}{1} 5891 & [0.0, 0.2] & $0.03\pm0.04$ & $0.17\pm0.04$ & $0.09\pm0.01$ \\
             & [0.2, 0.4] & $0.11\pm0.02$ & $0.29\pm0.02$ & $0.10\pm0.01$ \\
             & [0.4, 1.0] & $0.12\pm0.04$ & $0.35\pm0.04$ & $0.17\pm0.01$ \\
             & [1.0, 2.0] & $0.60\pm0.06$ & $0.84\pm0.05$ & $0.13^{+0.03}_{-0.02}$ \\
             & [2.0, 4.0] & $0.68\pm0.05$ & $0.99\pm0.04$ & $0.11\pm0.02$ \\
             & [4.0, 20.0] & $0.66^{+0.25}_{-0.24}$ & $0.98\pm0.16$ & $0.17^{+0.03}_{-0.02}$ \\
 \ion{Na}{1} 5897 & [0.0, 0.2] & $0.04\pm0.04$ & $0.15\pm0.04$ & $0.07\pm0.01$ \\
             & [0.2, 0.4] & $0.12\pm0.02$ & $0.26\pm0.02$ & $0.08\pm0.01$ \\
             & [0.4, 1.0] & $0.16\pm0.04$ & $0.35\pm0.04$ & $0.14\pm0.01$ \\
             & [1.0, 2.0] & $0.54\pm0.05$ & $0.71\pm0.04$ & $0.11\pm0.02$ \\
             & [2.0, 4.0] & $0.60\pm0.04$ & $0.84\pm0.03$ & $0.09^{+0.02}_{-0.01}$ \\
             & [4.0, 20.0] & $0.63\pm0.20$ & $0.83\pm0.13$ & $0.13\pm0.02$ \\
\enddata
\end{deluxetable}

\begin{figure}[ht]
 \includegraphics[width=0.5\textwidth]{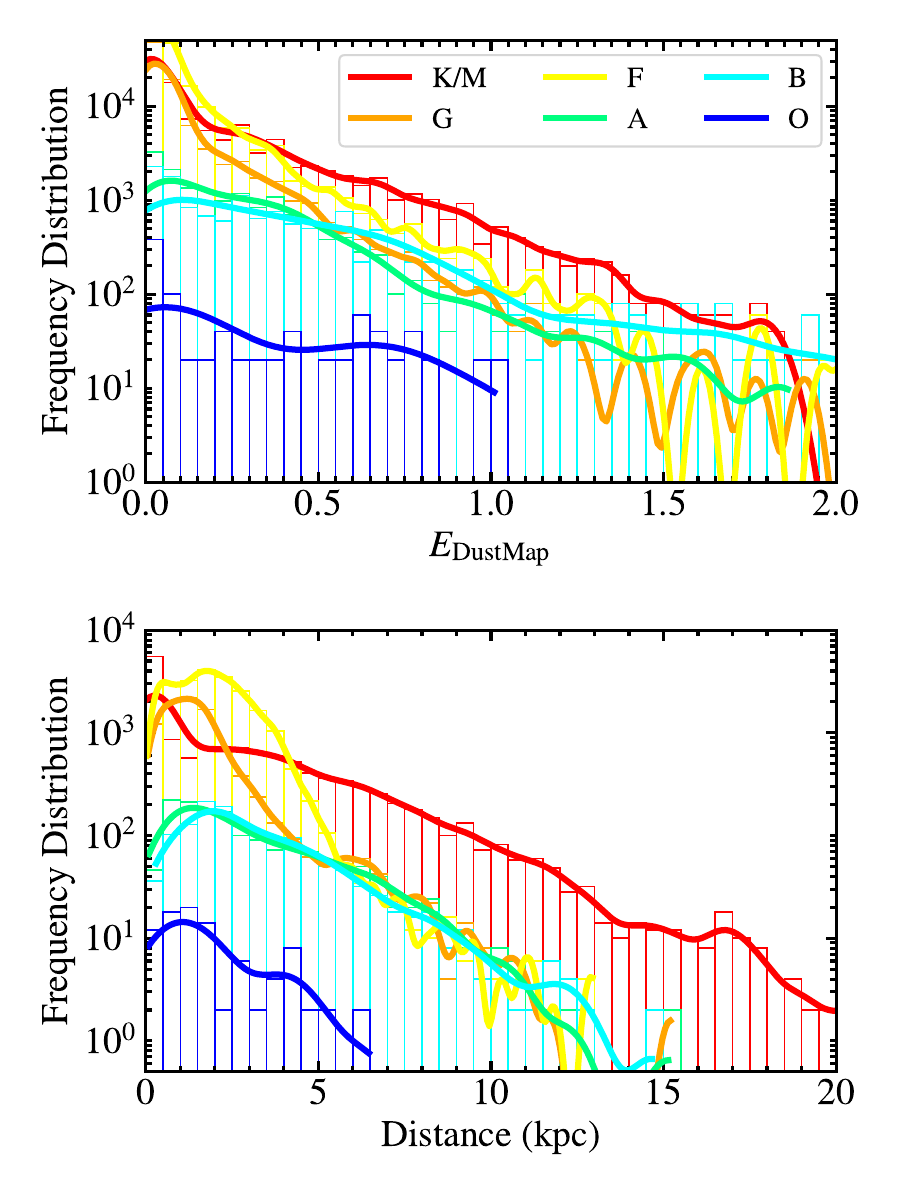}
\caption{{\it Top:} Frequency distributions of $E_{\rm DustMap}$ values for 23,991 MaStar targets having $E_{\rm DustMap}\ge0$, binned by spectral type as indicated in the legend.  The smooth curves show continuous probability density curves corresponding to each histogram.
{\it Bottom:} Frequency distributions of stellar distance for the same sample. \label{fig:hist_dist_ebv}}
\end{figure}

\begin{figure*}[ht]
 \includegraphics[width=\textwidth,trim={0 0.5cm 0 1cm},clip]{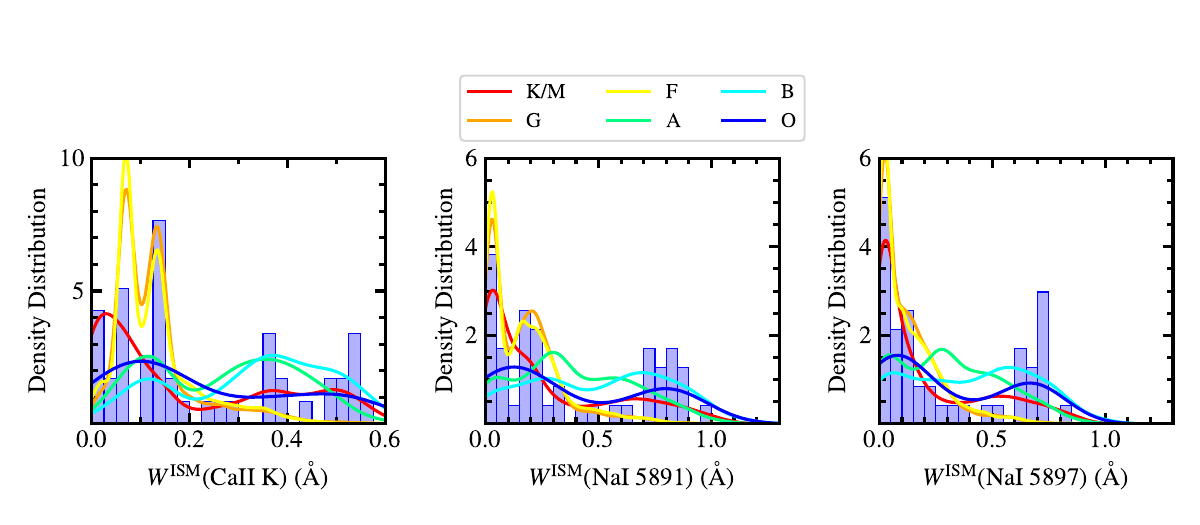}
\caption{The distribution of equivalent widths arising from interstellar absorption for the MaStar stellar sample, as predicted by the modeling described in Section~\ref{subsec:wr-distance-model}.  Distributions of $W^{\rm ISM}$(\ion{Ca}{2} K), $W^{\rm ISM}$(\ion{Na}{1} 5891), and $W^{\rm ISM}$(\ion{Na}{1} 5897) values are shown in the left, middle, and right panels, respectively.  We show the continuous probability density curves that represent these distributions for all spectral types, and include the corresponding histograms for O-type stars in blue.  \label{fig:hist_corr_ews}}
\end{figure*}

\subsection{Implied ISM Absorption Strengths for the MaStar Sample}

We now consider the implications of these statistics and models for the stellar sightlines in the MaStar sample.  We have found a strong dependence of the absorption strength of interstellar material on both stellar distance and dust reddening.  We thus begin by exploring the distributions of these latter quantities as a function of spectral type in the full sample of MaStar targets with high-quality stellar parameter constraints (including 24,162 stars).  We adopt $T_{\rm eff}$ ranges for each spectral type as listed in Table~\ref{tab.spectraltypes} and established for main-sequence stars by \citet[][see their Table 8]{HabetsHeintze1981}.  Figure~\ref{fig:hist_dist_ebv} shows the frequency distributions of $E_{\rm DustMap}$ values, estimated from the \citet{Green2019} dust maps as described above (top panel), for each of these spectral type categories.  The bottom panel shows the corresponding frequency distributions of stellar distances, for which we adopt the Gaia Early Data Release 3 photogeometric distance estimates of \citet{Bailer-Jones2021}.  Comparing these distributions among different spectral types, we see that, e.g., the O and B stars have higher median $E_{\rm DustMap}$ values than do the F or G stars in this sample.  These differences will be reflected in the distributions of ISM absorption strengths predicted by our modeling.

\begin{deluxetable}{lcc}
\tablewidth{700pt}
\tabletypesize{\footnotesize}
\tablecaption{Adopted Spectral Type $T_{\rm eff}$ Ranges\label{tab.spectraltypes}}
\tablehead{
\colhead{Spectral Type} & \colhead{minimum $T_{\rm eff}$} &
\colhead{maximum $T_{\rm eff}$}
}
\startdata
O & 29,900 K & \nodata \\
B & 9700 K & 29,900 K \\
A & 7610 K & 9700 K \\
F & 5950 K & 7610 K \\
G & 5200 K & 5950 K \\
K/M & \nodata & 5200 K \\
\enddata
\end{deluxetable}

We then use these $E_{\rm DustMap}$ and stellar distance values (as well as the Galactic latitude for a subset of the stars), in combination with the median $W^{\rm ISM}$ values and best-fit models generated as described above, to compute the implied absorption strength ($W^{\rm ISM}$) along each stellar sightline.  
In detail, for those sightlines having $E_{\rm DustMap}<0.01$, we simply adopt the median $W^{\rm ISM}$ value of the appropriate literature subsample (drawing from those shown in Figure~\ref{fig:ew_lowEBV_dist}).  For sightlines with larger amounts of reddening, we apply the appropriate best-fit model to calculate $W^{\rm ISM}$.
The distributions of these values for each transition (again separated by spectral type) are shown in Figure~\ref{fig:hist_corr_ews}. Here, we see that the $W^{\rm ISM}$(\ion{Na}{1}) values for O and B stars are roughly evenly distributed over the range $0-1$ \AA, whereas K/M spectral types have median $W^{\rm ISM}$(\ion{Na}{1}) values of 0.15 \AA\ and 0.09 \AA\ for the 5891 \AA\ and 5897 \AA\ transitions, respectively.

\begin{figure*}[ht]
 \includegraphics[width=\textwidth,trim={0.8cm 0.8cm 0.8cm 0.8cm},clip]{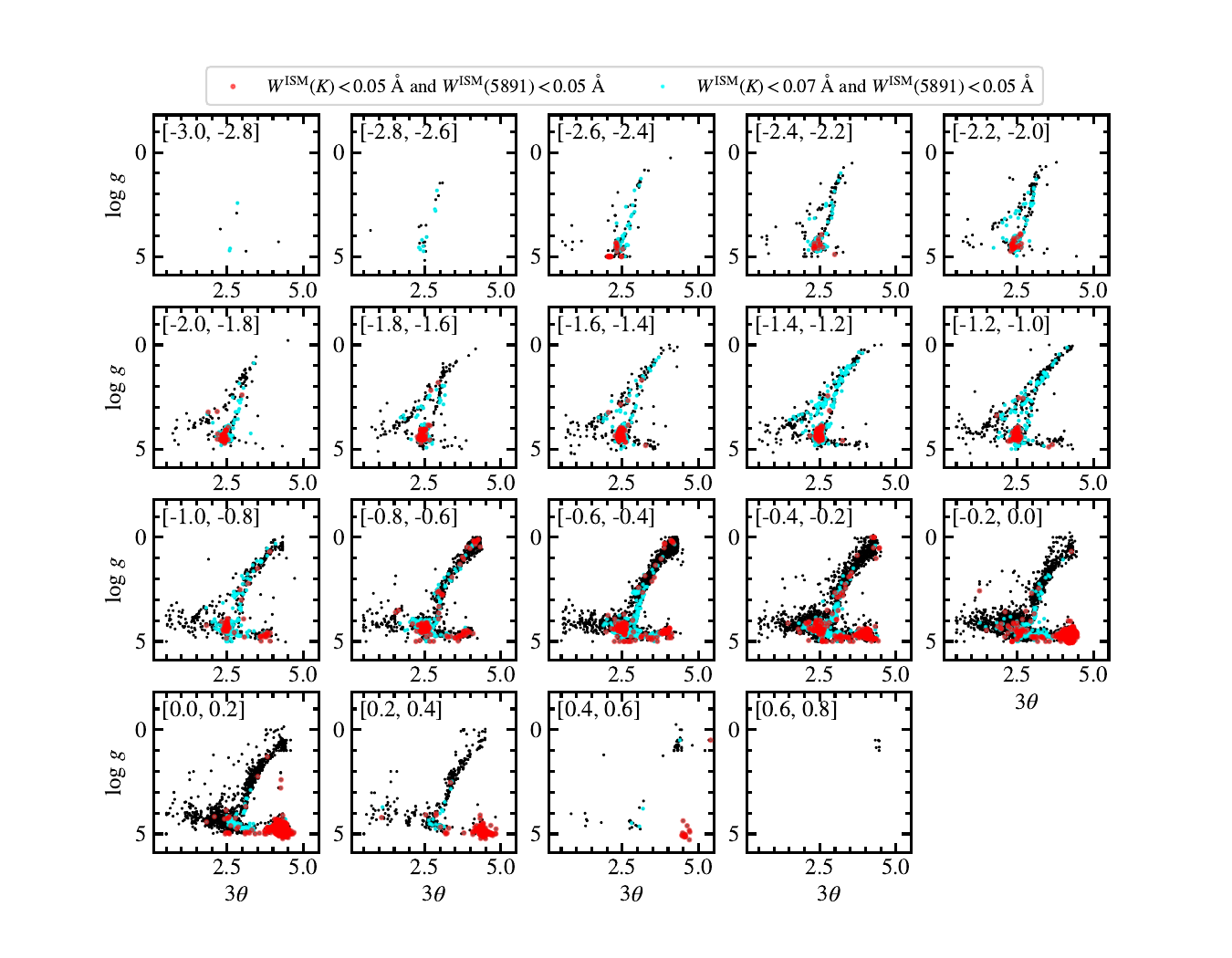}
\caption{Stellar parameter distributions for the ${\approx}$24,000 unique stars in the MaStar sample (black).  Each panel shows objects having [Fe/H] values between those listed at the top left.  Red points indicate those stars with predicted ISM absorption strengths $W^{\rm ISM}$(\ion{Ca}{2} K) $<0.05$ \AA\ and $W^{\rm ISM}($\ion{Na}{1} $5891) < 0.05$ \AA.  The combination of cyan and red points shows stars with $W^{\rm ISM}($\ion{Na}{1} $5891) < 0.05$ \AA\ and $W^{\rm ISM}$(\ion{Ca}{2} K) $<0.07$ \AA.
The latter set of stars samples much of this parameter space well; however, they are rare among hot stars ($3\theta < 2$), and are sparse along the red giant branch at solar and supersolar metallicities. \label{fig:thetaZlogg_NaIcut}}
\end{figure*}

In preparation for our effort to reduce the impact of interstellar absorption on the MaStar library, we also consider the detailed distribution of those stars with very low levels of predicted $W^{\rm ISM}$ in stellar parameter space.  We make use here of the parameter $\theta = 5040~\mathrm{K} / T_{\rm eff, med}$, and will in most cases be considering the quantity $3\theta$, as it has a very similar dynamic range to that of both surface gravity ($\log g$) and [Fe/H] for this sample. 
In Figure~\ref{fig:thetaZlogg_NaIcut}, we show the $\log g$ vs.\ $3\theta$ distribution of the full MaStar sample (including all stars in the stellar parameter catalog \texttt{mastar-goodstars-v3\_1\_1-v1\_7\_7-params-v2.fits} with valid parameter entries), divided into 19 bins in the calibrated median [Fe/H].\footnote{[Fe/H] values are calibrated to those derived by the APOGEE Stellar Parameters and Abundances Pipeline (J.\ Holtzman et al.\ {\it in preparation}) for stars with complementary APOGEE observations as described in \url{https://www.sdss4.org/dr17/mastar/mastar-stellar-parameters/}.}  Those stars with predicted $W^{\rm ISM}$(\ion{Ca}{2} K) $<0.05$ \AA\ and $W^{\rm ISM}$(\ion{Na}{1} 5891) $<0.05$ \AA, and which also have well-constrained stellar parameters (with $\sigma_\theta/\theta < 0.1$, $\sigma_{\log g} < 1$, and $\sigma_{\rm [Fe/H]} < 0.5$), are highlighted in red.  This subset includes only 2084 stars and fails to sample much of the $\log g - 3\theta - \rm [Fe/H]$ parameter space occupied by the full MaStar sample (and is instead strongly dominated by cool, high-metallicity dwarfs).  To improve our sampling of parameter space, we relax our selection criteria to include stars with slightly stronger predicted \ion{Ca}{2} K absorption:
$W^{\rm ISM}$(\ion{Ca}{2} K) $<0.07$ \AA\ and $W^{\rm ISM}$(\ion{Na}{1} 5891) $<0.05$ \AA.  This latter sample is indicated by the combination of cyan and red points,   and includes a total of 6408 sightlines.  This larger subset covers much of the desired $\log g - 3\theta - \rm [Fe/H]$ parameter space, and we therefore choose to adopt these spectra as ``low-ISM'' sightlines.

Under the assumption that the intrinsic \ion{Ca}{2} and \ion{Na}{1} profiles of stars with similar stellar parameter values will likely also be very similar on average, we can draw on these ``low-ISM'' spectra to predict the shape of these intrinsic profiles for much of the remainder of the sample.  
This approach is similar in spirit to a technique that has been successfully adopted in studies aiming to extract the absorption strength of diffuse interstellar bands from high-resolution stellar spectroscopy \citep[e.g.,][]{Kos2013,Gaia2023,Vogrincic2023}; however, here we rely solely on stellar parameters for our intrinsic profile prediction, rather than a minimization of spectral residual differences \citep{Kos2013}.  
We justify this choice in the following section, and then describe our approach to implementing our intrinsic profile prediction.

\section{Correcting \ion{Ca}{2} and \ion{Na}{1} Absorption Profiles with ISM Contamination} \label{sec:correction}

\subsection{The Intrinsic Dispersion of \ion{Ca}{2} and \ion{Na}{1} D Absorption Profiles in Stars with Similar Stellar Parameters }\label{subsec:intrinsic_dispersion}

\begin{figure*}[ht]
 \includegraphics[width=\textwidth]{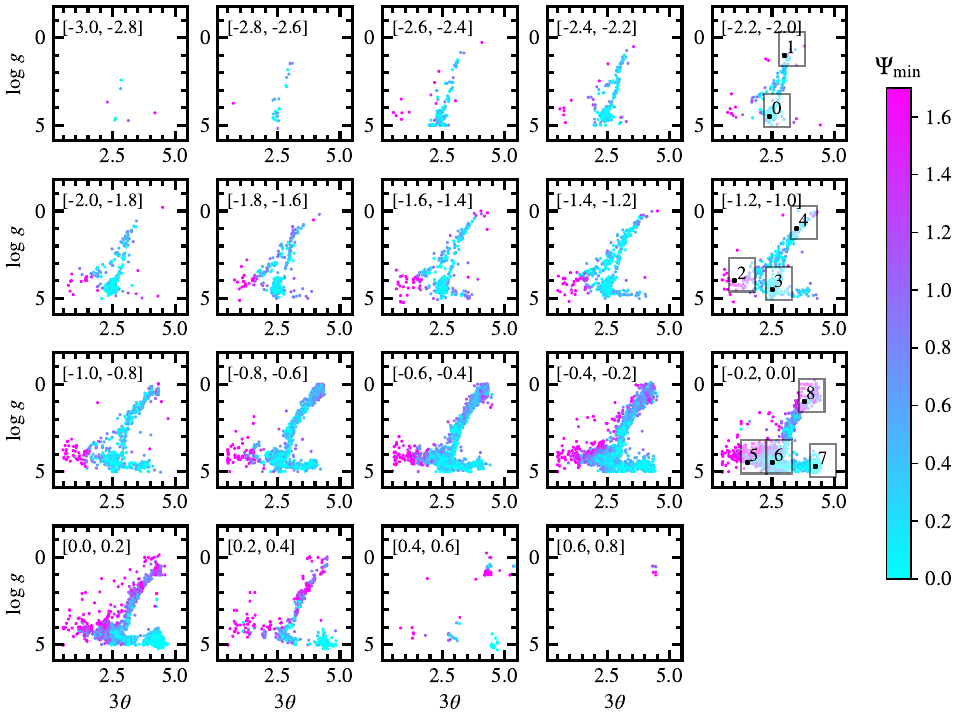}
\caption{Stellar parameter distributions for the full MaStar sample.  Each panel shows objects having [Fe/H] values between those listed at the top left.  Stars are color-coded by $\Psi_{\rm min}$, a measure of their distance in this parameter space from the nearest low-ISM sightline.  The black squares labeled with integers in the right-most column indicate the locations in parameter space chosen to investigate the intrinsic scatter in \ion{Ca}{2} and \ion{Na}{1} D profiles as described in Section~\ref{subsec:intrinsic_dispersion}.  \label{fig:psimin_NaI}}
\end{figure*}

To establish subsamples of stars that have sufficiently similar stellar parameters such that we may consider their \ion{Ca}{2} K and \ion{Na}{1} D profiles interchangeable, we calculate the 3D distance between every star ($i$) and all stars in the low-ISM sample ($j$) as follows:

\begin{eqnarray}
\nonumber
\Psi(i,j) = \Biggl\{\left (\frac{\log g_i - \log g_j}{\sigma_{\log g_{i,j}}} \right )^2
+
\left (\frac{\theta_i - \theta_j}{\sigma_{\theta_{i,j}}} \right )^2\\
\nonumber
+ 
\left (\frac{{\rm [Fe/H]}_i - {\rm [Fe/H]}_j}{\sigma_{{\rm [Fe/H]}_{i,j}}} \right )^2
\Biggr\}^{1/2},
\end{eqnarray}

\noindent 
with the values of $\sigma_{\log g_{i,j}}$, $\sigma_{\theta_{i,j}}$, and $\sigma_{{\rm [Fe/H]}_{i,j}}$ equal to the corresponding uncertainties on the parameters of the two stars ($i,j$) added in quadrature.  For each star $i$, we find the minimum value of this distance across all $j$ sightlines, $\Psi_{\rm min}(i)$.  In Figure~\ref{fig:psimin_NaI}, we show the same distributions of stellar parameters as in Figure~\ref{fig:thetaZlogg_NaIcut}, and have color-coded each star by this latter value.  Among the 24,162 stars shown here, $55\%$ have $\Psi_{\rm min} < 0.2$, while just $8\%$ (1819) have $\Psi_{\rm min} > 1.0$.  This latter subset tends to either exhibit high effective temperatures (with $3\theta < 2)$, or to be located toward the tip of the red giant branch.  The maximum value of $\Psi_{\rm min}$ among our sample is 9.4.  This analysis indicates that it is important to assess the scatter in the intrinsic \ion{Ca}{2} and \ion{Na}{1} D line profiles of stars having 3D parameter distances in the range $0 < \Psi \lesssim 2$ in order to determine whether our low-ISM sightlines may be used to predict these profiles across the full MaStar sample.

To make this assessment, we first choose nine locations in this parameter space that are approximately representative of the parameter ranges exhibited by the full sample.  These locations are indicated with black squares in Figure~\ref{fig:psimin_NaI}.  We then identify all low-ISM stars having parameters that fall within a distance $\Psi<\Psi_{\rm thresh}$ of each location, where we set $\Psi_{\rm thresh} = 0.2, 0.4, 0.6, 0.8, 1.0, 1.2$, and 1.5 in successive iterations of this exercise.

\begin{figure*}[ht]
 \includegraphics[width=\textwidth,trim={0 1cm 0 1cm},clip]{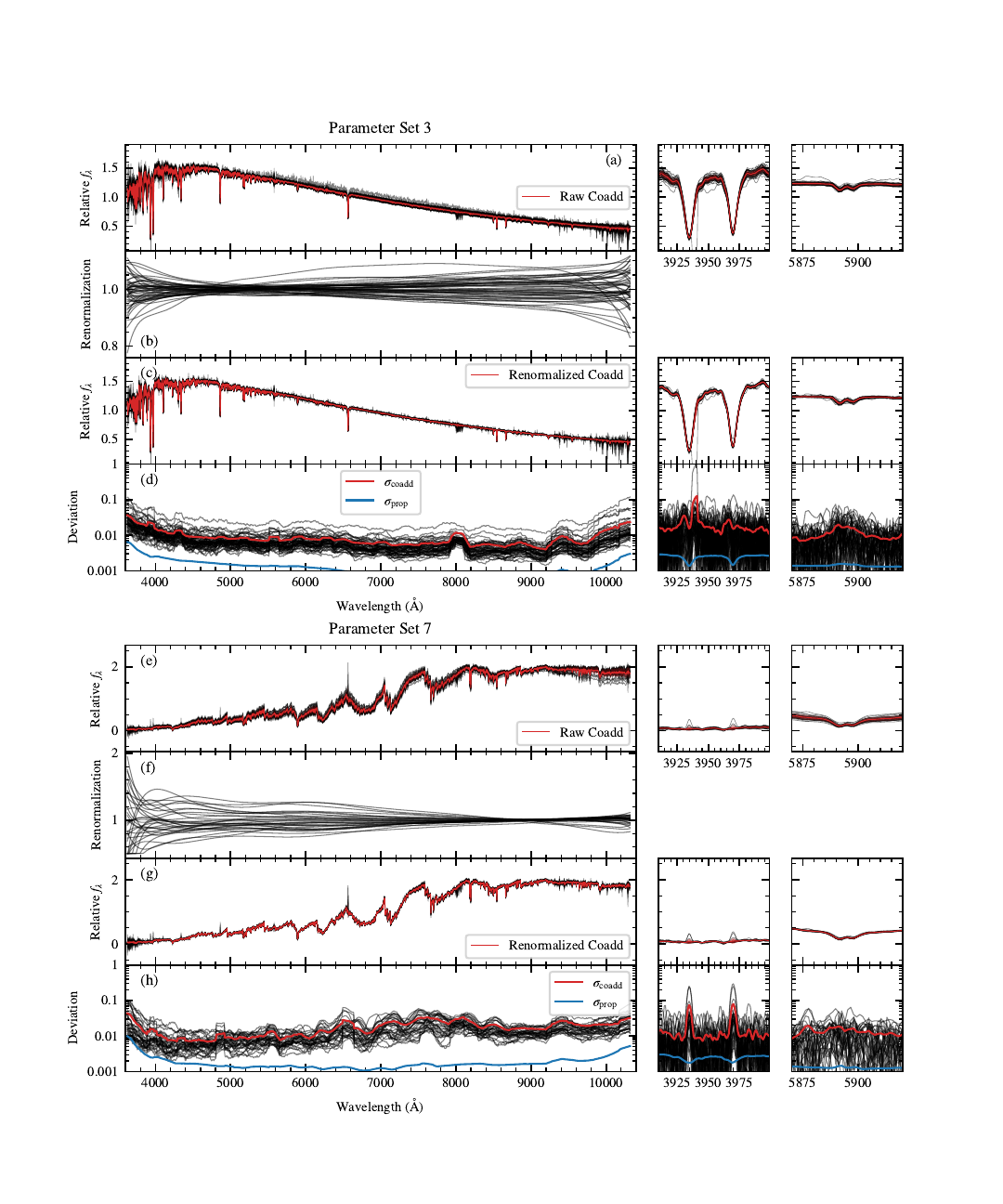}
\caption{Demonstration of our spectral coadding algorithm for stars selected within $\Psi_{\rm thresh}=0.4$ of the parameter locations labeled `3' (rows {\it (a)-(d)}) and `7' (rows {\it (e)-(h)}) in Figure~\ref{fig:psimin_NaI}.  The right-hand columns show the \ion{Ca}{2} and \ion{Na}{1} D regions of the spectra shown at left.  \emph{(a,e)} Spectra of all stars selected for this coadd, normalized by their mean flux values, are shown in black.  The average of these spectra, weighted by their median S/N, is shown in red.  \emph{(b,f)}  Fourteenth-order Legendre polynomial fits to the ratio of each normalized spectrum to the coadd shown above.  \emph{(c,g)}  Same as panels \emph{(a,e)} for individual spectra renormalized by their Legendre polynomial fits and then coadded in the same manner. \emph{(d,h)}  The absolute deviation of each renormalized spectrum from the final coadd is shown in black.  The uncertainty in the final coadd calculated via propagation of the individual spectral error arrays is shown in blue, and the total dispersion in the renormalized flux values from the final coadd is shown in red.  \label{fig:stack_demo}}
\end{figure*}

Then, for each location and each value of $\Psi_{\rm thresh}$, we compute the average of all selected low-ISM spectra.  For this analysis, and for all analyses to follow, we use the MaStar spectra which have been smoothed to a uniform line spread function (LSF) representative of the 99.5th percentile of the native LSFs across the survey.
 To coadd our spectral subsamples, we use the following approach:
\begin{enumerate}
    \item We first normalize each spectrum by its mean value, and then compute their median S/N-weighted average as a function of wavelength (shown in panels \emph{(a)} and \emph{(e)} in Figure~\ref{fig:stack_demo} for subsamples `3' and `7' constructed using $\Psi_{\rm thresh} = 0.4$).
    \item To reduce dispersion in the coadded sample due to slight differences in overall continuum shape, we divide each normalized spectrum by this coadd, and fit fourteenth-order Legendre polynomials to the results.  These polynomial models are shown in panels \emph{(b)} and \emph{(f)} of Figure~\ref{fig:stack_demo}.  Any objects with model values $>2$ or $<0.5$ are considered outliers and are removed from the subsample.
    \item We renormalize each spectrum by its fitted Legendre polynomial, and re-construct their average as described above (shown in panels \emph{(c)} and \emph{(g)} in Figure~\ref{fig:stack_demo}).
\end{enumerate}

We additionally compute the uncertainty in this final coadd via propagation of the errors in the individual spectra ($\sigma_{\rm prop}$), shown in blue in panels \emph{(d)} and \emph{(h)} of Figure~\ref{fig:stack_demo}, as well as the median S/N-weighted standard deviation of the flux across all spectra ($\sigma_{\rm coadd}$), shown in red in the same panels.

We then measure the equivalent widths and associated uncertainties of the \ion{Ca}{2} K and \ion{Na}{1} D features in the resulting sample of coadds.  To do so, we choose relatively feature-free spectral regions on either side of each profile in order to establish the continuum level (in the ranges $3920.0~\mathrm{\AA} < \lambda < 3925.0~\mathrm{\AA}$ and $3945.0~\mathrm{\AA} < \lambda < 3950.0~\mathrm{\AA}$ around \ion{Ca}{2} K; and in the ranges $5881.0~\mathrm{\AA} < \lambda < 5885.0~\mathrm{\AA}$ and $5904.0~\mathrm{\AA} < \lambda < 5908.0~\mathrm{\AA}$ around \ion{Na}{1} D).  We fit a linear model to the flux across these pixels, and use this model to locally continuum-normalize these spectral regions.  We then compute a boxcar equivalent width in the spectral window $3925.0~\mathrm{\AA} < \lambda < 3944.0~\mathrm{\AA}$ to measure the strength of \ion{Ca}{2} K ($W_{\rm coadd}$(\ion{Ca}{2} K)), and in the spectral window $5885.0~\mathrm{\AA}< \lambda < 5904.0~\mathrm{\AA}$ to measure the total strength of both transitions in the \ion{Na}{1} D doublet ($W_{\rm coadd}$(\ion{Na}{1} D)).  We likewise compute the uncertainty in these line strengths arising from the deviation in the flux profiles across each subsample ($\sigma_{W, \rm coadd}$, calculated from $\sigma_{\rm coadd}$).

\begin{figure*}[ht]
 \includegraphics[width=\columnwidth,trim={0cm 0cm 0.6cm 0cm},clip]{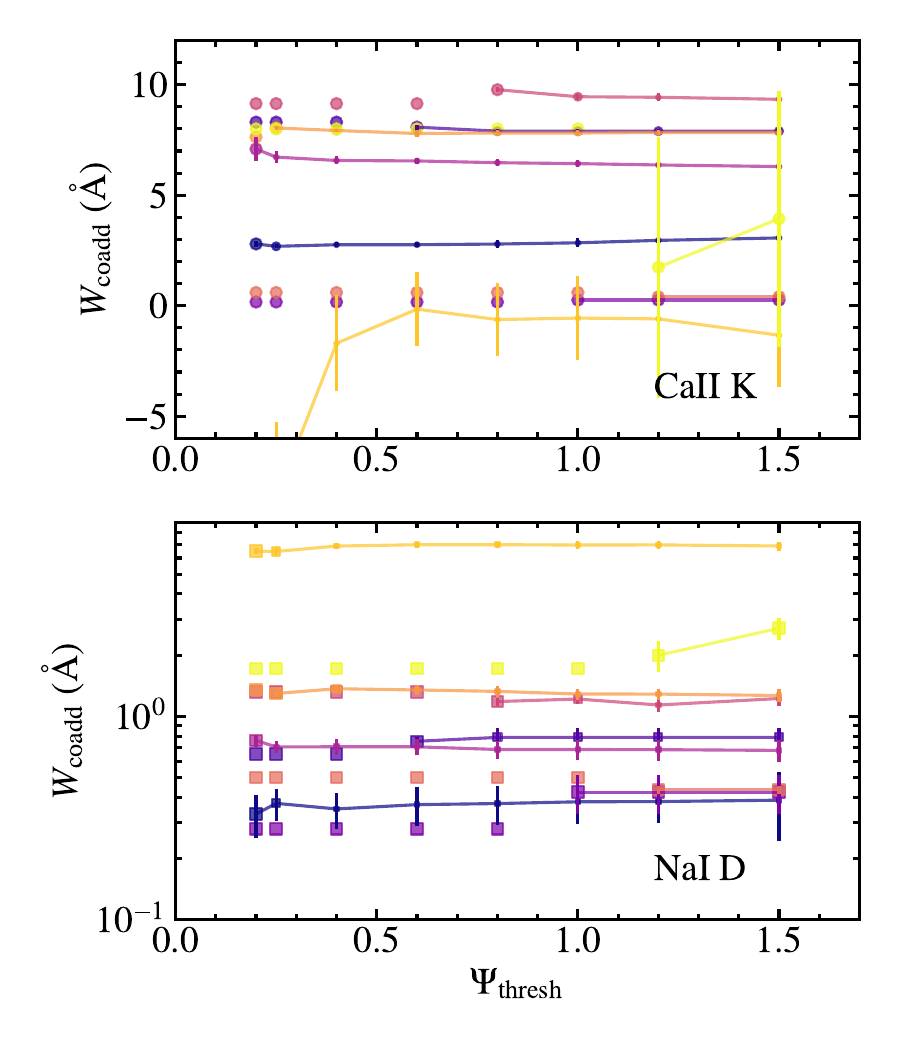}
 \includegraphics[width=\columnwidth,trim={0cm 0cm 0.6cm 0cm},clip]{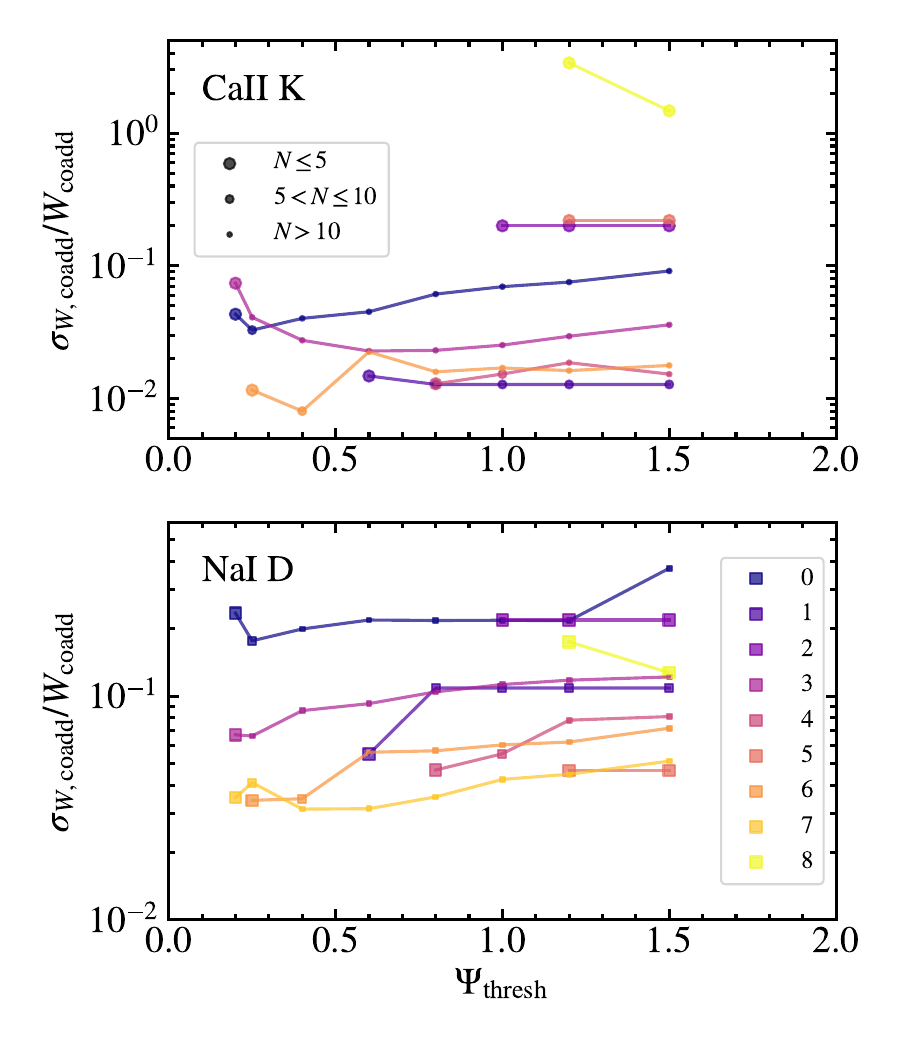}
\caption{\emph{Left panels:}  The equivalent width of the \ion{Ca}{2} K (top) and \ion{Na}{1} D (bottom) transitions in the spectral coadds described in Section~\ref{subsec:intrinsic_dispersion} as a function of $\Psi_{\rm thresh}$ (with $\Psi$ representing the 3D parameter space distance from the parameter space location of each coadd).  Points are color-coded according to their location in parameter space as indicated in the legend at bottom right.  The size of each point corresponds to the number of spectra in the coadd, as indicated by the legend in the upper right panel.  Symbols that are not connected by lines represent the equivalent width measured in the individual spectrum closest to the corresponding parameter space location.  Error bars indicate the uncertainty in $W_{\rm coadd}$ ($\sigma_{W,\rm coadd}$) due to the deviation in the flux levels across the subsample.
\emph{Right panels:} The ratio $\sigma_{W,\rm coadd} / W_{\rm coadd}$ for \ion{Ca}{2} K (top) and \ion{Na}{1} D (bottom).  Points are color-coded and sized as in the left-hand panels.  The dispersion in flux levels across these absorption profiles gives rise to a $\lesssim 10-20\%$ uncertainty in their equivalent widths in the vast majority of the coadds included here. \label{fig:assess_stacks}}
\end{figure*}

We show these results in Figure~\ref{fig:assess_stacks}.  The left-hand panels show the $W_{\rm coadd}$ measured in each spectral coadd as a function of $\Psi_{\rm thresh}$.  The point sizes indicate the number of spectra included in the coadd (see legend at upper right).  Symbols that are not connected by a colored line represent the equivalent width measured in the individual spectrum of the star closest to the appropriate parameter space location, and are included only in cases in which there are no other stars meeting the $\Psi < \Psi_{\rm thresh}$ criterion (such that we cannot construct a coadd).  Error bars indicate $\sigma_{W, \rm coadd}$.

Broadly speaking, these equivalent width values do not change significantly with increasing $\Psi_{\rm thresh}$.  The $W_{\rm coadd}$(\ion{Ca}{2} K) values of all but one of the coadds of sets 0, 2, 3, 4, 5, 6, and 8 differ from the corresponding equivalent width value shown at $\Psi_{\rm thresh} = 0.2$ by $\le2.2\sigma$. 
The spectra of the solar-metallicity cool dwarf stars in parameter set 7 exhibit only very weak emission in this spectral region and lack well-defined \ion{Ca}{2} profiles.
The coadds of parameter set 1 exhibit \ion{Ca}{2} K equivalent width differences of $0.2-0.4$ \AA\ (at $1.9\sigma-4.0\sigma$ significance) relative to that of the closest individual spectrum; however, the equivalent widths of the five coadds shown are consistent with each other within $<1.2\sigma$.

\ion{Na}{1} D exhibits similar behavior: the equivalent width values of sets 0, 1, 2, 3, 4, 5, 6 and 7 are all within $2.4\sigma$ of the equivalent width shown at $\Psi_{\rm thresh}=0.2$.  Set 8 exhibits modestly more significant differences of $0.8\sigma-2.9\sigma$.  However, we note that the spectra of stars in parameter set 8 lack a well-defined \ion{Na}{1} D profile and are dominated by molecular absorption bands. 

The right-hand panels of Figure~\ref{fig:assess_stacks} show the ratio $\sigma_{W,\rm coadd}/W_{\rm coadd}$ for the \ion{Ca}{2} K (top) and \ion{Na}{1} D (bottom) profiles in each of these spectral stacks.  Here, we have not included equivalent width measurements of individual spectra, and also exclude negative equivalent width measurements (which arise for \ion{Ca}{2} K in set 7).
In general, we see that this ratio tends to increase weakly with increasing $\Psi_{\rm thresh}$, but that it remains $\lesssim0.2$ in most cases. The coadds in set 8 are exceptions to these rules: in particular, the equivalent width of \ion{Ca}{2} K exhibits a scatter well above $>1$ \AA\ at $\Psi_{\rm thresh} > 1.0$, implying significant inconsistency in these line profiles.

In summary, we find that the coadded spectra of stars with stellar parameters that differ by $\Psi_{\rm thresh}$ values between $0.2$ and $1.0$ exhibit statistically consistent \ion{Ca}{2} K and \ion{Na}{1} D equivalent widths.  The dispersion in these equivalent width values ($\sigma_{W, \rm coadd}$) arising from differences in absorption profile shapes within each coadd is $<20\%$ in most subsamples. Moreover, this dispersion is typically minimized at lower values of $\Psi_{\rm thresh}$. These findings imply that we may use the \ion{Ca}{2} and \ion{Na}{1} D profiles of our low-ISM sample to replace those in stars with greater predicted foreground ISM absorption in cases in which we are able to identify low-ISM sightlines that are sufficiently close in stellar parameter space.

\subsection{Empirical Spectral Replacement}\label{subsec:def_psi}

We now use the findings laid out above to identify subsamples of low-ISM stellar sightlines that may be coadded to replace the relevant spectral regions for every star with $W^{\rm ISM}$(\ion{Ca}{2} K) $>0.07$ \AA\ or $W^{\rm ISM}$(\ion{Na}{1} 5891) $>0.05$ \AA\ in the MaStar sample.  For each such star, we first search for all low-ISM sightlines having $\Psi<0.2$.  If there are 10 or more low-ISM stars meeting this criterion, we consider this our ``final'' replacement subsample.  If there are fewer than 10 stars, we increase $\Psi_{\rm thresh}$ in increments of 0.1 until either (1) the subsample includes at least 10 stars or (2) $\Psi_{\rm thresh}=0.6$.  If the $\Psi_{\rm thresh}=0.6$  subsample includes at least five stars, we consider them our ``final'' replacement subsample.  If it includes fewer than five stars, we continue to increase $\Psi_{\rm thresh}$ in increments of 0.1 until (1) the subsample includes at least five stars or (2) $\Psi_{\rm thresh}=1.0$.  If there is a minimum of one star within $\Psi < \Psi_{\rm thresh} = 1.0$, we use that sample as our ``final'' replacement subsample.  If there are no stars within $\Psi < 1.0$, then we do not attempt a spectral replacement.  
Among the 23,771 stars with high-quality parameter values and available spectra, there are 6342 low-ISM sightlines, and 15,661 stars for which we were able to construct a replacement subsample in the manner described above.  This leaves 1768 stars with potentially high levels of ISM contamination that cannot be removed via this method.
This latter sample corresponds to those stars shown in magenta in Figure~\ref{fig:psimin_NaI}.

For each of the 15,661 stars for which ``empirical replacement'' is possible (referred to below as ``primary'' stars),  we first coadd the spectra of all stars in the corresponding replacement subsample in the same manner described in Section~\ref{subsec:intrinsic_dispersion}.  We then normalize the spectrum of the primary star by its mean value.  We divide this normalized spectrum by the initial spectral coadd (described in Step 2 above), and fit a fourteenth-order Legendre polynomial to the resulting spectral ratio.  We then renormalize the primary spectrum by this model so that its continuum level is well-matched to that of the final coadd.  We perform the complementary manipulations of the inverse variance, first multiplying it by the square of the mean value of the primary spectrum, and then by the square of the Legendre polynomial fit.

\begin{figure*}[ht]
 \includegraphics[width=\textwidth]{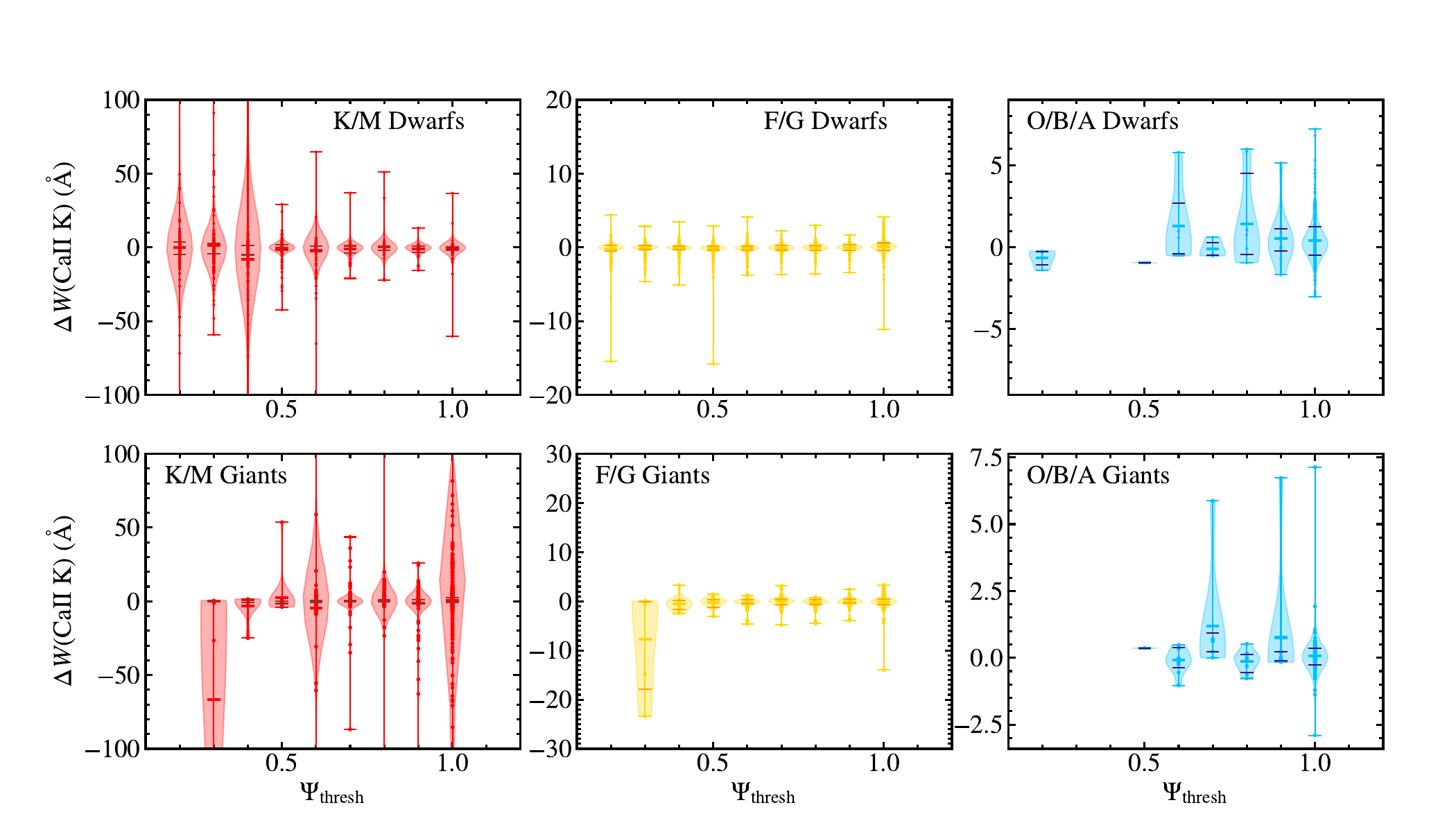}
\caption{The distributions of the change in the \ion{Ca}{2} K  equivalent width in spectra ``cleaned'' with a replacement coadd relative to the original stellar spectra ($\Delta W = W_{\rm orig} - W_{\rm coadd}$) as a function of $\Psi_{\rm thresh}$.  Only sightlines that are not considered ``low-ISM'' and for which we can construct replacement coadds are included.  The mean value of each distribution is indicated with a {matching red, yellow, or light blue} horizontal bar, and the minimum and maximum values correspond to the extremes for each subsample. {The widths of the filled contours scale with the frequency of the data along the $y$-axis.} 
{The 16th- and 84th-percentile values are shown with horizontal bars of a complementary color.}
 Stars have been grouped according to their location in $T_{\rm eff}-\log g$ space, with dwarf and giant stars having $\log g > 4$ and $\log g < 4$, respectively.  \label{fig:violinplots_CaIIK}}
\end{figure*}

\begin{figure*}[ht]
 \includegraphics[width=\textwidth]{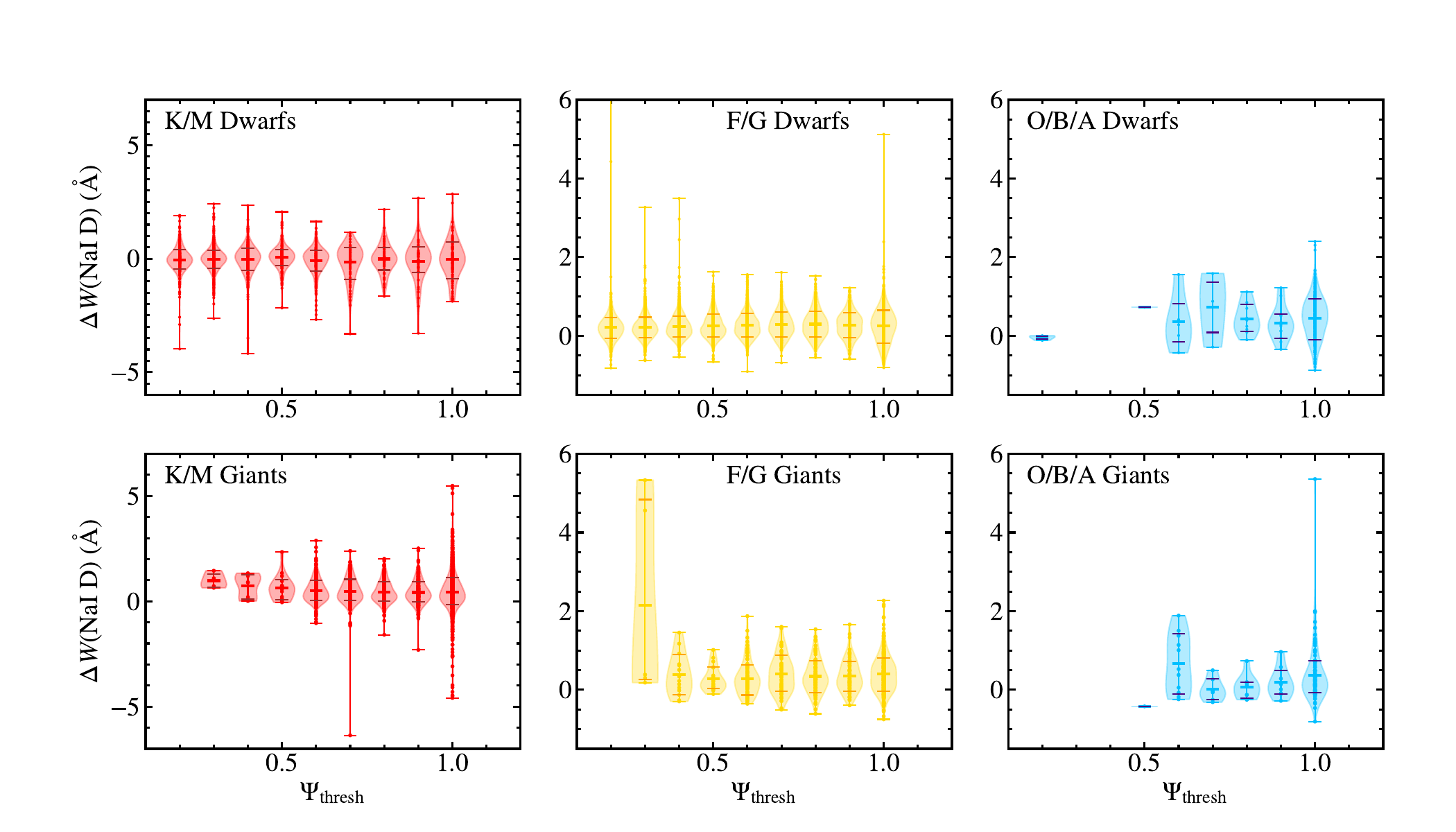}
\caption{Same as Figure~\ref{fig:violinplots_CaIIK}, for \ion{Na}{1} D.  \label{fig:violinplots_NaID}}
\end{figure*}

\begin{figure*}[ht]
 \includegraphics[width=\textwidth]{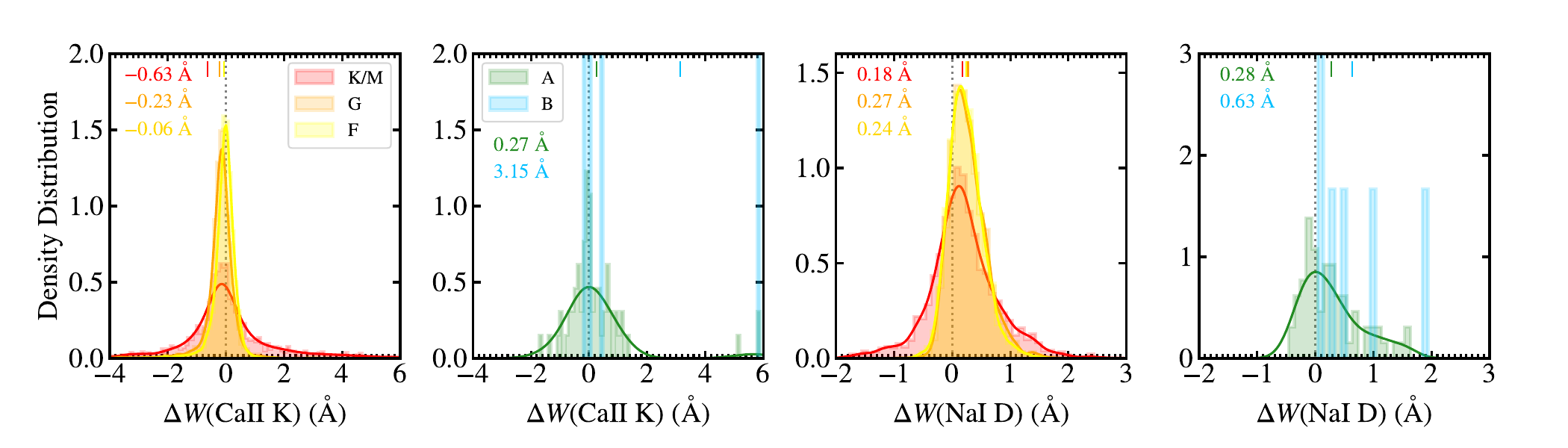}
\caption{The distributions of the change in \ion{Ca}{2} K (left-hand panels) and \ion{Na}{1} D (right-hand panels) equivalent widths in spectra ``cleaned'' with a replacement coadd relative to the original stellar spectra ($\Delta W = W_{\rm orig} - W_{\rm coadd}$).  Only sightlines that are not considered ``low-ISM'' and for which we can construct replacement coadds with $\Psi_{\rm thresh} < 1.0$ are included.  Distributions are color-coded by spectral type as indicated in the legend.  The smooth curves show continuous probability density distributions corresponding to each histogram.  
{The average value of each distribution is printed and indicated toward the top of each panel with a bar of the corresponding color.}
\label{fig:change_in_ews}}
\end{figure*}

We then replace the spectral regions $3912.0~\mathrm{\AA} < \lambda < 3995.0~\mathrm{\AA}$ and $5873.0~\mathrm{\AA} < \lambda < 5917.0~\mathrm{\AA}$ in the primary spectrum with the same regions in the coadd.  These windows are chosen to encompass both \ion{Ca}{2} H \& K and \ion{Na}{1} D in stars in which these profiles are very broad.  To avoid sharp features at the edges of these regions, we adopt a weighted sum of the primary and coadded spectra. The primary spectrum is weighted by the amplitude of the sum of two Gaussian functions ($G(\lambda)$): one Gaussian is centered at the left edge of the spectral window, and the other is centered at the right edge.  Both Gaussians are assigned dispersions $\sigma = 3$ \AA, such that $G(\lambda) = 0$ across much of the window. The coadded spectrum is weighted by $1-G(\lambda)$.  

After performing this replacement, we multiply the entire primary spectrum by its Legendre polynomial fit, as well as by its mean value, and have verified that this procedure returns all pixels outside of the \ion{Ca}{2} and \ion{Na}{1} D spectral windows to their original flux levels.  The renormalized variance of the primary star spectrum is likewise replaced with a sum of the square of the weighted $\sigma_{\rm coadd}$ array and the variance array itself.  The resulting inverse variance is then divided by the square of the Legendre polynomial fit, and again by the square of the mean value of the primary spectrum.  

In the case that the replacement subsample includes only one star, we simply adopt its flux and inverse variance (normalized by its mean value) as that used to replace the relevant spectral regions in the primary star.  We again apply the Gaussian tapering function described above to smooth the transition from one spectrum to the other.  We ensure the stars have comparable continuum shapes by fitting a Legendre polynomial to their ratio, and renormalizing the primary spectrum by this model (as described above).

We assess the results of this procedure by first measuring the equivalent widths of the relevant transitions in the original spectra ($W_{\rm orig}$) and in the spectra that have been ``cleaned'' via replacement with a coadd ($W_{\rm coadd}$), using the same spectral windows and approach to continuum fitting as described in Section~\ref{subsec:intrinsic_dispersion}.  We then compute the difference $\Delta W = W_{\rm orig} - W_{\rm coadd}$ for each star.  In Figure~\ref{fig:violinplots_CaIIK}, we show the distributions of these values for the \ion{Ca}{2} K transition as a function of $\Psi_{\rm thresh}$, excluding those stars in the ``low-ISM'' subsample.  Each panel includes stars in a different region of $\log g - T_{\rm eff}$ space: the top three panels include dwarf stars with $\log g > 4$ at  cool ($T_{\rm eff,med}<5200$ K), medium (F/G; $5200~\mathrm{K} < T_{\rm eff,med} < 7610~\mathrm{K}$), and hot (O/B/A; $T_{\rm eff,med} > 7610$ K) temperatures, and the bottom three panels include giant stars with $\log g < 4$ in the same temperature categories.  We find that the $\Delta W$(\ion{Ca}{2} K) distributions for cool dwarfs and giants are very broad even at $\Psi_{\rm thresh} < 0.5$, with numerous catastrophic failures resulting in $\Delta W$(\ion{Ca}{2} K) values of $\approx \pm 50-100$ \AA.  Medium and hot stars overall have significantly more narrow $\Delta W$(\ion{Ca}{2} K) distributions, but nevertheless include a handful of spectra with $|\Delta W$(\ion{Ca}{2} K)$|$ $> 3$ \AA, which is unlikely to arise from ISM contamination.  We also note that the averages of these distributions (indicated by the {red, yellow, and light blue} horizontal dashes) are not systematically $> 0$ \AA\ for either cool- or medium-temperature stars.

Figure~\ref{fig:violinplots_NaID} shows the same set of distributions for the \ion{Na}{1} D doublet.  These distributions all have much smaller dispersions than the analogous \ion{Ca}{2} K distributions, and for medium-temperature and hot stars, have mean values $> 0$ \AA.  We also note that the incidence of catastrophic failures (with $\Delta W$(\ion{Na}{1} D) $< -1$ \AA) increases sharply at $\Psi_{\rm thresh} = 1$ in several of these subsamples.

We next show the distributions of these values as a function of the spectral type in Figure~\ref{fig:change_in_ews}.  
Here again, only those stars for which a replacement coadd may be constructed and that are not ``low-ISM'' sightlines are included.  We have also excluded all sightlines for which coadds were constructed adopting a $\Psi_{\rm thresh}=1.0$ to reduce the incidence of catastrophic failures in our empirical replacement sample.  The second and fourth panels from the left in this figure do not include histograms for O-type stars, as there are no such stars that satisfy all of these criteria.

Even having excluded those coadds with $\Psi_{\rm thresh} = 1.0$, the mean $\Delta W$(\ion{Ca}{2} K) for our K/M and G samples are $-0.6$ \AA\ and $-0.2$ \AA, and is very close to zero  for F types.  
The A and B spectral type distributions have means of $0.3$ \AA\ and $3.2$ \AA, respectively.  We have also looked closely at sightlines that are outliers in the A- and B-star distributions at $\Delta W$(\ion{Ca}{2} K) $ > 4~\mathrm{\AA}$,
for which the original spectrum exhibits a much stronger K transition relative to that of its replacement coadd than could arise from ISM contamination.  We posit that this scenario arises because \ion{Ca}{2} H \& K are particularly sensitive to temperature at $T_{\rm eff} \approx $ 10,000 K (see Figure~\ref{fig:maraston_ews}), with the K transition decreasing in strength from $\approx 2.0$ \AA\ to $0.2$ \AA\ over the $T_{\rm eff}$ range from 8000 K to 12,000 K.  Moreover, the median and maximum uncertainties on $T_{\rm eff, med}$ for stars in this temperature range are $316$ K and $7861$ K, respectively, meaning that the low-ISM stars satisfying our $\Psi_{\rm thresh}$ criteria may in some cases have temperatures spanning this range.

The distributions showing the change in the \ion{Na}{1} D absorption strength indicate that its equivalent width has been reduced in the vast majority of the cleaned spectra.  The mean value of $\Delta W$(\ion{Na}{1} D) falls in the range $0.2-0.3$ \AA\ for K/M, G, and F stars, and is 0.3 \AA\ and 0.6 \AA\ for A and B stars, respectively.  This trend is qualitatively consistent with the predictions shown in Figure~\ref{fig:hist_corr_ews}.  

\subsection{Spectral Replacement with Theoretical Stellar Templates}

We now consider an alternative approach to removing ISM contamination from these sightlines.  This is of particular importance for those stars that are isolated in parameter space, such that no replacement subsample could be identified.  
From examination of Figure~\ref{fig:psimin_NaI}, it is evident that many of these stars have high effective temperatures (i.e., $3\theta < 2$ with $T_{\rm eff,med} > 7560$ K). Such O, B, and A stars will dominate stellar continuum models of strongly star-forming galaxies, along with any extant absorption due to foreground Milky Way ISM.  Moreover, the analysis presented in Figure~\ref{fig:hist_corr_ews} demonstrates that these hot stars are likely subject to the strongest ISM contamination among all stars in the MaStar sample, as they tend to exhibit a high degree of reddening overall (see also Figure~\ref{fig:hist_dist_ebv}).

At the same time, theoretical modeling of the atmospheres of stars in this temperature range is not subject to complications arising from significant line blanketing and the interplay of molecular opacities \citep[e.g.,][]{BOSZ2017,Byrne2023}.  We therefore
choose to explore the intrinsic \ion{Ca}{2} and \ion{Na}{1} D profiles predicted from model stellar atmospheres for all stars having $T_{\rm eff,med} >$ 7,500 K. 

\citet{BOSZ2017} have computed and made public a large suite of stellar spectra predicted using the ATLAS-APOGEE ATLAS9 model atmosphere database \citep{Meszaros2012}.  The suite is generated from a grid of models covering the temperature range $3500~\mathrm{K} \le T_{\rm eff} \le $ 35,000 K, a metallicity range $\rm -2.5 \le [M/H] \le 0.75$, a range in $\alpha$ abundance of $\rm -0.25 \le [\alpha/M] \le 0.5$, and a range in carbon abundance of $\rm -0.75 \le [C/M] \le 0.25$.  The resulting spectra cover a wavelength range of $1000~\mathrm{\AA} < \lambda < 32\mu$m at a native resolution $\mathcal{R}=$ 300,000.  \citet{BOSZ2017} demonstrated that $\chi^2$ minimization of the residuals between 
this model spectroscopy and that 
of the CALSPEC standard stars yields low values in the range $\chi^2 = 0.1-3$, and yields best-fit $T_{\rm eff}$ values that are fully consistent with those based on the models of \citet{CastelliKurucz2003}.  

\begin{figure*}[ht]
 \includegraphics[width=\textwidth,trim={0cm 0.5cm 0cm 0.5cm},clip]{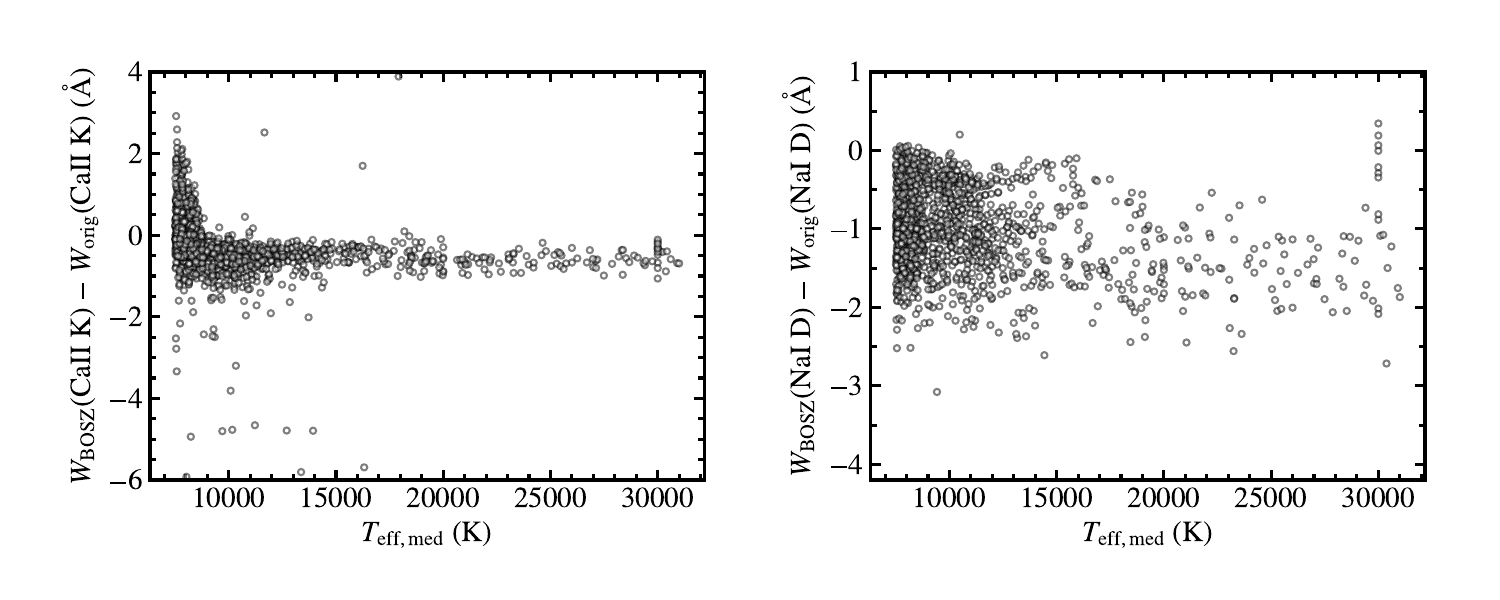}
 \includegraphics[width=\textwidth,trim={0cm 0.3cm 0cm 0.8cm},clip]{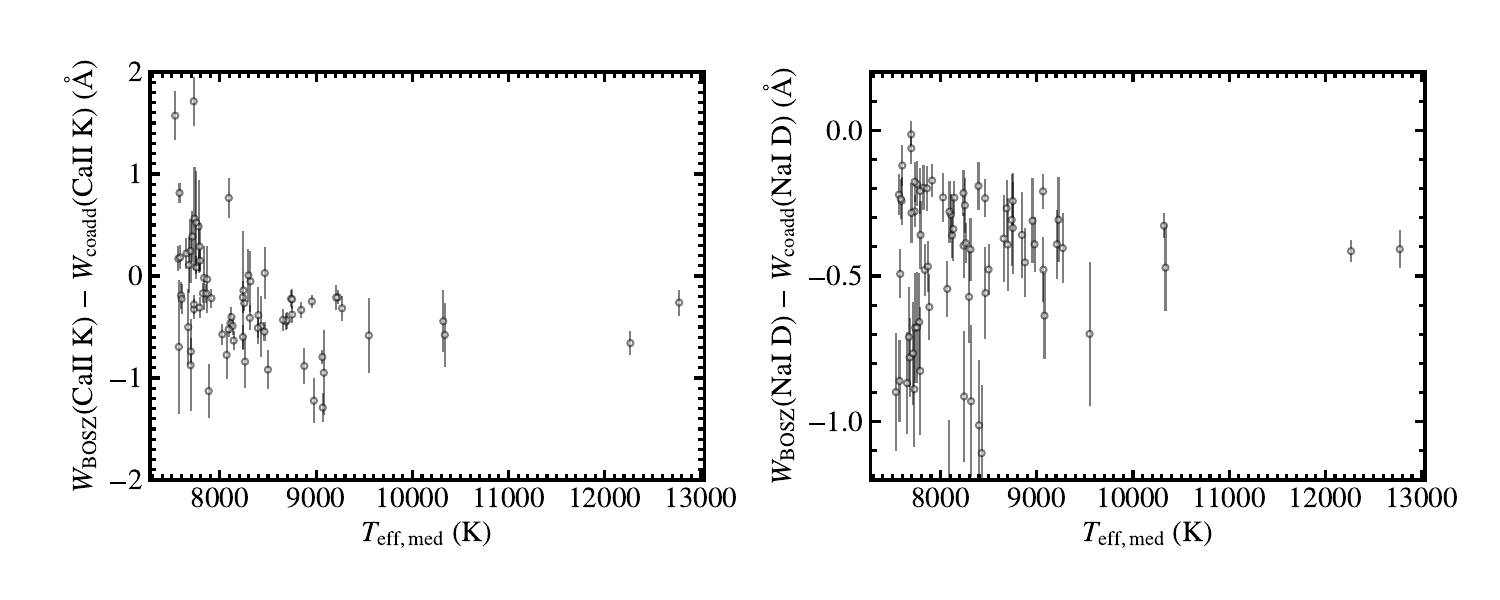}
\caption{\emph{Top panels:}  The difference in the $W$ measured for \ion{Ca}{2} K (left) and \ion{Na}{1} D (right) in the spectra ``cleaned'' with BOSZ theoretical templates and that measured in the original spectra, plotted vs.\ $T_{\rm eff, med}$. \emph{Bottom panels:} Same as above, for the difference in $W$ measured in the 
spectra cleaned with BOSZ templates and those measured in the corresponding empirical replacement coadds.  {The error bars represent the uncertainties on $W_{\rm coadd}$.}  \label{fig:theory_vs_emp}}
\end{figure*}

We retrieve the subset of these theoretical spectra that were generated from model atmospheres with $T_{\rm eff} \ge 7500$ K and that have been smoothed to a resolution $\mathcal{R}=2000$ from the MAST archive.\footnote{\url{https://archive.stsci.edu/prepds/bosz/}} 
The $T_{\rm eff}$ and $\log g$ sampling of this portion of the grid of models is variable, with the range $7500~\mathrm{K} \le T_{\rm eff} \le 8000~\mathrm{K}$ sampled in increments of $250$ K with surface gravities over the range $1 \le \log g \le 5$ in increments of 0.5.  The temperature range
$8250~\mathrm{K} \le T_{\rm eff} \le$ 12,000 K is sampled in increments of $250$ K with surface gravities over the range $2 \le \log g \le 5$ in increments of 0.5.  The temperature range 12,500 K $\le T_{\rm eff} \le$ 20,000 K is sampled in increments of $500$ K at $\log g = 3.0, 3.5, 4.0, 4.5$, and $5.0$, and the temperature range 21,000 K $\le T_{\rm eff} \le$ 30,000 K is sampled in increments of $1000$ K at $\log g = 4.0, 4.5$, and $5.0$.  Together, these comprise 246 grid points in $T_{\rm eff}$ and $\log g$.  We make use of models across the full range in metallicity and $\alpha$-abundance listed above, but consider only those with $\rm [C/M] = 0$.  With both the aforementioned metallicity and $\alpha$-abundance ranges sampled in increments of 0.25 dex, the nominal size of the grid of theoretical spectra we consider is $246 \times 13 \times 4 = 12792$.  In practice, there are additional models available from MAST that range in temperature up to 35,000 K, but that are not described in \citet{BOSZ2017}.  Considering these additions, we make use of 13,466 models in total.

To match each of the hot MaStar stars requiring ``cleaning'' with the appropriate model stellar spectrum, we use the APOGEE-calibrated values of $[Z/\rm H]_{\rm med}$ and $\rm [\alpha/Fe]_{\rm med}$ available from the MaStar stellar parameter catalog to compute $[\alpha/Z]_{\rm med} = \mathrm{[\alpha/Fe]_{med} + [Fe/H]_{med}} - [Z/\rm H]_{med}$.  For each star, we then compute the Euclidean distance between its values of $3\theta, \log g_{\rm med}$, $[Z/\rm H]_{med}$, and $[\alpha/Z]_{\rm med}$, and the values of $3\theta, \log g$, [M/H], and $\rm [\alpha/M]$ at each model grid point.  We select the model with the minimum value of this distance as our ``replacement'' spectrum.  

We use the function \texttt{match\_spectral\_resolution} available as part of the MaNGA Data Analysis Pipeline Python package\footnote{\url{https://sdss-mangadap.readthedocs.io/en/latest/}} \citep{Westfall2019} to smooth the model spectra to the observed spectral resolution, and then rebin the smoothed spectra to match the pixel sampling of the MaStar stars.  We continuum-normalize each model and corresponding observed spectrum in a similar manner to that described in Section~\ref{subsec:def_psi}, fitting a Legendre polynomial to their ratio, and then dividing the MaStar spectrum by the polynomial model.  We replace the spectral regions $3912.0~\mathrm{\AA} < \lambda < 3990.0~\mathrm{\AA}$ and $5873.0~\mathrm{\AA} < \lambda < 5917.0~\mathrm{\AA}$ in the data with the resampled, smoothed model spectrum, again weighting by a Gaussian with $\sigma=3$ \AA\ at the window edges to avoid introducing sharp spectral features.  We then reverse our normalization procedure in order to return all pixels outside of these replacement windows to their original flux values.  We do not adjust the values of the inverse variance array during this process.

We measure the strength of the \ion{Ca}{2} K and \ion{Na}{1} D features in the spectra cleaned using this method using the same continuum normalization procedure and spectral windows described in Section~\ref{subsec:intrinsic_dispersion}.  We show the offset between these line strengths and those measured from the original spectra in the top panels of Figure~\ref{fig:theory_vs_emp} as a function of $T_{\rm eff, med}$.  The \ion{Na}{1} D equivalent widths are nearly universally weaker in the BOSZ spectra across this temperature range, by up to ${\approx} 2.5$ \AA, consistent with the presence of significant interstellar \ion{Na}{1} in the original spectra.  The same is true for \ion{Ca}{2} K for stars at $T_{\rm eff, med} >$  9000 K; however, at lower effective temperatures, a significant fraction of the BOSZ matched templates have stronger \ion{Ca}{2} K absorption than the original spectrum.  This implies that either the BOSZ modeling fails to capture the behavior of this transition at temperatures below 9000 K, and/or that there are small offsets between the true effective temperature and the $T_{\rm eff, med}$ determined for some stars that result in a mismatch between the \ion{Ca}{2} absorption profile predicted in the matched template and the observed profile.

For comparison, we also show the offsets between the equivalent widths measured in the BOSZ-cleaned spectra and those measured in the corresponding empirical replacement coadds with $\Psi_{\rm thresh} < 1.0$ in the bottom panels of this figure. There are 73 stars shown in the top panels for which we could construct an empirical replacement coadd, all of which are included in the bottom panels.
Because the replacement coadds are composed of ``low-ISM'' sightlines, we expect these offsets to be smaller than those shown in the top panels.  This is the case for both \ion{Ca}{2} K and  \ion{Na}{1}, with most $W_{\rm BOSZ} - W_{\rm coadd}$ values falling between $-1$ and $+0.8$ \AA\  or $-1$ and 0 \AA, respectively.  Moreover, we see the same rise in $W_{\rm BOSZ}$(\ion{Ca}{2} K) $-$ $W_{\rm coadd}$(\ion{Ca}{2} K) values
below $T_{\rm eff, med} \approx 9000$ K that is evident in the top panel, suggesting that our empirical replacement coadds are likewise dominated by the significant intrinsic dispersion in stellar \ion{Ca}{2} profiles in these cooler stars.  {For the eleven stars with $T_{\rm eff, med} > 9000$ K in these panels, the mean values of $W_{\rm BOSZ} - W_{\rm coadd}$ for \ion{Ca}{2} K and \ion{Na}{1} are $-0.57$ \AA\ and $-0.43$ \AA, respectively.  The mean values of the uncertainty $\sigma_{W,\rm coadd}$ are 0.20 \AA\ and 0.11 \AA, implying that the empirical replacement coadds tend to overestimate the equivalent widths of these transitions relative to BOSZ by ${\sim}3-4\sigma_{W,\rm coadd}$.}
{This could suggest either (1) that even our “low-ISM” sample, from which the empirical replacement coadds are constructed, suffers some level of ISM contamination, or (2) that the BOSZ theoretical template spectra systematically underestimate the intrinsic absorption in these transitions, even at $T_{\rm eff,med}>9000$ K.}

\subsection{The ``Clean'' MaStar Spectroscopy}\label{subsec:clean_spectra}

\begin{figure}[ht]
 \includegraphics[width=\columnwidth,trim={0cm 0cm 0.5cm 2.0cm},clip]{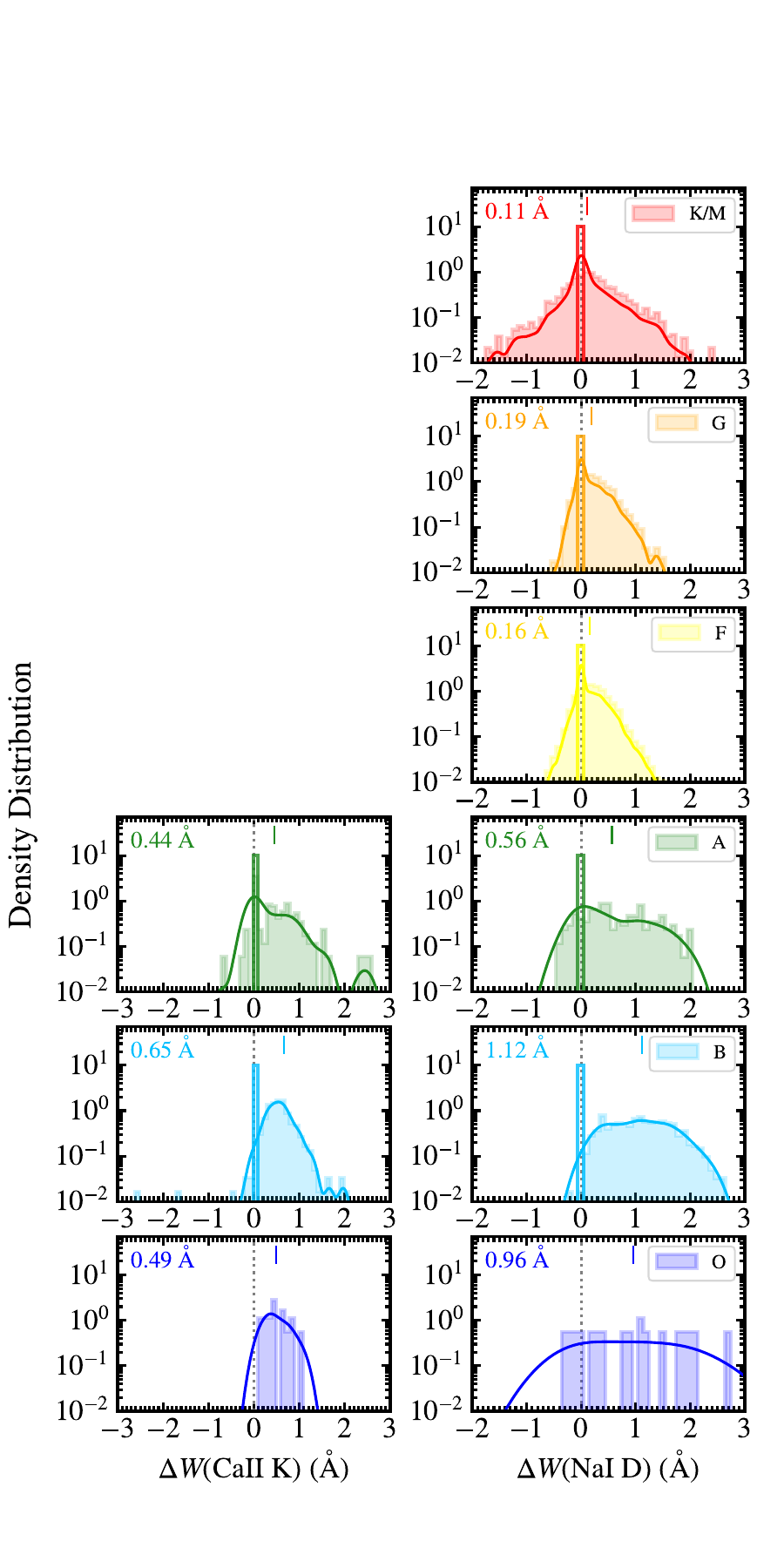}
\caption{The distributions of the change in the \ion{Ca}{2} K (left-hand panels) and \ion{Na}{1} D (right-hand panels) equivalent widths relative to the original stellar spectra ($\Delta W = W_{\rm orig} - W_{\rm clean}$) for our final cleaned sample.  All sightlines that are not considered ``low-ISM'', and for which we are able to generate either an empirical or BOSZ spectral replacement, are included in each transparent histogram.  The number of low-ISM sightlines of a given spectral type is indicated with the narrow bar at $\Delta W = 0$ \AA. Distributions are color-coded by spectral type as indicated in the legends.  The smooth curves show continuous probability densities.  The colored markers toward the top of each panel indicate the average value of the full distribution (including low-ISM sightlines).  {These values are also printed at the top left of each panel.} \label{fig:empandbosz_change_in_ews}}
\end{figure}

To assemble a ``clean'' spectroscopic sample from the edited spectra described above, we proceed as follows:

\begin{enumerate}
\item For all stars having $T_{\rm eff,med} \le 9000~\mathrm{K}$, we use our empirical replacement spectra (constructed with $\Psi_{\rm thresh} < 1$) to replace their \ion{Na}{1} D profiles where available.  We do not make any changes to their \ion{Ca}{2} profiles.
    \item For stars having $T_{\rm eff, med} > 9000$ K, we use empirical replacement spectra to replace their \ion{Ca}{2} and \ion{Na}{1} D profiles if they are available.  We instead use our BOSZ-cleaned spectra if we could not construct an empirical replacement sample with $\Psi_{\rm thresh} < 1$ for the star.  
\end{enumerate}

We have assembled this clean sample into a FITS file that is identical in form to that of \texttt{mastar-combspec-v3\_1\_1-v1\_7\_7-lsfpercent99.5.fits}, with the addition of two flags indicating how each spectrum has been treated as described in Table~\ref{tab:fitsflags}, and with our estimates of $W^{\rm ISM}$.  {This FITS file is available via Zenodo at \dataset[doi:10.5281/zenodo.14014915]{https://doi.org/10.5281/zenodo.14014915}.}

We again assess the overall impact of these changes on the MaStar spectroscopy
by measuring $\Delta W = W_{\rm orig} - W_{\rm clean}$ for each star that we now consider to be ``clean'' (i.e., with flag values of 0, 1, 2 or 3).  We show the distributions of these values for each spectral type in
Figure~\ref{fig:empandbosz_change_in_ews}.  The transparent histogram in each panel shows the distribution of $\Delta W$ values for sightlines for which we replace their \ion{Ca}{2} K and \ion{Na}{1} D profiles using either coadded low-ISM spectra or a BOSZ theoretical template.  The vertical bar at $\Delta W = 0$ \AA\ indicates the number of low-ISM sightlines of each spectral type.  The short vertical hash mark indicates the average $\Delta W$ of the full distribution (including low-ISM and cleaned sightlines).  $\Delta W$(\ion{Ca}{2} K) distributions for K/M, G and F stars are not shown because we have chosen not to use spectral replacements for \ion{Ca}{2} in this temperature range (see above).  

The distributions of $\Delta W$(\ion{Ca}{2} K) now all have positive mean values (of $0.4$, $0.7$ and $0.5$ \AA, for A, B, and O types, respectively), and include only five sightlines in total with $\Delta W$(\ion{Ca}{2} K) $< -1.0$ \AA.  
These distributions also include many more sightlines with $W$(\ion{Ca}{2} K) $> 0.5$ \AA\ relative to those shown in Figure~\ref{fig:change_in_ews} due to the inclusion of the BOSZ-cleaned sample.

The distributions showing the change in \ion{Na}{1} D absorption strength indicate that its equivalent width has been reduced in the vast majority of the cleaned spectra.  The mean value of $\Delta W$(\ion{Na}{1} D) falls in the range $0.1-0.2$ \AA\ for K/M, G, and F stars, and increases to 0.6, 1.1, and 1.0 \AA\ for A, B, and O stars, respectively.  This trend is again qualitatively consistent with the predictions shown in Figure~\ref{fig:hist_corr_ews}.  We discuss some of the implications of these findings for analyses of the \ion{Na}{1} D profile in external galaxy spectra in Section~\ref{sec:discussion}.

\begin{deluxetable*}{clcc}
\tablecaption{Treatment of \ion{Ca}{2} and \ion{Na}{1} D Spectral Regions\label{tab:fitsflags}}
\tabletypesize{\footnotesize}
\tablehead{
\colhead{Flag Value} & \colhead{Description} & \colhead{Number of \ion{Ca}{2} Profiles} & \colhead{Number of \ion{Na}{1} Profiles}}
\startdata
    0 & low-ISM sightline (no replacement) & 6342 & 6342\\
    1 & theoretical spectral replacement & 738 & 738\\
    2 & empirical spectral replacement & 14 & 12,110\\
    3 & empirical spectral replacement is available but not used for K/M/G/F/cool A stars & 12,109 & \nodata \\
    4\tablenotemark{a} & {supersolar low-ISM sightline (no replacement)} & 84 & 84\\
    5\tablenotemark{a}  & supersolar empirical replacement & 0  & 13\\
    10 & sightline with likely ISM contamination & 4484 & 4484\\
\enddata
\tablenotetext{a}{{Flag values of 4 indicate stars having $\rm [Fe/H] > 0$, $W^{\rm ISM}$(\ion{Ca}{2} K) $<0.4$ \AA, $W^{\rm ISM}$(\ion{Na}{1} 5891) $<0.15$ \AA, and either $W^{\rm ISM}$(\ion{Ca}{2} K) $>0.07$ \AA\ or $W^{\rm ISM}$(\ion{Na}{1} 5891) $>0.05$ \AA\ (our supersolar low-ISM sample)}.  Flag values of 5 indicate stars having $\rm [Fe/H] > 0$, for which we have replaced their \ion{Na}{1} D profiles with coadditions of stars with similar parameters drawn from the supersolar low-ISM sample.  Both of these subsamples are used in our construction of SSP model spectra, as described in Section~\ref{subsec:MaStar-SSPs}.  }
\end{deluxetable*}

\subsection{Construction of Cleaned Hierarchially-Clustered Spectral Templates}\label{subsec:hc_templates}

To facilitate the use of these cleaned products for the continuum modeling of external galaxy spectra (e.g., in tandem with the \texttt{pPXF} method; \citealt{Cappellari2011}), we construct a stellar template library based on the hierarchical clustering analysis described in \citet[][see their Section 5.2.1]{Abdurrouf2022}.  In brief, this analysis uses \texttt{pPXF} to fit pairs of MaStar spectra, where one spectrum is optimized to fit the other.  The ``distance'' used in the hierarchical-clustering analysis is defined as the rms of the optimized difference between the spectra in the pair.  The details of the clustering analysis differ between those used in the DR15/DR16 analysis of the MaNGA data \citep[][Section 5]{Westfall2019}, and the more recent approach described by \citet[][Section 5.2.1]{Abdurrouf2022} used to create the set referred to as the \texttt{MASTAR-HC-v2} templates. We use the latter approach here.\footnote{The \texttt{MASTAR-HC-v2} templates and the table identifying the spectra in each cluster by their MaNGA IDs are provided with the \texttt{mangadap} Python package; see \url{https://github.com/sdss/mangadap/tree/4.2.0/mangadap/data/spectral\_templates/mastarhc\_v2}.}  In particular, the association of each star with each cluster is kept mostly intact (with some exceptions noted below); however, we have updated the stacking approach and included our ISM corrections, as follows.

Our first step in preparing to create new versions of these templates is to apply an extinction correction to each stellar spectrum.  We adopt the \citet{Fitzpatrick2019} extinction curve derived for $R_V = 3.1$, taking advantage of the tabulation available with the \texttt{dust\_extinction} Python package\footnote{\url{https://dust-extinction.readthedocs.io/}}.  For stars with {\it Gaia} colors $G_{\rm BP}-G_{\rm RP}<0.8$ \citep{GaiaDR3}, we use the extinction values ($A_V$) calculated in the course of the stellar template fitting and parameter analysis of \citet{Lazarz2022}.  For stars with redder colors, we instead adopt the extinction value implied by its 3D dust map reddening and Equation~\ref{eq:av_edustmap}.  

We then create two new versions of the \texttt{HC} templates.  The first set is constructed from the original, extinction-corrected stellar spectra, but excludes all stars with a flag value ${>}3$ (i.e., all stars for which we could not construct a ``cleaned" spectrum, {and which are not low-ISM sightlines}).  This subsample does not contain any stars assigned to cluster IDs 0, 6, 8, 13, 29, 35, or 88, leaving us with a total sample of 58 templates.  We further caution that our newly constructed templates for clusters 9, 12, 26, 30, 66, 84, and 96 each contain only one star.  We have performed a visual inspection to verify that these templates are in all cases very similar to the original \texttt{MASTAR-HC-v2} versions.

Finally, we repeat this procedure using the clean version of the spectra assigned to each cluster, creating a new \texttt{MASTAR-HC-NoISM} template set.\footnote{These templates are publicly available at \url{https://github.com/sdss/mangadap/tree/4.2.0/mangadap/data/spectral\_templates/mastarhc\_v2\_noism}.  They may be used with the \texttt{mangadap} Python package for galaxy continuum modeling by setting the \texttt{eline\_fits.fit.templates} key \texttt{$=$ `MASTARHC2-NOISM'}.}  We show the resulting \ion{Ca}{2} K and \ion{Na}{1} profiles in orange in Figure~\ref{fig:hc_velplots} and in Appendix~\ref{sec:appendix-hctemplates}.  We show the matching templates created from the original stellar spectra in black for comparison.  The clusters are sorted by their median $T_{\rm eff, med}$ values, ordered from hot to cool.  The orange vs.\ black profiles exhibit the largest differences in the templates containing the hottest stars ({shown in Figure~\ref{fig:hc_velplots}}), consistent with the trends exhibited in the $\Delta W$ values shown in Figure~\ref{fig:empandbosz_change_in_ews}.  The differences in the \ion{Ca}{2} K transitions between these two sets of profiles become negligible for nearly all templates having median stellar effective temperatures $\mathrm{med}(T_{\rm eff,med}) < 8100$ K.  {Subtle differences} persist in the black vs.\ orange \ion{Na}{1} D profiles above $\mathrm{med}(T_{\rm eff,med}) \approx 3500$ K (shown in Appendix~\ref{sec:appendix-hctemplates}).

\begin{figure*}[ht]
  \includegraphics[width=\textwidth]{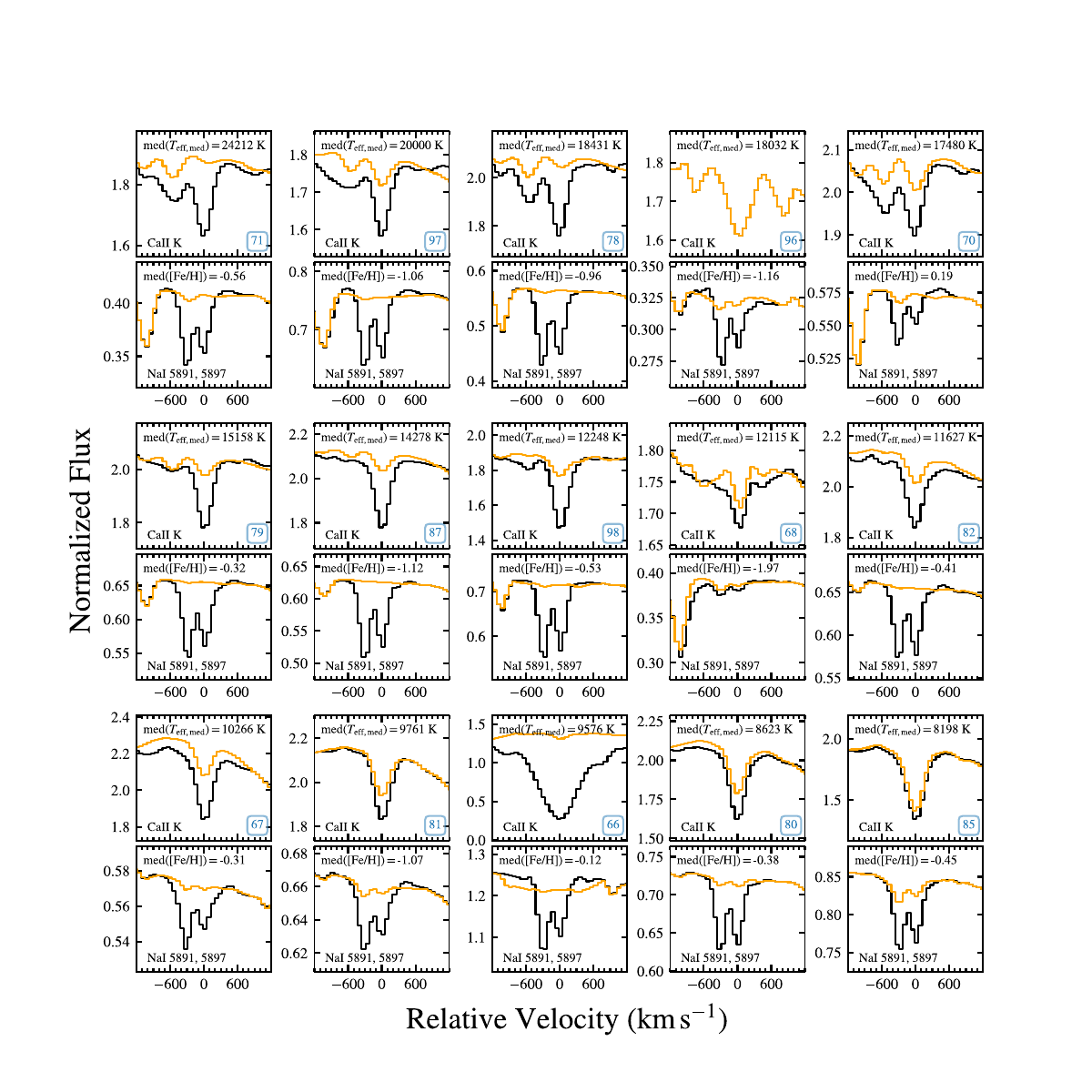}
\caption{Comparison of hierarchically-clustered MaStar spectral templates constructed without correction for ISM absorption (black) and from our ``cleaned'' spectral sample (orange).
Each pair of stacked panels shows the same template at the locations of the \ion{Ca}{2} K and \ion{Na}{1} $\lambda \lambda 5891, 5897$ transitions on top and bottom, respectively.  Velocities are computed relative to the \ion{Ca}{2} K $\lambda3934$ and \ion{Na}{1} $\lambda 5897$ rest wavelengths. 
The templates have been ordered according to the median $T_{\rm eff,med}$ value of the stars used in each.  This value, along with the median $\rm [Fe/H]$ value, is noted in each panel pair.  The cluster ID of each template is indicated in blue.  The remaining templates are included in Appendix Figure~\ref{fig:hc_velplots_2}.  
\label{fig:hc_velplots}}
\end{figure*}

\subsection{Construction of Simple Stellar Population Model Spectra}\label{subsec:MaStar-SSPs}

{\subsubsection{Methodology}}
Analyses of external galaxy spectra aiming to constrain stellar population ages, metallicities, abundance patterns, or the IMF often rely on spectral templates constructed from stellar population models \citep[e.g.,][and many others]{CidFernandes2005,Ocvirk2006,Perez2013,Conroy2014,Gonzalez-Delgado2015,Cappellari2017,Wilkinson2017,Goddard2017,Conroy2018,Feldmeier-Krause2021}.  To facilitate the use of our cleaned spectra in such analyses, we follow the methods described in \citet{Maraston2020} to construct a suite of ``cleaned'' SSP model spectra.  Summarizing briefly, as in \citet{Maraston2005} and \citet{Maraston2020}, we adopt isochrones and stellar tracks from \citet{Cassisi1997a}, \citet{Cassisi1997b}, and \citet{Cassisi2000} for ages older than ${\sim}30$ Myr, \citet[][]{Schaller1992} for younger ages, and \citet{Girardi2000}.  Fuel consumption during the thermally pulsating asymptotic giant branch phase is calibrated empirically.  Realistic horizontal branch morphologies are generated by assuming Reimers-type mass-loss along the red giant branch as in  \citet{Maraston2005}.  We adopt the fuel consumption approach to calculate the energetic contribution of each post-main-sequence phase to a given SSP, assuming it is proportional to the amount of fuel available for burning in that phase \citep[e.g.,][]{RenziniBuzzoni1986,Maraston1998}.

In constructing our SSPs, we adopt the calibrated median stellar parameters for the MaStar sample {as described in Section~\ref{subsec:MaStar_Wr} (R. Yan et al., in preparation).  These parameters differ from those adopted in \citet{Maraston2020}, which made use of parameters included in the first release of the MaStar library \citep{Yan2019,Chen2020} and those derived from theoretical spectral fitting by \citet{Hill2022}.}
We correct the spectra for extinction as described in \citet{Yan2019}.  We also establish the energy scale of each stellar spectrum as described in \citet{Maraston2020}, first normalizing each spectrum by its average flux at ${\sim}5550$ \AA, and then rescaling it to the average ${\sim}5550$ \AA\ luminosity of a theoretical stellar spectrum with the same stellar parameters.  As in \citet{Maraston2005} and \citet{Maraston2020}, we use the \citet{Lejeune1997} theoretical library for this renormalization.  
Finally, we separate the MaStar stars into several different metallicity bins centered at $\rm [Z/H] = -1.35, -0.33, 0.0$, and $+0.35$ (corresponding to the metallicities of the SSP grid calculated in \citealt{Maraston2005}) with approximate widths of $\pm 0.1$ dex.  Within each metallicity bin, empirical spectra are assigned an evolutionary phase (e.g., main sequence, red giant branch).  Then, we calculate the spectrum representing a given $T_{\rm eff}$ and $\log g$ via  interpolation of the logarithmic fluxes of stars within the relevant phase as a function of $\log T_{\rm eff}$ and $\log g$.  These interpolated spectra are combined into representative SSPs with ages spanning 3 Myr to 15 Gyr.  We assume a Salpeter IMF for these calculations.  

{\subsubsection{The Addition of a Supersolar Low-ISM Sample}}

In examining the $T_{\rm eff}-\log g$ distributions of stars having ``cleaned'' spectra in the $\rm [Z/H] = +0.35$ bin, we found that {we lacked any stars having $T_{\rm eff} > 6000~\rm K$, as well as any stars sampling the red giant branch at $\log g < 2$.  We were therefore unable to construct SSPs using the procedure described above.}  This is unsurprising given the small numbers of low-ISM stars with $\rm 0.2 < [Fe/H] < 0.4$ and $\log g < 3$ or $3\theta < 2$ shown in Figure~\ref{fig:thetaZlogg_NaIcut}.  To improve our sampling of these regions of parameter space, we relaxed our criteria defining low-ISM stars to include those with $\rm [Fe/H] > 0$, $W^{\rm ISM}$(\ion{Ca}{2} K) $< 0.4$ \AA, and $W^{\rm ISM}$(\ion{Na}{1} 5891) $< 0.15$ \AA.  This ``supersolar low-ISM'' sample includes 84 stars that were previously assigned flag values of 10.  

We use this supersolar low-ISM sample to construct empirical spectral replacement coadds as described in Section~\ref{subsec:def_psi} for as many additional stars as possible.  We are able to construct coadds with $\Psi_{\rm thresh} < 1$ for 13 more stars, and we use these coadds to replace their \ion{Na}{1} D spectral regions.  We find that this supplementary sample of high-[Z/H] stars {fully samples the red giant branch and high-$T_{\rm eff}$ portions of} the corresponding isochrones, and we include them in our construction of our $\rm [Z/H] = +0.35$  SSP spectral templates.
We have also indicated all supersolar low-ISM stars and all stars with supersolar low-ISM empirical replacements with flag values of 4 and 5 in the FITS file containing our cleaned spectra (described in Section~\ref{subsec:clean_spectra} and Table~\ref{tab:fitsflags}).  \\

{\subsubsection{``Cleaned'' SSP \ion{Ca}{2} and \ion{Na}{1} D Absorption Profiles}}

\begin{figure*}[ht]
\includegraphics[width=\textwidth,trim={0 1cm 0 1cm},clip]{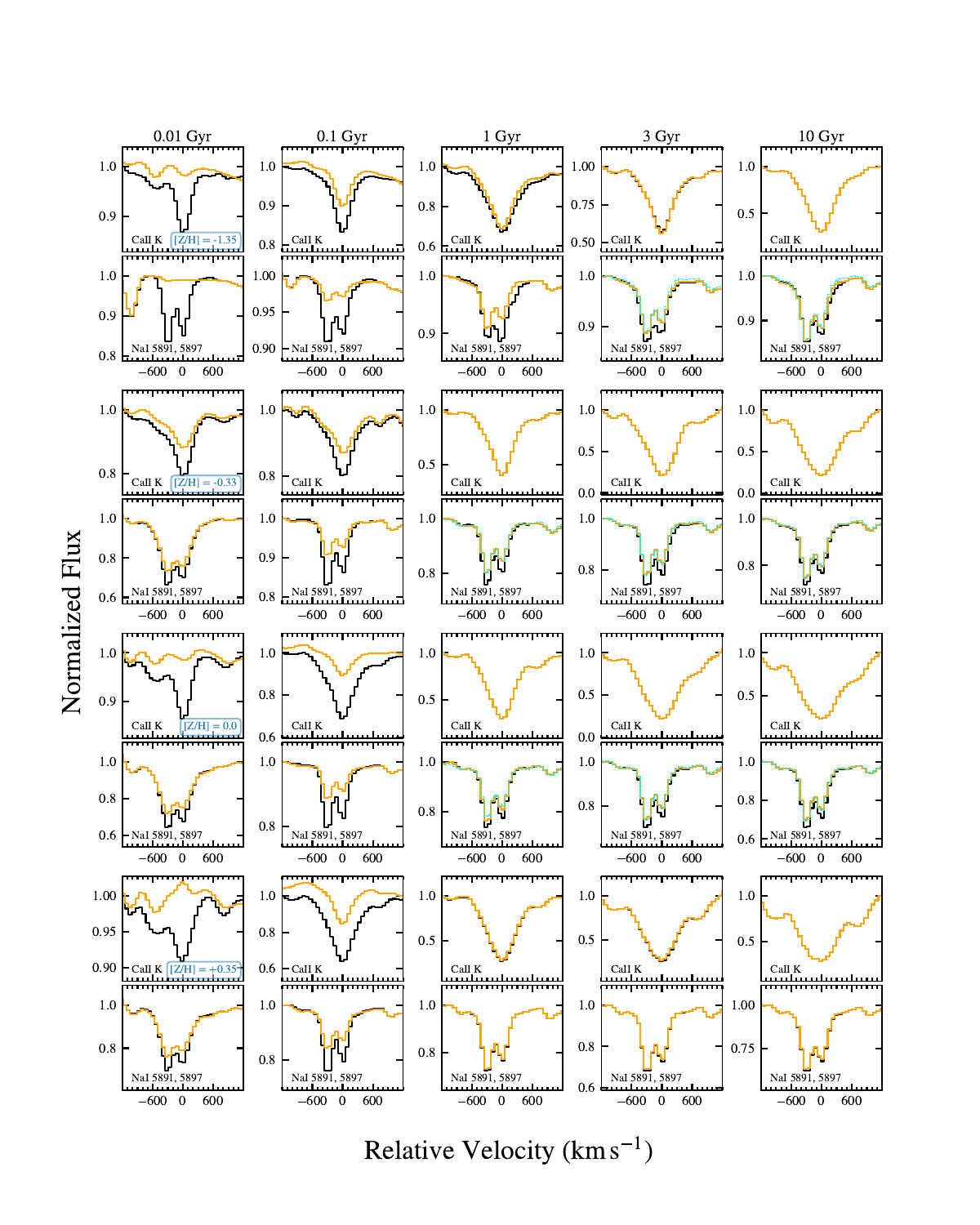}
\caption{Comparison of MaStar SSP spectral templates constructed without correction for ISM absorption (black) and from our ``cleaned'' spectral sample (orange).  {Results for SSP templates constructed solely from low-ISM stars as described in Section~\ref{subsubsec:low-ISM-SSPs} are shown with dashed cyan lines.}
Each pair of stacked panels shows the same template at the locations of the \ion{Ca}{2} K and \ion{Na}{1} $\lambda \lambda 5891, 5897$ transitions on top and bottom, respectively.  Velocities are computed relative to the \ion{Ca}{2} K $\lambda3934$ and \ion{Na}{1} $\lambda 5897$ rest wavelengths. 
Models increase in age from left to right, as labeled above the top panels, and increase in [Z/H] from top to bottom, as indicated in blue in the left-most \ion{Ca}{2} K panels.
\label{fig:SSP_velplots}}
\end{figure*}

As for the hierarchically-clustered templates described above, we generate two versions of each SSP spectrum.  The first is constructed from the original, extinction-corrected MaStar stellar spectra, but includes only stars having flag values $\le 5$.  The second\footnote{These templates are publicly available at \url{https://doi.org/10.5281/zenodo.14807331}.  A subset of these templates with ages $= 0.003, 0.01, 0.03, 0.1, 0.3, 1.0, 3.0, 9.0,$ and 14 Gyr is available at \url{https://github.com/sdss/mangadap/tree/4.2.0/mangadap/data/spectral\_templates/mastar\_ssp\_noism\_v1.0}.  They may be used with the \texttt{mangadap} Python package for galaxy continuum modeling by setting the \texttt{eline\_fits.fit.templates} key \texttt{$=$ `MASTARSSP-NOISM'}.}  is constructed from the ``cleaned" versions of the extinction-corrected stellar spectra.   We show the profiles of the resulting \ion{Ca}{2} K and \ion{Na}{1} spectral regions for a subset of these SSPs in orange in 
Figure~\ref{fig:SSP_velplots}.  The matching SSP templates created from the original stellar spectra are shown in black.  {Note that in a few cases, the orange profile lies underneath a dashed cyan profile showing an SSP spectrum constructed solely from low-ISM stars, described in the following subsection.}  As expected, the differences between the orange vs.\ black profiles tend to be the greatest at the youngest ages and lowest metallicities.  The \ion{Na}{1} profiles for the 10 Myr old SSPs (both original and cleaned) exhibit stronger absorption than those with 100 Myr ages, particularly for metallicities $\rm [Z/H] \ge -0.33$.  This is likely due to the dominant contribution of red supergiant stars in the \citet{Schaller1992} 10 Myr isochrones.  
The \ion{Ca}{2} K profiles exhibit minimal differences for ages ${\gtrsim}1$ Gyr; however, the black \ion{Na}{1} profiles are evidently affected by ISM contamination even at an age of 10 Gyr at solar and subsolar metallicities.  The differences between the original and cleaned \ion{Na}{1} profiles in our supersolar SSPs are negligible; however, this is because the supersolar SSPs are dominated by stars in the supersolar low-ISM sample, rather than those for which the \ion{Na}{1} spectral region has been replaced.  Given the relatively high degree of ISM contamination we have predicted for the overall sample of high-metallicity stars, we posit that supersolar SSP templates constructed without regard for this contamination would be even more severely affected than those constructed for solar metallicity.  We also caution that because we have relaxed our low-ISM criteria to construct the ``cleaned'' supersolar SSP templates, they too likely suffer from some degree of ISM contamination (i.e., at a level $W^{\rm ISM}$(\ion{Na}{1} 5891) $< 0.15$ \AA).
We discuss the implications of all of these findings for constraints on stellar ages, Na abundances, and the IMF from stellar population modeling below in  Sections~\ref{subsec:ISM_affects_SSPs} and \ref{subsec:ISM_impacts_age_IMF}.\\

{\subsubsection{SSP Construction from Low-ISM Stars}\label{subsubsec:low-ISM-SSPs}

To test the robustness of our cleaning procedure, we have constructed an additional set of SSP templates using only our original ``low-ISM'' sample of stars.  The parameter space sampling of low-ISM stars is insufficient for the construction of SSP spectra with $\rm [Z/H] = +0.35$; however, we were able to use the procedure described above to generate SSP templates for metallicity $\rm [Z/H] = -1.35$ at ages $> 2$ Gyr, and for metallicities $\rm [Z/H] = -0.33$ and 0.0 at ages $> 0.5$ Gyr.  We show some of the resulting \ion{Na}{1} profiles in Figure~\ref{fig:SSP_velplots} with dashed cyan lines.  These profiles are very similar, and in some cases nearly identical, to those of the cleaned templates.  The \ion{Na}{1} D equivalent width values measured from these low-ISM templates are also very similar to those of the cleaned SSP spectra described above, with a maximum offset $\lvert \Delta W \rvert$ of 0.17 \AA, a maximum fractional offset $\lvert \Delta W/W_{\rm clean}\rvert$ of 9\%, and a mean fractional offset of $3-4\%$.  We conclude from this comparison that our procedure for ISM removal is indeed robust, and does not introduce unwanted artifacts into our final cleaned SSP templates.

}

\section{Discussion}\label{sec:discussion}

\begin{figure*}[ht]
 \includegraphics[width=\textwidth,trim={0.7cm 0.3cm 0cm 0cm},clip]{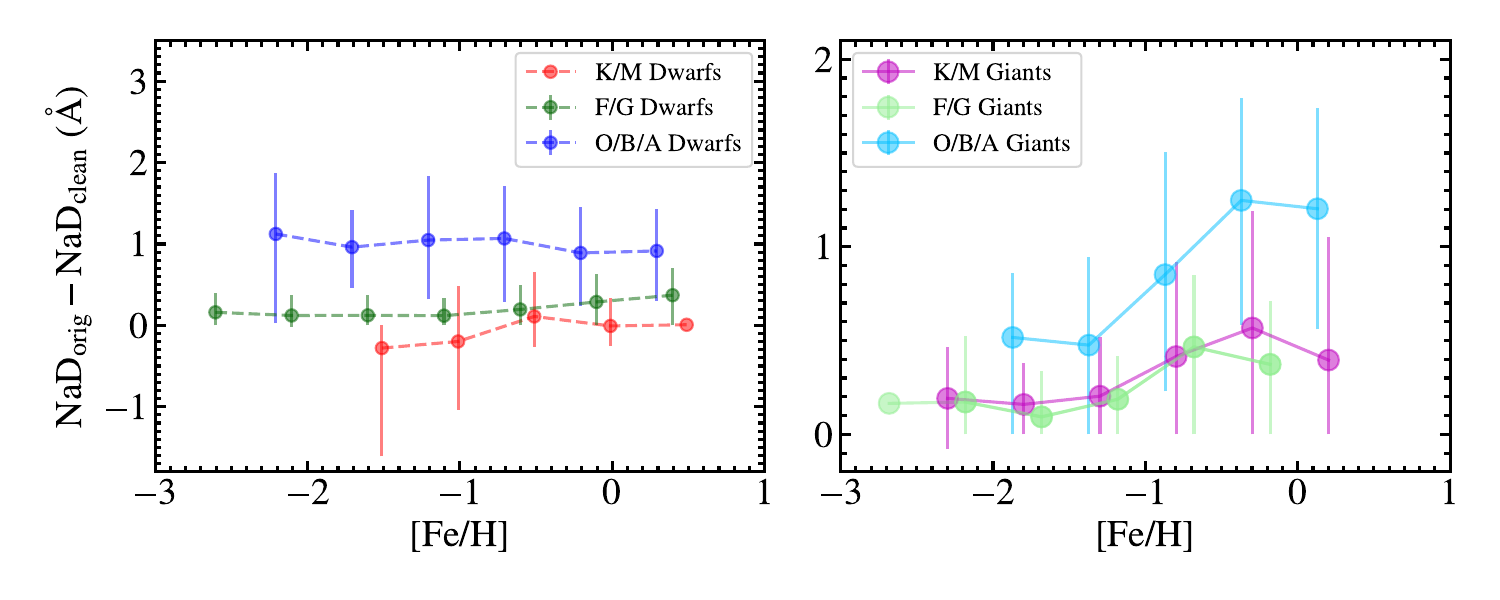}
\caption{The difference between the NaD spectral index measured in the original MaStar spectra and that measured in the cleaned spectra vs.\ stellar [Fe/H].  \emph{Left:} $\rm NaD_{orig}-NaD_{clean}$ for dwarf stars (with $\log g > 4$) divided into three temperature bins. Each point shows the average value of this offset in a bin of width $\Delta \rm [Fe/H] = 0.5$.  The error bars indicate the 16th- and 84th-percentile values of these offsets within each bin.  \emph{Right:} Same as that shown at left, for giant stars having $\log g < 4$.  \label{fig:NaD_spectral_index}}
\end{figure*}

\begin{figure}[ht]
\includegraphics[width=\columnwidth,trim={0.7cm 0.3cm 0.7cm 0cm},clip]{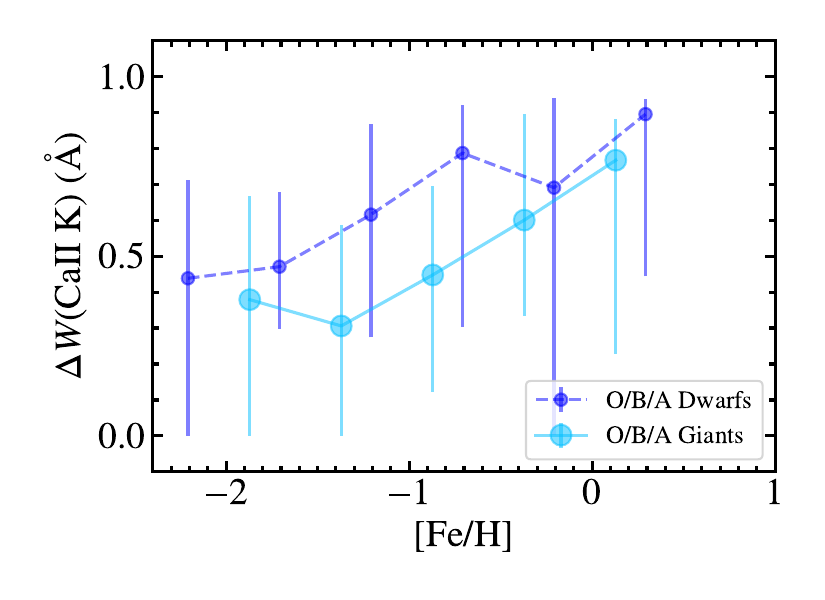}
\caption{The difference between $W$(\ion{Ca}{2} K) measured in the original MaStar spectra and that measured in the cleaned spectra vs.\ stellar [Fe/H] for O/B/A-type stars.  Results for dwarf stars (with $\log g > 4$) are shown with small dark blue circles, and results for giant stars are shown with large light blue circles.  
Each point shows the average value of this offset in a bin of width $\Delta \rm [Fe/H] = 0.5$.  The error bars indicate the 16th- and 84th-percentile values of these offsets within each bin.  \label{fig:CaK_spectral_index}}
\end{figure}

{The strengths of metal and hydrogen absorption lines in the spectra of external galaxies have long been understood to trace the elemental abundances and ages of the underlying stellar populations \citep[e.g.,][]{Burstein1984,Burstein1986,Worthey1992,Worthey1994,Worthey2014,Trager1998,Kuntschner2000,Thomas2005,Bernardi2006,Johansson2012}.  
The \ion{Na}{1} D stellar absorption feature has traditionally been quantified by the ``NaD" Lick index \citep{Trager1998}, and serves as an indicator of the Na abundance relative to Fe \citep[e.g.,][]{Thomas2003,Thomas2011,Parikh2018,Parikh2021}. Recent studies of this feature in the nearby galaxy population have concluded that massive systems (both early-type galaxies and massive spirals) are typically significantly Na-enhanced (with [Na/Fe] up to $+0.6-0.8$ dex; \citealt{McConnell2016,Alton2017,vanDokkum2017,Vaughan2018,Parikh2018}), while low-mass galaxies ($\log M_*/M_{\odot} < 10$) have Na abundances close to solar \citep{Parikh2018,Parikh2019,Parikh2021}.  The NaD index has also been used in conjunction with the surface-gravity-sensitive ``NaI'' spectral index at ${\sim} 8200$ \AA\ to constrain the slope of the IMF in early-type galaxies \citep[e.g.,][]{Martin-Navarro2015,Martin-Navarro2023,Parikh2018,
LaBarbera2019,
Feldmeier-Krause2021, Gu2022}.
A common approach to linking the strength of this feature in a given stellar population to its [Na/Fe] relies on the measurement of NaD for individual stellar template spectra in a stellar library of choice.  NaD is then calibrated to [Na/Fe] via the construction of SSP models that establish the weighting of the index strengths of a representative set of stars for a given IMF, age, and elemental abundance pattern \citep[e.g.,][]{Worthey1994,WortheyOttaviani1997,Maraston1998,Thomas2003,Johansson2010}.
}

\subsection{Relation Between Interstellar Absorption Effects and Stellar $T_{\rm eff}$, $\log g$, and [Fe/H]}\label{subsec:disc-specindx}

We now consider the degree to which the NaD spectral index strengths of stars in the MaStar sample are affected by the Milky Way ISM.  While none of the aforementioned studies rely on the MaStar stellar library to calibrate the relation between spectral indices and elemental abundances (and instead, more recent studies have used the MILES library constructed by \citealt{Sanchez-Blazquez2006} or one of its many extensions; e.g., \citealt{Vazdekis2010}, \citealt{Conroy2018}), we argue that the Milky Way ISM will likely affect any such calibration if it is based on an empirical library, and if the stars were selected without regard for their interstellar reddening, distance, or Galactic latitude.  
We have measured the NaD index using the \citet{Trager1998} definition\footnote{The NaD index is defined to be the equivalent width measured in the spectral region $5876.875~\mathrm{\AA} < \lambda_{\rm rest} < 5909.375~\mathrm{\AA}$.  The continuum is determined from a linear fit to the mean flux measured in two pseudocontinuum regions on either side of the feature (i.e., $5860.625~\mathrm{\AA} < \lambda_{\rm rest} < 5875.625~\mathrm{\AA}$ and $5922.125~\mathrm{\AA} < \lambda_{\rm rest} < 5948.125~\mathrm{\AA}$).} for both the original spectrum ($\rm NaD_{orig}$) and cleaned spectrum ($\rm NaD_{clean}$) of every star in MaStar with a flag value ${\le} 3$.  The results are summarized in  Figure~\ref{fig:NaD_spectral_index}.  The left-hand panel shows the difference in these values for ``dwarf'' stars having $\log g > 4$.  The dwarf sample is further divided into ``K/M'' ($T_{\rm eff,med}<5200$ K), ``F/G'' ($5200~\mathrm{K} < T_{\rm eff,med} < 7610~\mathrm{K}$), and ``O/B/A'' ($T_{\rm eff,med} > 7610$ K) subsamples.  
Each point shows the average value of this difference in the relevant subsample across a metallicity bin of width $\Delta \rm [Fe/H] = 0.5$.  We include both low-ISM sightlines and sightlines for which we have successfully generated replacement spectral regions in this mean.  The error bars show the 16th- and 84th-percentile values within each bin.  The right-hand panel shows the same measurements for giant stars having $\log g < 4$.  Each bin shown contains at least five stars.

This figure demonstrates that the strength of NaD in cool dwarfs is not significantly affected by ISM absorption in an aggregate sense.  However, the dispersion in these offsets can be large (${>}0.5$ \AA), suggesting that individual stars may occasionally be subject to strong ISM absorption. The lowest-[Fe/H] bin exhibits a relatively large negative mean offset (of $-0.3$ \AA); however, we caution that it contains only 10 stars, three of which are low-ISM sightlines.
F/G dwarfs exhibit a typical offset of $0.1-0.4$ \AA, while O/B/A dwarfs exhibit more extreme offsets of $0.9-1.1$ \AA.  
K/M giants exhibit mean offsets of $0.2-0.6$ \AA;
F/G giants exhibit mean offsets in the range $0.1-0.5$ \AA; and O/B/A giants exhibit offsets in the range $0.5-1.2$ \AA.

In summary, we find that the strength of NaD is systematically overestimated in the original MaStar sample due to ISM contamination.  The degree of this overestimation is largest for hot stars, but persists at a level of $\sim0.2-0.6$ \AA\ for cool giants.  This in turn suggests that SSP models built from the original MaStar library to calibrate the relationship between NaD and [Na/Fe] would yield index strengths that are larger than those intrinsic to the stars for a given [Na/Fe].  If this relationship is used in combination with NaD measurements of observed spectra of external galaxies, {assuming the galaxies do not themselves contain significant interstellar \ion{Na}{1}}, it would yield [Na/Fe] abundances that are systematically underestimated.

For completeness, we show the difference in the \ion{Ca}{2} K line strength between the original and cleaned spectra of O/B/A dwarfs and giants as a function of the stellar metallicity in Figure~\ref{fig:CaK_spectral_index}.\footnote{While a spectral index (CaHK) defined to assess absorption strength in this spectral region was introduced by \citet{Serven2005} in their analysis of line indices in elliptical galaxies, we find that it is not well-suited for use on hot stars due to their strong Balmer absorption features falling in the index pseudocontinuum regions.}  Metallicity bins are constructed in the same manner as for Figure~\ref{fig:NaD_spectral_index}, and each contains at least five stars. 
 O/B/A dwarfs exhibit mean equivalent width offsets of $0.4-0.9$ \AA, while hot giants exhibit slightly more modest offsets in the range of $0.3-0.8$ \AA.  These offsets furthermore appear to increase with increasing stellar [Fe/H], as is evident for the NaD offsets exhibited by giant stars (i.e., shown in the right-hand panel of Figure~\ref{fig:NaD_spectral_index}).  As highly temperature-sensitive transitions, \ion{Ca}{2} H \& K are not typically used to constrain Ca abundances via stellar population synthesis analysis (although they may be used to aid in the selection of extremely metal-poor stars; e.g., \citealt{Starkenburg2017,Youakim2017}).  Instead, because the strengths of \ion{Ca}{2} H \& K increase dramatically with decreasing temperature in A stars, they
 have been useful as a diagnostic of stellar population age in young stellar populations (${\lesssim} 1$ Gyr; \citealt{Rose1985,Leonardi1996,WildCaHK2007}).  SSP models constructed from the original MaStar library would overestimate the strength of \ion{Ca}{2} H \& K at a given age (Figures~\ref{fig:SSP_velplots} and \ref{fig:CaK_spectral_index}).  The best-fitting combination of SSP model templates for an observed galaxy spectrum would therefore imply an age that is younger than the galaxy's true age, {under the assumption that the observed galaxy does not exhibit significant interstellar \ion{Ca}{2} absorption.  We further discuss the validity of this assumption in Section~\ref{subsec:ISM_impacts_age_IMF}.}  

{We also note that the construction of SSP templates or [Na/Fe] calibrations from stars selected to have very close distances, or
 very low values of $E(B-V)$ and high Galactic latitudes, would likely reduce their ISM contamination to a degree similar to that seen above.  That is, a selection of stars with distances $<0.1$ kpc, or with $0.1~\mathrm{kpc} < D < 0.3~\mathrm{kpc}$, $E_{\rm DustMap} < 0.01$, and $b > 40^{\circ}$, yields 1177 low-ISM stars and only 61 stars that do not fall into our low-ISM sample.  
 A simple selection of stars with $E_{\rm DustMap} < 0.01$, on the other hand, yields 5222 low-ISM stars and 2117 that do not fall into our low-ISM sample.  SSP templates constructed from such a selection would likely have lower ISM contamination than those constructed without regard for reddening, but would not be as ``clean'' as the stellar sample we have constructed above.  
The degree of ISM contamination for SSP templates or abundance calibrations constructed from such a sample -- or any stellar library -- may be estimated using our model for $W^{\rm ISM}(\log E_{\rm DustMap}, b, D)$ described in Section~\ref{subsec:wr-distance-model}.}

\subsection{Interstellar Absorption Effects in Simple Stellar Population Model Spectra}\label{subsec:ISM_affects_SSPs}

\begin{figure}[ht]
\includegraphics[width=\columnwidth,trim={0.7cm 0.3cm 0cm 0cm},clip]{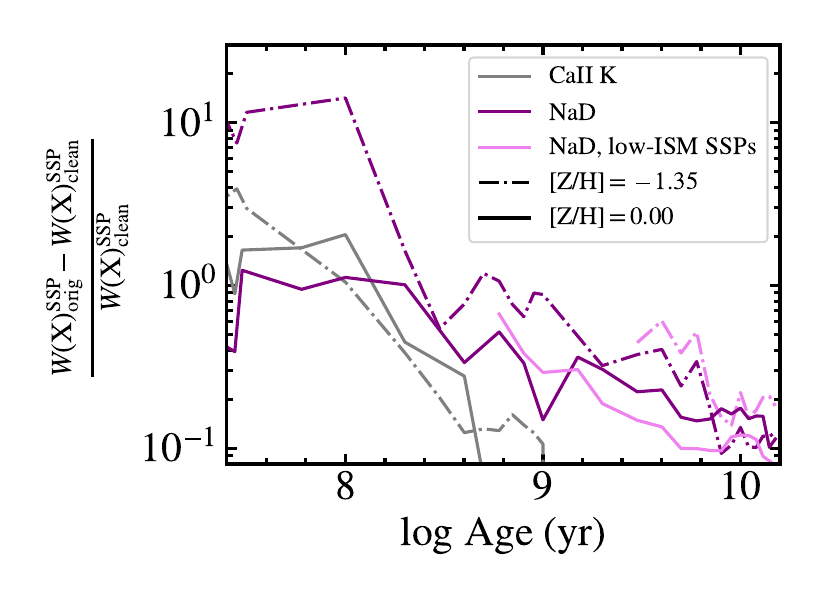}
\caption{
Fractional enhancement due to interstellar absorption of the equivalent width of \ion{Ca}{2} K (gray) and the NaD spectral index (purple) for solar-metallicity (solid) and $\rm [Z/H] = -1.35$ (dashed-dotted) SSPs.  {The violet curves show the fractional NaD enhancement estimated relative to the SSPs constructed solely from low-ISM stars described in Section~\ref{subsubsec:low-ISM-SSPs}.}
We measure enhancements of ${\gtrsim}100\%$ for young stellar populations in both transitions, and measure NaD enhancements ${\gtrsim}10\%$ across the full range of ages at solar metallicity.
\label{fig:SSPs}}
\end{figure}

Here we assess 
the significance of the effects of interstellar \ion{Ca}{2}  and \ion{Na}{1} absorption as a function of the stellar population age and metallicity using the MaStar SSP spectra constructed as described in Section~\ref{subsec:MaStar-SSPs}.  In detail, we compare the strength of either $W$(\ion{Ca}{2} K) or NaD measured from the SSP templates constructed from the original spectra,  $W\mathrm{(X)_{orig}^{SSP}}$, with the strength of the same feature measured from the corresponding ``cleaned" SSPs ($W\mathrm{(X)_{clean}^{SSP}}$).
In Figure~\ref{fig:SSPs}, 
we show the fractional enhancement in these features due to Milky Way ISM, ($W\mathrm{(X)_{orig}^{SSP}}-W\mathrm{(X)_{clean}^{SSP}})/W\mathrm{(X)_{clean}^{SSP}}$, as a function of the SSP age for 
populations with $\rm [Z/H] = -1.35$ and $0.0$.  {For completeness, we also show NaD fractional enhancements calculated relative to the SSPs constructed solely from low-ISM stars (described in Section~\ref{subsubsec:low-ISM-SSPs}) for ages ${>} 2$ Gyr at $\rm [Z/H] = -1.35$, and for ages ${>} 0.5$ Gyr at solar metallicity (i.e., $W\mathrm{(X)_{orig}^{SSP}}-W\mathrm{(X)_{lowISM}^{SSP}})/W\mathrm{(X)_{lowISM}^{SSP}}$ ).}

We find that interstellar \ion{Ca}{2} absorption enhances the strength of $W$(\ion{Ca}{2} K)$^{\rm SSP}$ by between 20\% and 200\% at solar metallicity for ages $<400$ Myr due to the dominant contribution of hot stars in such young stellar populations.  At $\rm [Z/H] = -1.35$, a similar level of \ion{Ca}{2} K enhancement is evident, but weakens to $<20\%$ at ages $\gtrsim 300$ Myr.  Because we do not alter the \ion{Ca}{2} H\&K profiles of stars cooler than $T_{\rm eff,med} =$ 9000 K, we do not predict any artificial enhancement for these transitions in SSPs dominated by later spectral types (at ages ${\gtrsim}1$ Gyr).  

We predict more significant enhancements of NaD across the full range of ages explored. 
At young ages, we predict a  ${\gtrsim}100\%$ enhancement in the NaD spectral index strength, which persists until ages of ${\sim} 600$ Myr for the low-metallicity SSPs. At solar metallicity, we predict enhancements of ${\gtrsim}50\%$ for nearly all ages between $30$ and $600$ Myr.  In general, we predict more significant fractional enhancements of NaD for our subsolar SSPs relative to the solar-metallicity case due to the increased importance of hot dwarfs and cool giants in the former, and because the intrinsic NaD is weaker in lower-metallicity stars.  However, even in the solar-metallicity case at older ages, we predict artificial enhancements of NaD of ${\sim}20-30\%$ in the age range $1-4$ Gyr, and of ${>}10\%$ at {$4 - 15$ Gyr.   
If we instead use the SSPs constructed from low-ISM stars as our reference, we predict artificial enhancements of NaD of ${\sim}20\%$ at ages $0.5-1.5$ Gyr, and of ${>}10\%$ at $1.5-12$ Gyr.}

\subsection{The Impact of Interstellar Absorption on Analyses of Stellar Population Age, Sodium Abundance, and the IMF}\label{subsec:ISM_impacts_age_IMF}

{\subsubsection{Artificial Enhancements as a Function of Age}}

Nearby extragalactic stellar populations are of course substantially more complex than the SSPs explored above \citep[e.g.,][]{
FranxIllingworth1990,Kauffmann2003,CidFernandes2005,Perez2013,Sanchez-Blazquez2014,Roig2015,Gonzalez-Delgado2015,Goddard2017,Parikh2021}. Typical  late-type galaxies are understood to be comprised of both old (${\sim}5-6$ Gyr), metal-poor and younger (${\sim}1$ Gyr), metal-rich stellar populations, and exhibit negative lightweighted age gradients consistent with ``inside out'' formation \citep[e.g.,][]{Perez2013,Gonzalez-Delgado2014,Gonzalez-Delgado2015,Goddard2017,Parikh2021}.  
Early-type galaxies are well-understood to be dominated by old stars (with ages ${\gtrsim} 10$ Gyr), but can contain a subdominant younger population (${\gtrsim}3$ Gyr old) toward their centers \citep{Trager1998,
Kuntschner2000,Mehlert2003,Thomas2005,Bernardi2006,Clemens2006,Spolaor2010,Greene2015,Goddard2017}.  We therefore expect that absorption originating in the Milky Way's ISM would systematically increase the strength of NaD by at least $10-30\%$  in MaStar stellar population synthesis models of late-type galaxies if constructed from the original library, and that the magnitude of this effect would increase with radius as the overall stellar age decreases.  Stellar population synthesis models of early-type galaxies would exhibit at least a $10\%$ enhancement of NaD due to contaminating ISM absorption, with a greater enhancement associated with any younger stars at their centers.  

Stellar population synthesis analyses that must suffer the most significant impacts from Milky Way ISM are those focusing on galaxies with a starburst or post-starburst component.  
Post-starburst galaxies are often identified by a combination of strong Balmer absorption and a lack of emission lines signifying ongoing star formation \citep[e.g.,][]{DresslerGunn1983,CouchSharples1987,Zabludoff1996,WildCaHK2007,Kaviraj2007}.  In the local Universe, these spectral features have been found to arise from starburst events that occurred within the past ${\sim}1$ Gyr and generated $20-60\%$ of the galaxy's stellar mass \citep{Kaviraj2007}.  The light from these recent bursts will contribute a yet larger fraction to the galaxies' rest-frame optical emission, implying that their spectra will be dominated by stellar populations for which stellar synthesis models constructed from the original MaStar library would have a ${\gtrsim}20\%$ enhancement in the strength of \ion{Ca}{2} K due to the ISM, and a ${\gtrsim}50\%$ enhancement of NaD.

Many modern analyses that use the stellar population synthesis technique to assess the star formation histories, chemical abundances, and/or IMF of extragalactic systems do so by fitting linear combinations of SSP templates to observed spectra covering the full optical range \citep[e.g.,][]{CidFernandes2005,Ocvirk2006,Conroy2014,Conroy2018,Cappellari2017,Wilkinson2017,Feldmeier-Krause2021}.  This full spectrum fitting can in principle take advantage of the simultaneous fitting of numerous spectral features to constrain physical quantities, reducing the importance of accurate modeling of those few transitions affected by the ISM.  However, as noted above, constraints on the ages of recent starburst events are driven primarily by 
 the \ion{Ca}{2} H\&K transitions for studies that lack corroborating rest-frame UV photometry \citep[e.g.,][]{Rose1985,Leonardi1996,Leonardi2003,WildCaHK2007,Kaviraj2007}.  

{\subsubsection{Implications for [Na/Fe] Abundances in Early-type Galaxies}}

{[Na/Fe] may likewise be constrained via the simultaneous analysis of numerous other metal-line transitions in rest-frame optical spectroscopy of early-type galaxies, as the abundance of Na has a significant impact on the electron pressure in cool stellar atmospheres \citep{ConroyvanDokkum2012a}.
However, such constraints are heavily dominated by \ion{Na}{1} D, as it is significantly stronger than many of these other transitions, including the surface-gravity-sensitive \ion{Na}{1} $\lambda \lambda 8183, 8195$ features \citep{TJM2011}, and the \ion{Na}{1} $1.14\mu$m and $2.21\mu$m transitions in the near-IR \citep{ConroyvanDokkum2012a}.  It also has a weak dependence on the IMF slope \citep{ConroyvanDokkum2012b,Spiniello2012,LaBarbera2013,Parikh2018}. }
{A ${\gtrsim}10\%$ enhancement of NaD due to ISM in stellar population model spectra of early-type galaxies would imply a yet more extreme Na abundance enhancement than is currently estimated for these systems (i.e., beyond the ${\sim}+0.6-0.8$ dex reported by \citealt{Alton2017}, \citealt{vanDokkum2017}, \citealt{Parikh2018}, and \citealt{Vaughan2018}).  
In detail, the SSP models of \citet{ConroyvanDokkum2012a} predict that varying [Na/Fe] by $\pm0.3$ dex for a 13.5 Gyr stellar population with a Chabrier IMF would change the strength of the NaD index by ${\approx}+30\%/-20\%$ relative to the value predicted at solar metallicity.  This suggests that a ${>}10\%$ correction to SSP model NaD indices would result in a ${\sim}+0.1-0.2$ dex enhancement in [Na/Fe], and thus an overall Na abundance of ${\sim}+0.7-1.0$ dex.}
In the case that there is a younger component toward the centers of these galaxies \citep{Goddard2017}, the effect of interstellar contamination would vary systematically with radius.  
These studies have further noted that the enhanced [Na/Fe] abundances they estimate are inconsistent with the Type II supernova yields calculated by \citet{WoosleyWeaver1995}, and 
are not observed in individual stars in the Milky Way \citep{Bensby2014,Bensby2017}. 
However, the updated yield calculations by \citet{Kobayashi2006} are in better accord with early-type galaxy abundances, as they imply that [Na/Fe] increases strongly with increasing progenitor metallicity while remaining consistent with observed Galactic stellar abundance ratios \citep{Kobayashi2020}.
The prospect that early-type galaxy [Na/Fe] measurements are underestimated due to ISM contamination may therefore be fully consistent with the Na enrichment levels expected for such massive systems.\\

{\subsubsection{Complications from External Galaxy ISM Absorption}}

{The effects of these systematics are further complicated by the possible presence of cool \ion{Ca}{2}- and/or \ion{Na}{1} D-absorbing material in the external galaxies being studied.  Starburst galaxies are indeed well-known to exhibit both \ion{Ca}{2} and \ion{Na}{1} interstellar absorption \citep{Heckman2000,Rupke2005a,Zych2007,ChenTremonti2010,Concas2019,RobertsBorsani2020,Veilleux2020,Rubin2022}, with stronger absorption associated with higher star formation rates \citep[e.g.,][]{ChenTremonti2010,Straka2015,Rubin2022}.  In particular, interstellar $W($\ion{Ca}{2} K) values of ${\sim}0.3-1$ \AA\ are observed to arise in hosts with SFRs $\gtrsim 1~M_{\odot}~\rm yr^{-1}$ \citep{Rubin2022}.  These absorption strengths are comparable to the equivalent width enhancement due to Milky Way interstellar \ion{Ca}{2} (i.e., $W$(\ion{Ca}{2} K)$^{\rm SSP}_{\rm orig} - W$(\ion{Ca}{2} K)$^{\rm SSP}_{\rm clean}$ values fall in the range $0.5-1.5$ \AA\ for young SSPs).  We therefore expect that in practice, the effect of the presence of Milky Way ISM in stellar population synthesis analyses of starbursting systems is partially or completely compensated by the presence of ISM in the host.  That is, an analysis that fits ``contaminated'' SSP templates to observed \ion{Ca}{2} H\&K profiles, ignoring the effects of the ISM altogether, will not underestimate the galaxy's age to the extent implied by Figure~\ref{fig:SSPs}, and indeed may overestimate its age in some cases.     

Early-type galaxies, on the other hand, are commonly assumed to exhibit minimal interstellar \ion{Na}{1} D absorption \citep[e.g.,][]{ConroyvanDokkum2012b,Spiniello2012,LaBarbera2013,Parikh2018,Gu2022,Lonoce2023,denBrok2024,Parikh2024,Maksymowicz-Maciata2024}.  \citet{Parikh2018} tested this assumption by deriving the spatially-resolved $E(B-V)$ profile for SDSS-IV/MaNGA galaxies from spectra coadded in radial bins (and having stellar masses $9.9 < \log M_*/M_{\odot} < 10.8$).  These authors measured values of ${\sim}0.1$ mag toward the galaxy centers and ${\lesssim}0.05$ mag toward the outskirts.  The \citet{Poznanski2012} relation, derived for the Milky Way's ISM and halo, implies a \ion{Na}{1} absorption strength of ${\approx}0.7$ \AA\ for $E(B-V)=0.1$.  While this relation may not be appropriate for the physical conditions in the ISM of early-type galaxies, this absorption strength nevertheless exceeds the offset $\rm NaD_{\rm orig} - NaD_{\rm clean}$ we observe for ${\sim}10$ Gyr old SSPs (${\sim} 0.3-0.5$ \AA).  On the other hand, a Herschel Space Observatory survey of dust emission at 250, 350, and 500$\mu$m across ${\sim}22.6$K ellipticals detected significant emission in only $13\%$ of the sample \citep{Lesniewska2023}.  Early-type galaxies have also been surveyed for both neutral hydrogen and molecular gas, with approximately two-thirds of field early types exhibiting \ion{H}{1} masses greater than a few times $10^6 M_{\odot}$, and fewer than 10\% of cluster galaxies detected to the same limit \citep{Oosterloo2010}.  Molecular gas (traced by CO) is likewise detected in ${\approx}20\%$ of the $\rm ATLAS^{3D}$ and MASSIVE samples \citep{Young2011,Davis2019}, to a gas mass detection limit of 0.1\% of the stellar mass.  This molecular material is moreover observed to be concentrated toward the centers of the systems \citep[e.g.,][]{Alatalo2013,Davis2013,Ruffa2019}.  Given that both neutral and molecular hydrogen could harbor the \ion{Na}{1} ion, it is therefore likely that at least some early types exhibit interstellar \ion{Na}{1} D absorption that, when fit with ``contaminated'' SSP templates, would compensate for the tendency to underestimate [Na/Fe].  This possibility should be considered in all stellar population synthesis analyses that include \ion{Na}{1} D.}\\

{\subsubsection{Implications for IMF Slope Constraints}}

[Na/Fe] measurements in turn play a role in anchoring spectroscopic constraints on the IMF slope.  There are several surface-gravity-sensitive transitions in the rest-frame optical and near-IR that have been explored as IMF indicators, including \ion{Na}{1} $\lambda\lambda 8183,8195$, the FeH $\lambda 9916$ Wing-Ford band, the \ion{Ca}{2} $\lambda\lambda 8498, 8542, 8662$ triplet \citep{WingFord1969,SpinradTaylor1971,Whitford1977,FaberFrench1980,vanDokkumConroy2010,vanDokkumConroy2012,Martin-Navarro2015}, and $\rm TiO_2~\lambda 6230$ \citep{Spiniello2012}.  Many recent studies have focused their analyses on the former three sets of transitions, as Na, Fe, and Ca abundances may be tightly constrained from complementary (and IMF-insensitive) transitions \citep[e.g.,][]{vanDokkumConroy2010,Smith2012,ConroyvanDokkum2012a,LaBarbera2013,Parikh2018,LaBarbera2019,Gu2022}.  The strength of both \ion{Na}{1} $\lambda \lambda 8183, 8195$ and FeH increase with the fraction of dwarf stars (as well as with Na and Fe abundances), while the strength of the Ca triplet increases with the fraction of giant stars.  Our suggestion that the presence of ISM contamination in empirical stellar libraries leads to a systematic underestimation of the Na abundance {(assuming the absence of significant interstellar \ion{Na}{1} in the host galaxy)} implies that, when accounted for, a less bottom-heavy (i.e., more Milky Way-like) IMF will be needed to explain the observed strengths of \ion{Na}{1} $\lambda \lambda 8183, 8195$ in massive ellipticals.  {We further posit that the same (or a greater) degree of ISM contamination could be present in SSP models of lower-mass ellipticals, for which a Milky Way-like IMF is typically derived \citep[e.g.,][]{ConroyvanDokkum2012b,LaBarbera2013,Parikh2018}.  By the same token, the fraction of dwarf stars in these models may need to be reduced in order to avoid the overprediction of \ion{Na}{1} $\lambda \lambda 8183, 8195$ for such systems, pushing the models toward bottom-light IMFs.  More detailed modeling is required to test whether ISM contamination has had a statistically significant impact on these spectroscopic IMF constraints.}

The case for a steeper IMF slope in more massive early-type systems by no means rests solely on [Na/Fe] and \ion{Na}{1}.   Several independent, corroborating lines of evidence for a variable IMF come from studies relying on dynamical mass constraints from gravitational lensing or galaxy kinematics \citep[e.g.][]{Treu2010,ThomasJ2011,Cappellari2012,Lyubenova2016}. 
\citet{ConroyvanDokkum2012b} found evidence for significant IMF variation even after excluding \ion{Na}{1} lines from their analysis; however, they also found that their derived mass-to-light ratios changed by ${\gtrsim}50\%$ as a result.
\ion{Na}{1} D and \ion{Na}{1} $\lambda \lambda 8183, 8195$ thus
play a crucial role in setting IMF constraints, including for studies that make use of full spectrum fitting, in part because FeH has confounding sensitivities to [Fe/H], age, and $\rm [\alpha/Fe]$ \citep{ConroyvanDokkum2012b,LaBarbera2013,Parikh2018}.  
We argue that precision constraints on the IMF slope that rely on analysis of \ion{Na}{1} transitions must account for the systematic effects we have elucidated.

\section{Conclusion}

The cool material that pervades the Milky Way's ISM has long been understood to give rise to \ion{Ca}{2} $\lambda \lambda 3934, 3969$ and \ion{Na}{1} $\lambda \lambda 5891, 5897$ absorption in optical spectroscopy of stars and QSOs \citep{Hobbs1969,Hobbs1974,Crawford1992,Sembach1993,Welty1996,Welsh2010,Poznanski2012,Murga2015,Bish2019}.  We have quantified the impact of interstellar absorption in these transitions on the spectra of 23,771 stars comprising the SDSS-IV MaNGA Stellar Library (\citealt{Yan2019,Chen2020,Abdurrouf2022,Hill2022,Imig2022,Lazarz2022}).  MaStar includes over an order of magnitude more stars than any other extant empirical stellar library, and thus enables population synthesis that adequately captures the diversity of the MaNGA sample of $\sim$10,000 nearby galaxies \citep{Yan2019}.

Our analysis leverages high-resolution spectroscopic observations of interstellar \ion{Ca}{2} and \ion{Na}{1} absorption from \citet{Sembach1993}, \citet{MunariZwitter1997}, and \citet{Welsh2010} to develop a model of the equivalent widths of these transitions as a function of stellar distance, Galactic latitude, and the dust reddening of the stellar sightline.  We apply this model to the MaStar sample, making use of stellar distances and reddening values available from \citet{Green2019} and \citet{Bailer-Jones2021}.  We find that the predicted equivalent widths of ISM absorption are roughly uniformly distributed across the ranges $0~\mathrm{\AA} < W^{\rm ISM}$(\ion{Ca}{2} K) $ < 0.6~\mathrm{\AA}$, $0~\mathrm{\AA} < W^{\rm ISM}$(\ion{Na}{1} 5891) $ < 1.0~\mathrm{\AA}$, and $0~\mathrm{\AA} < W^{\rm ISM}$(\ion{Na}{1} 5897) $ < 0.7~\mathrm{\AA}$ for stars with effective temperatures $T_{\rm eff} > 7610$ K, whereas cooler stars are predicted to have median interstellar absorption strengths of $W^{\rm ISM}$(\ion{Ca}{2} K) $=0.11$ \AA, $W^{\rm ISM}$(\ion{Na}{1} 5891) $=0.15$ \AA, and $W^{\rm ISM}$(\ion{Na}{1} 5897) $=0.09$ \AA.  

We then use this simple model to identify a subset of 6342 ``low-ISM'' stars for which the interstellar contamination level is minimal ($W^{\rm ISM}$(\ion{Ca}{2} K) $<0.07$ \AA\ and $W^{\rm ISM}$(\ion{Na}{1} D) $<0.05$ \AA).  For 12,110 of the remaining stars, we remove interstellar contamination from each star by identifying a subset of the low-ISM sample with similar stellar parameters, coadding the spectra in this subset, and replacing the \ion{Na}{1} D profile in the affected spectrum with that in the coadd.  We execute the same replacement of \ion{Ca}{2} H\&K only in the very small subset of these stars with $T_{\rm eff} > $ 9000 K, as we found that this replacement does not yield any systematic reduction in the \ion{Ca}{2} K equivalent widths in cooler stars.  For those 738 stars for which we could not identify a low-ISM replacement subset that was sufficiently close in stellar parameter space and which have $T_{\rm eff} >$ 9000 K, we select well-matched stars from the theoretical stellar library of \citet{BOSZ2017} to replace the \ion{Ca}{2} H \& K and \ion{Na}{1} D spectral regions.  

This procedure results in a mean reduction in $W$(\ion{Ca}{2} K) of 0.4, 0.7, and 0.5 \AA\ for A, B, and O spectral types, and a mean reduction in $W$(\ion{Na}{1} D) of $0.1-0.2$ \AA\ for stars with $T_{\rm eff} < 7610$ K, and of 0.6, 1.1, and 1.0 \AA\ for A, B, and O spectral types.  
We additionally find that the degree of ISM contamination of \ion{Na}{1} D is larger in giant stars (having $\log g < 4$), and that it systematically increases with stellar [Fe/H] in these giants.
{Our catalog of ``cleaned'' stellar spectra is publicly available via Zenodo with \dataset[doi:10.5281/zenodo.14014915]{https://doi.org/10.5281/zenodo.14014915}.}  We also make public a hierarchically-clustered stellar template library constructed from our
cleaned MaStar spectra that is suitable for continuum modeling of external galaxy spectroscopy.\footnote{The hierarchically-clustered templates are available at \url{https://github.com/sdss/mangadap/tree/4.2.0/mangadap/data/spectral\_templates/mastarhc\_v2\_noism}.}  

Finally, we demonstrate the impact of this interstellar absorption on stellar population analyses by constructing simple stellar population (SSP) templates from both the ``cleaned'' and original stellar spectra.  {We find that MaStar SSPs constructed from the original spectra overestimate the strength of \ion{Ca}{2} K absorption by ${\gtrsim} 20\%$ in relatively young (${\lesssim} 400$ Myr) stellar populations at solar metallicity.} We further demonstrate that the presence of interstellar \ion{Na}{1} absorption in SSP templates constructed from the original MaStar spectra implies (1) a dramatic overestimate of the strength of stellar \ion{Na}{1} D absorption in starbursting systems (by ${\gtrsim}50\%$); and (2) an
overestimate of the strength of the NaD index in older stellar populations (${\gtrsim} 10$ Gyr) by ${\gtrsim}10\%$.  The former effect would systematically reduce the equivalent width attributed to the host's ISM, inflows, and outflows in ``down-the-barrel'' analyses of \ion{Na}{1} D kinematics \citep[e.g.,][]{ChenTremonti2010,RobertsBorsani2019,Avery2022}, and would do so to a more extreme degree in the youngest starburst systems.  The latter effect would lead to a systematic underestimation of [Na/Fe] in early-type galaxies {(under the assumption that the galaxies themselves do not harbor significant interstellar \ion{Na}{1})}, potentially weakening the requirement for a steep IMF slope to match near-IR \ion{Na}{1} absorption line strengths \citep[e.g.,][]{ConroyvanDokkum2012b,LaBarbera2013,Martin-Navarro2015,Parikh2018}.

Our findings suggest that the Milky Way's ISM is latent in any empirical stellar library which lacks the spectral resolution to distinguish stellar from interstellar absorption \citep[e.g.,][]{LeBorgne2003,Valdes2004,Sanchez-Blazquez2006}.  Our model for interstellar absorption equivalent widths may be used to estimate the magnitude of these effects on previous analyses.  
Moreover, our ``cleaned'' SSP templates {are publicly available}\footnote{The full set of SSP templates is available at \url{https://doi.org/10.5281/zenodo.14807331}.  A subset of the SSP templates that have been modified for use with the MaNGA Data Analysis Pipeline is available at \url{https://github.com/sdss/mangadap/tree/4.2.0/mangadap/data/spectral_templates/mastar_ssp_noism_v1.0}.}, and may now be used to 
quantitatively assess the systematic effects of interstellar \ion{Ca}{2} and \ion{Na}{1} absorption on stellar age, [Na/Fe], and IMF slope constraints.

\section*{Acknowledgments}

The authors are grateful for support for this project from NSF grants AST-1715630, AST-1715898, and AST-2009417.  
R.Y.\ acknowledges the support of two grants from the Research Grants Council of the Hong Kong Special Administrative Region, China (Project No: CUHK 14303123, CUHK 14302522) and a grant from the National Science Foundation of China (No.12373008). R.Y.\ also acknowledges support from the Hong Kong Global STEM Scholar scheme, by the Hong Kong Jockey Club Charities Trust through the JC STEM Lab of Astronomical Instrumentation.

We thank Dan Welty for sharing his Milky Way absorption line data with us.  We wish to also thank the MaNGA team for their extraordinary efforts to obtain and process this exceptionally high-quality dataset.  It is a pleasure to thank Scott Trager and Else Starkenburg for sharing their extensive expertise in stellar spectroscopy which led to improvements in this manuscript.  K.H.R.R. thanks the astrophysicists at the University of Pittsburgh for their warm hospitality during her sabbatical stay, during which much of this manuscript was written, and thanks Evan Schneider and Jessica Werk for numerous helpful conversations and writing support.  {The authors also wish to acknowledge the anonymous referee, who provided valuable feedback that improved this work.} 

Funding for the Sloan Digital Sky 
Survey IV has been provided by the 
Alfred P. Sloan Foundation, the U.S. 
Department of Energy Office of 
Science, and the Participating 
Institutions. 

SDSS-IV acknowledges support and 
resources from the Center for High 
Performance Computing  at the 
University of Utah. The SDSS 
website is www.sdss4.org.

SDSS-IV is managed by the 
Astrophysical Research Consortium 
for the Participating Institutions 
of the SDSS Collaboration including 
the Brazilian Participation Group, 
the Carnegie Institution for Science, 
Carnegie Mellon University, Center for 
Astrophysics Harvard \& 
Smithsonian, the Chilean Participation 
Group, the French Participation Group, 
Instituto de Astrof\'isica de 
Canarias, The Johns Hopkins 
University, Kavli Institute for the 
Physics and Mathematics of the 
Universe (IPMU) / University of 
Tokyo, the Korean Participation Group, 
Lawrence Berkeley National Laboratory, 
Leibniz Institut f\"ur Astrophysik 
Potsdam (AIP),  Max-Planck-Institut 
f\"ur Astronomie (MPIA Heidelberg), 
Max-Planck-Institut f\"ur 
Astrophysik (MPA Garching), 
Max-Planck-Institut f\"ur 
Extraterrestrische Physik (MPE), 
National Astronomical Observatories of 
China, New Mexico State University, 
New York University, University of 
Notre Dame, Observat\'ario 
Nacional / MCTI, The Ohio State 
University, Pennsylvania State 
University, Shanghai 
Astronomical Observatory, United 
Kingdom Participation Group, 
Universidad Nacional Aut\'onoma 
de M\'exico, University of Arizona, 
University of Colorado Boulder, 
University of Oxford, University of 
Portsmouth, University of Utah, 
University of Virginia, University 
of Washington, University of 
Wisconsin, Vanderbilt University, 
and Yale University.

\clearpage
\appendix

\section{Telluric Contamination of the \ion{Na}{1} D Spectral Region}\label{sec:appendix-telluric}

While absorption features arising in Earth's atmosphere are most prominent at wavelengths $\lambda > 6800$ \AA, there are several comparatively weak telluric features in the wavelength range $5885~\mathrm{\AA} < \lambda_{\rm air} < 5906~\mathrm{\AA}$\footnote{\url{https://sites.astro.caltech.edu/~tb/makee/}} \citep[e.g.,][]{Lallement1993,Chen2014,Sandford2023}.  At the $\mathcal{R}\sim1800$ resolution of MaStar, these features are fully blended with the lines of the \ion{Na}{1} D doublet and cannot be removed using traditional techniques for telluric correction. As described in \citet{Yan2016} and \citet{Yan2019}, telluric correction was performed for MaStar by fitting a high-order cubic basis spline (B-spline) function to the ratio of fluxes observed in standard stars to a matching theoretical stellar template.  However, these high-frequency corrections were performed only in spectral regions with severe telluric contamination (all redward of 6842 \AA).  In all other spectral regions, the flux calibration vector was derived in two steps.  First, a large sample of standard star observations was coadded and divided by a matching theoretical template.  Complex spectral regions (including Balmer transitions, \ion{Ca}{2} H\&K, and \ion{Na}{1} D) were masked, and then a B-spline with break points spaced every 10 pixels was fit to this ratio to derive an average calibration vector.  A time-dependent correction to this curve was determined by fitting a lower-order B-spline (with break points spaced every 160 pixels) to the standard star spectra on individual plates (again normalized by matching spectral templates), and applied in addition to the average calibration vector.  

\begin{figure*}[h]
\centering
\includegraphics[width=4.5in,trim={0cm 0cm 0cm 0cm},clip]{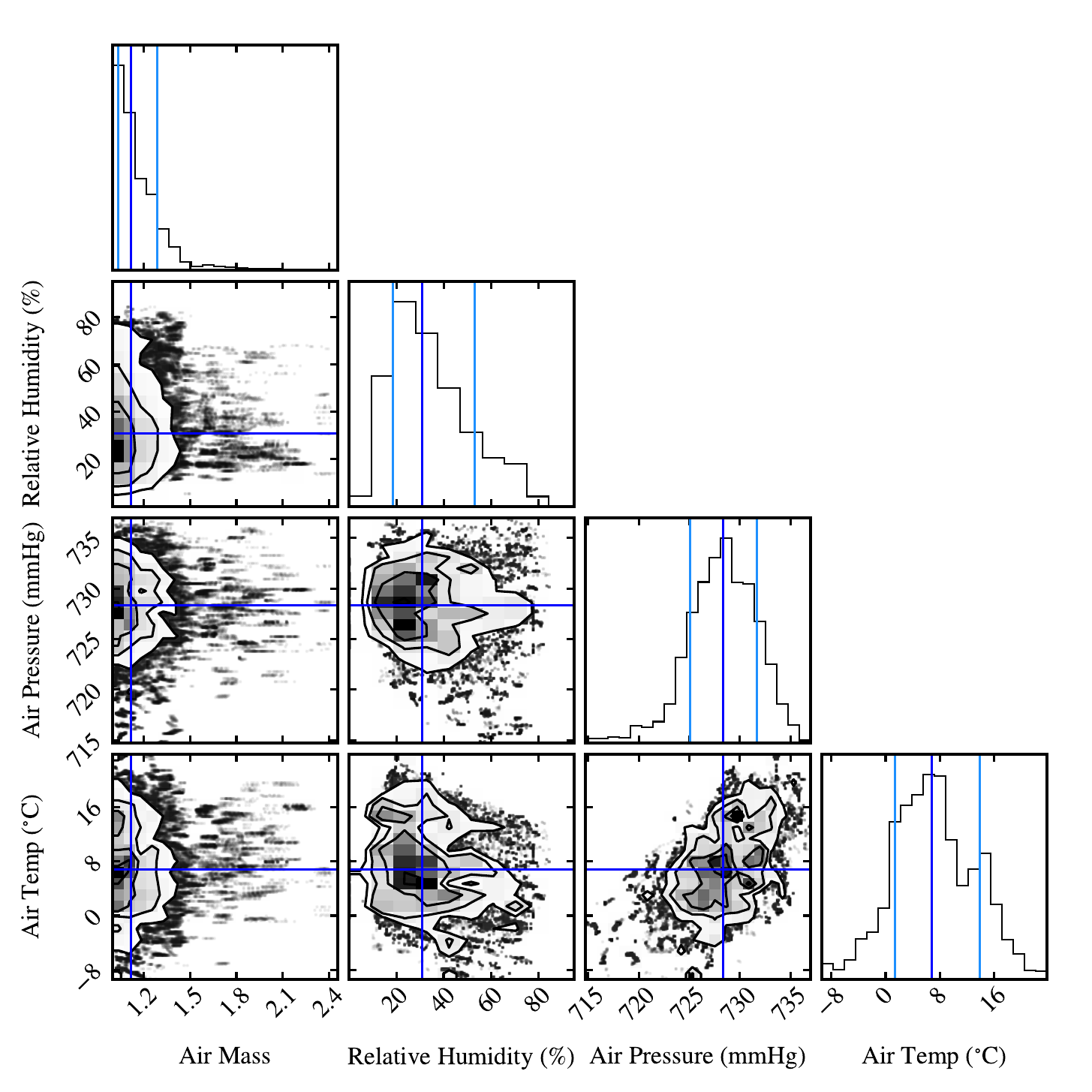}
\caption{Distributions of the atmospheric conditions for all high-quality exposures used in ``good'' visit spectra in the MaStar library.  This includes all spectra with no \texttt{MASTAR\_QUAL} flags set. The median of each distribution is indicated in dark blue, with the $16$th- and $84$th-percentile values indicated in light blue.  
\label{fig:mastar_cond}}
\end{figure*}

This procedure in effect leaves all telluric absorption features in the final stellar spectra near \ion{Na}{1} D.  We investigate the potential impact this may have on our analysis as follows.  First, because telluric absorption is a strong function of the atmospheric conditions (i.e., air temperature, air pressure, relative humidity) and air mass, we assemble this information for each of the high-quality exposures comprising the ``good'' MaStar visit spectra\footnote{This includes all exposures with the \texttt{USED\_IN\_VISIT} flag set, for all visit spectra with \texttt{MJDQUAL} bits set to indicate high quality.} as described in the SDSS-IV DR17 documentation.\footnote{https://www.sdss4.org/dr17/mastar/mastar-spectra/}
We show the distributions of these conditions in Figure~\ref{fig:mastar_cond}.  The median values of the air mass, relative humidity, air pressure, and air temperature are 1.12, $31\%$, 728 mmHg, and $6.8^{\circ}$C, respectively.

\begin{figure*}[ht]
\includegraphics[width=\textwidth,trim={0cm 0cm 0cm 0cm},clip]{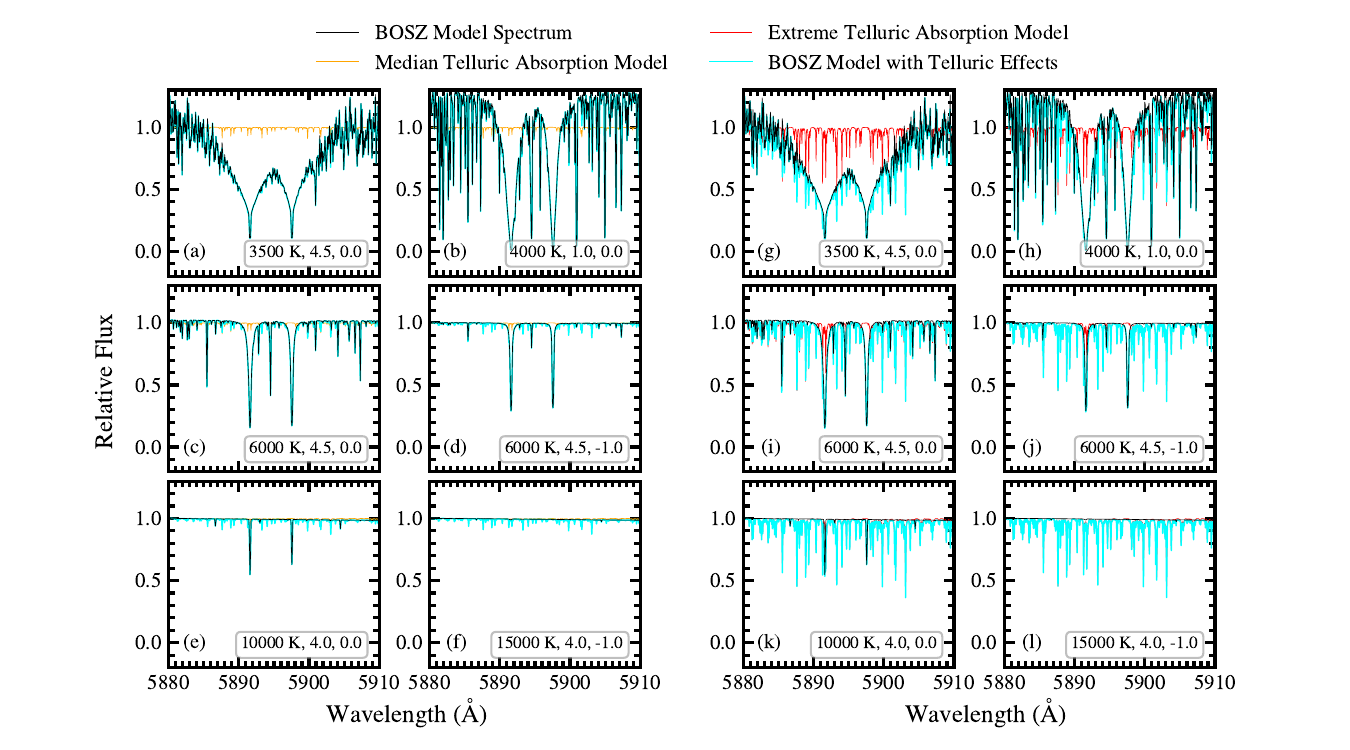}
\caption{\emph{Panels (a)-(f):} The atmospheric transmission spectrum predicted by TelFit/LBLRTM for Apache Point Observatory
for an air mass $=1.1$, 30\% relative humidity, atmospheric pressure $=730$ mmHg, and an air temperature of $5.5^{\circ}$C is shown in orange.  The black spectra show a selection of six BOSZ theoretical stellar templates calculated at $\mathcal{R} = 300,000$.  The $T_{\rm eff}$, $\log g$, and [M/H] values for each template are printed in the lower right of each panel.  We have rebinned the telluric model to match that of the BOSZ templates, and show the product of the two in cyan. \emph{Panels (g)-(l):}  An APO TelFit/LBLRTM atmospheric  transmission spectrum for an air mass $=1.6$, $70\%$ relative humidity, atmospheric pressure $=730$ mmHg, and an air temperature of $20.5^{\circ}$C is shown in red.  The black spectra are the same as in the corresponding panel at left.  The cyan spectrum in each panel shows the product of the theoretical template and this more extreme telluric absorption.
\label{fig:tell_spec_hires}}
\end{figure*}

\begin{figure*}[ht]
\includegraphics[width=\textwidth,trim={0cm 0cm 0cm 0cm},clip]{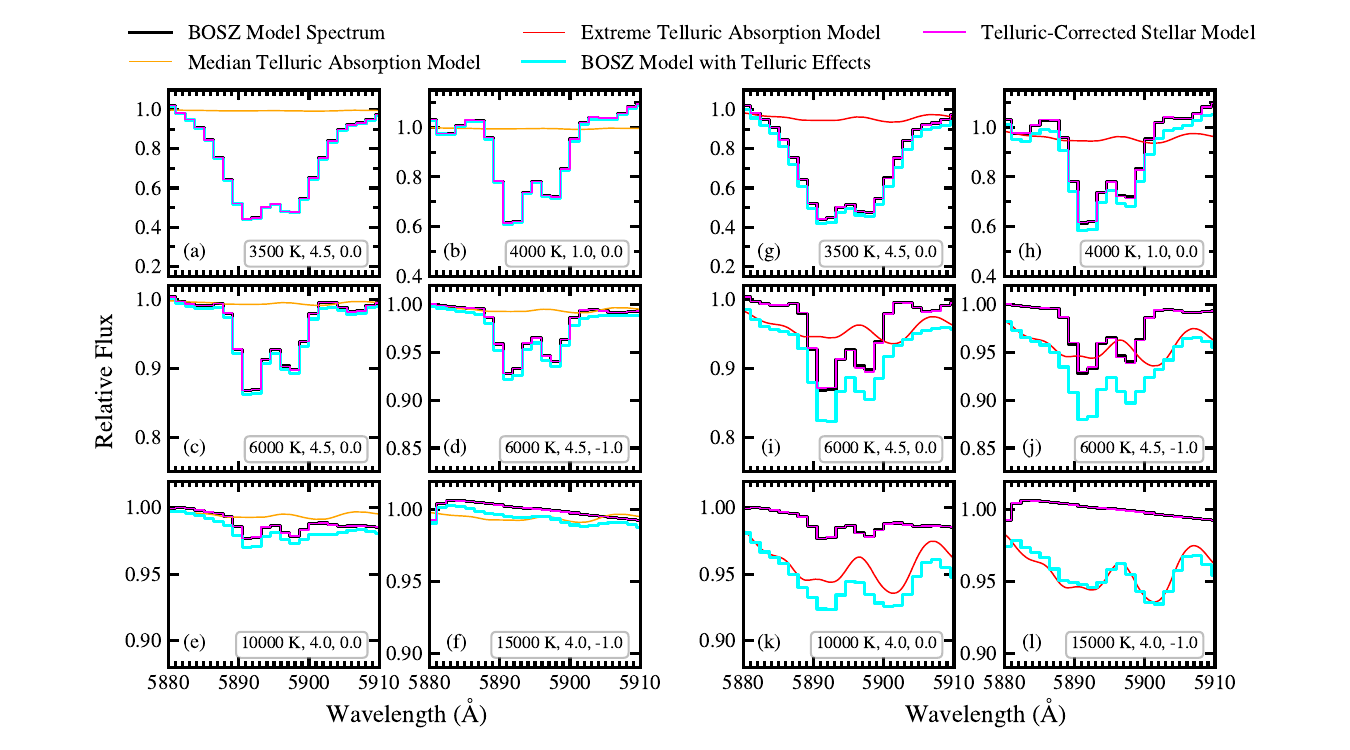}
\caption{The BOSZ model spectra (black), telluric absorption models (orange and red), and combined spectra (cyan) shown in Figure~\ref{fig:tell_spec_hires}, smoothed to the resolution of the MaStar 99.5th-percentile LSF.  The magenta spectra show the ratio of each smoothed combined spectrum and the smoothed telluric absorption model.  The magenta spectra lie nearly exactly on top of the original BOSZ models, indicating that this simple approach to telluric correction is effective.  
\label{fig:tell_spec_lores}}
\end{figure*}

We then use the TelFit Python package\footnote{https://telfit.readthedocs.io/en/latest/} to calculate a grid of telluric atmospheric transmission spectra appropriate for these conditions.  TelFit is a wrapper for the LBLRTM FORTRAN code \citep{Clough1992,Clough2005}, a radiative transfer modeling code that draws from the
HITRAN database of molecular spectroscopic parameters \citep{HITRAN2020}.  The grid contains atmospheric transmission spectra for humidities in the range $10-70$\%, air pressures in the range $720-735$ mmHg, air temperatures in the range $-4.5^{\circ} \rm C - +20.5^{\circ} C$, and air masses between 1.0 and 1.7. These ranges encompass the 2.5th- and 97.5th-percentile values of the distributions shown in Figure~\ref{fig:mastar_cond}.

We show two examples of these predictions in Figure~\ref{fig:tell_spec_hires}.  In the left-hand set of panels (\emph{(a)-(f)}), a model that assumes an air mass $=1.1$, atmospheric pressure $=730$ mmHg, 30\% relative humidity, and an air temperature of $5.5^{\circ}$C is shown in orange.  For comparison, we also show a selection of six BOSZ
theoretical stellar spectra calculated with $\mathcal{R} = 300,000$ 
for a broad range of stellar parameters (approximately following the locations in parameter space numbered in Figure~\ref{fig:psimin_NaI}).  
The product of each of these models and the atmospheric transmission (telluric) spectrum is in principle representative of the observed spectrum in these atmospheric conditions.  We have rebinned the telluric spectrum to match that of the BOSZ templates, and show their products in cyan.  A preponderance of weak telluric features across this spectral region, greater in strength than the stellar \ion{Na}{1} D absorption in one case, is evident.  In the right-hand set of panels (\emph{(g)-(l)}), we show a telluric absorption model calculated assuming more extreme conditions: we adopt an air mass $=1.6$, $70\%$ relative humidity, and an air temperature of $20.5^{\circ}$C.  Such atmospheric conditions were very rare during the course of the MaStar observations, but are representative of the extremes of the distributions shown in Figure~\ref{fig:mastar_cond}.  
The product of each BOSZ template and this telluric model is again shown in cyan.  In this case, the telluric absorption contributes significant equivalent width, and dominates over that produced by the two hottest stars.

To estimate the impact of these features on the \ion{Na}{1} D equivalent widths we observe in the MaStar spectra, we repeat the following procedure for each of these stellar templates at each telluric model grid point.  First, we smooth the BOSZ models and the product of the BOSZ and telluric models to the 99.5th-percentile spectral resolution curve of MaStar.  We then rebin the BOSZ and product  spectra to a wavelength bin width matching that of the MaStar catalog (i.e., $1.357~\rm \AA~pix^{-1}$ near \ion{Na}{1} D). The results of this exercise for the telluric models displayed in Figure~\ref{fig:tell_spec_hires} are shown in Figure~\ref{fig:tell_spec_lores}.

We then establish the continuum level around the \ion{Na}{1} D feature by fitting a linear model to the flux in the spectral windows $5881.0~\mathrm{\AA} < \lambda < 5885.0~\mathrm{\AA}$ and $5904.0~\mathrm{\AA} < \lambda < 5908.0~\mathrm{\AA}$ (as is done in our MaStar analysis), noting that this level is affected by the overall strength of the telluric absorption.  Finally, we measure the boxcar equivalent width in the spectral window $5885.0~\mathrm{\AA} < \lambda < 5904.0~\mathrm{\AA}$ and compute the difference between that measured in the product spectrum and that measured in the original spectrum of each star ($\Delta W_{\rm tell}$).

\begin{figure}[ht]
\centering
\includegraphics[width=4.5in,trim={0cm 0cm 0cm 0cm},clip]{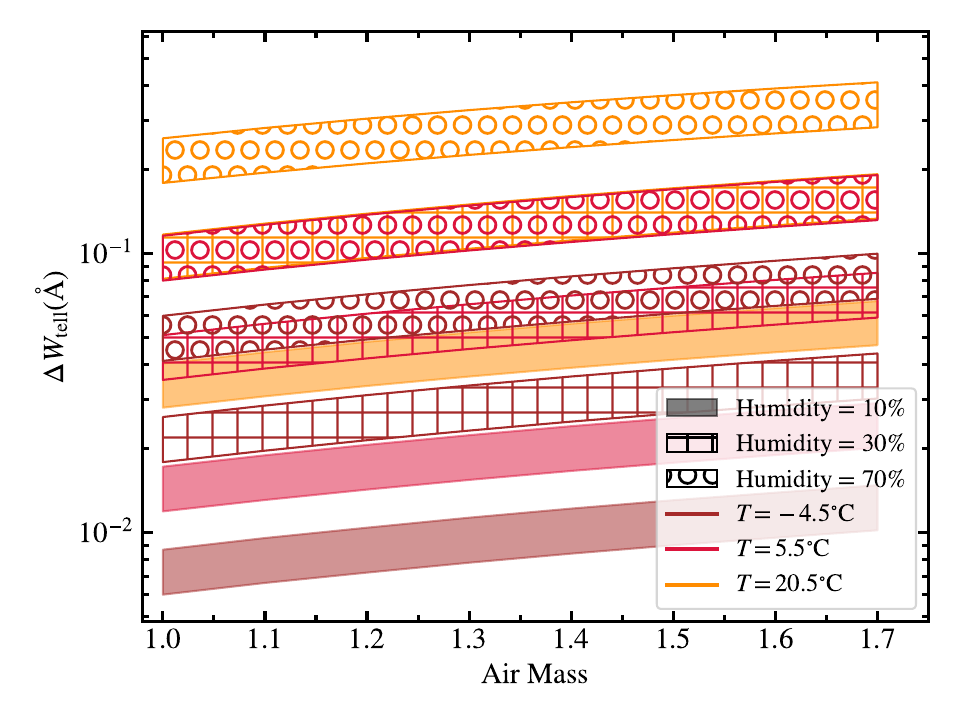}
\caption{Equivalent width enhancement due to telluric absorption within the \ion{Na}{1} D spectral window predicted by TelFit/LBLRTM for APO.  All models shown adopt atmospheric pressures of 730 mmHg.  
Each contour shows the full range of equivalent width enhancements exhibited by the BOSZ stellar templates analyzed in this section as a function of the air mass.  
Solid, hatched, and circle-hatched contours show results for models assuming $10$\%, $30$\%, and $70$\% relative humidity, respectively.  At each of these humidity levels, we show results for models adopting air temperatures of $-4.5^{\circ}$C in brown, $5.5^{\circ}$C in red, and $20.5^{\circ}$C in orange.  
\label{fig:tell_am}}
\end{figure}

We summarize these results in Figure~\ref{fig:tell_am}.  
Each contour indicates the range in $\Delta W_{\rm tell}$ values exhibited by the six BOSZ stellar templates for telluric models with the specified relative humidity and air temperature as a function of the air mass.  
We find that at very low humidities (10\%), the strength of these features remains $< 0.1$ \AA\ at all air masses and temperatures covered in the grid.  Higher humidities at warm temperatures can yield a significantly stronger telluric contribution (e.g., up to $0.2-0.4$ \AA\ for $70\%$ humidity at $20.5^{\circ}$C).  
However, the model that best represents the median atmospheric conditions during the MaStar observations (with 30\% relative humidity, an air temperature of $5.5^{\circ}$, and an air pressure of 730 mmHg) produces a $\Delta W_{\rm tell} \lesssim 0.06$ \AA\ at air masses $<1.2$.  
Among the 204,185 exposures included in Figure~\ref{fig:mastar_cond}, 4.6\% were obtained at ${>}70\%$ relative humidity, and 2.1\% were obtained at ${>}70\%$ relative humidity with air temperatures ${>} 5.5^{\circ}$C.  Fewer than 1\% of all exposures were taken in air temperatures ${>}20.5^{\circ}$C, and all of these latter exposures were obtained with relative humidities ${<}39\%$.

Together, these findings imply that the contribution of telluric absorption to our measured \ion{Na}{1} equivalent widths is negligible in comparison to the effect of ISM contamination for most of the MaStar sightlines, but that telluric absorption may be significant in a small minority of cases.  
At the same time, the degree of telluric absorption should not have any dependence on the reddening or distance to the target stars, as is the case for interstellar absorption.  Telluric absorption will therefore introduce an additional, very slight systematic enhancement of \ion{Na}{1} equivalent widths across our sample as a whole.  

We do not attempt to correct for this enhancement. However, we have found that it is possible to recover the intrinsic \ion{Na}{1} profile of a star when observed at low spectral resolution in the case that the telluric absorption model is known.  For each telluric model $+$ BOSZ template combination shown in Figure~\ref{fig:tell_spec_lores} (in cyan), we smooth the telluric model to the same spectral resolution as the data, and then divide the product spectrum by the smoothed telluric.  The results are shown with the magenta spectra, which in all cases overlap completely with the original, smoothed BOSZ spectra.  While this approach is not mathematically sound (as the observed spectrum is convolved with the LSF after the starlight is absorbed by the atmosphere), it nevertheless yields the same \ion{Na}{1} equivalent widths as are measured in the original smoothed BOSZ spectra to within ${<}0.03$ \AA.  It should therefore be feasible to correct the \ion{Na}{1} region of low-resolution spectroscopy for telluric effects by constraining the atmospheric absorption model using ``clean'' spectral regions at longer wavelengths.

\clearpage 

\section{\ion{Ca}{2} and \ion{Na}{1} in Hierarchically-Clustered MaStar Templates}\label{sec:appendix-hctemplates}

In Figure~\ref{fig:hc_velplots_2} we present the \ion{Ca}{2} K and \ion{Na}{1} profiles for the hierarchically-clustered templates described in Section~\ref{subsec:hc_templates} which were not shown in Figure~\ref{fig:hc_velplots}.  

\begin{figure*}[th]
\includegraphics[width=1.0\textwidth]{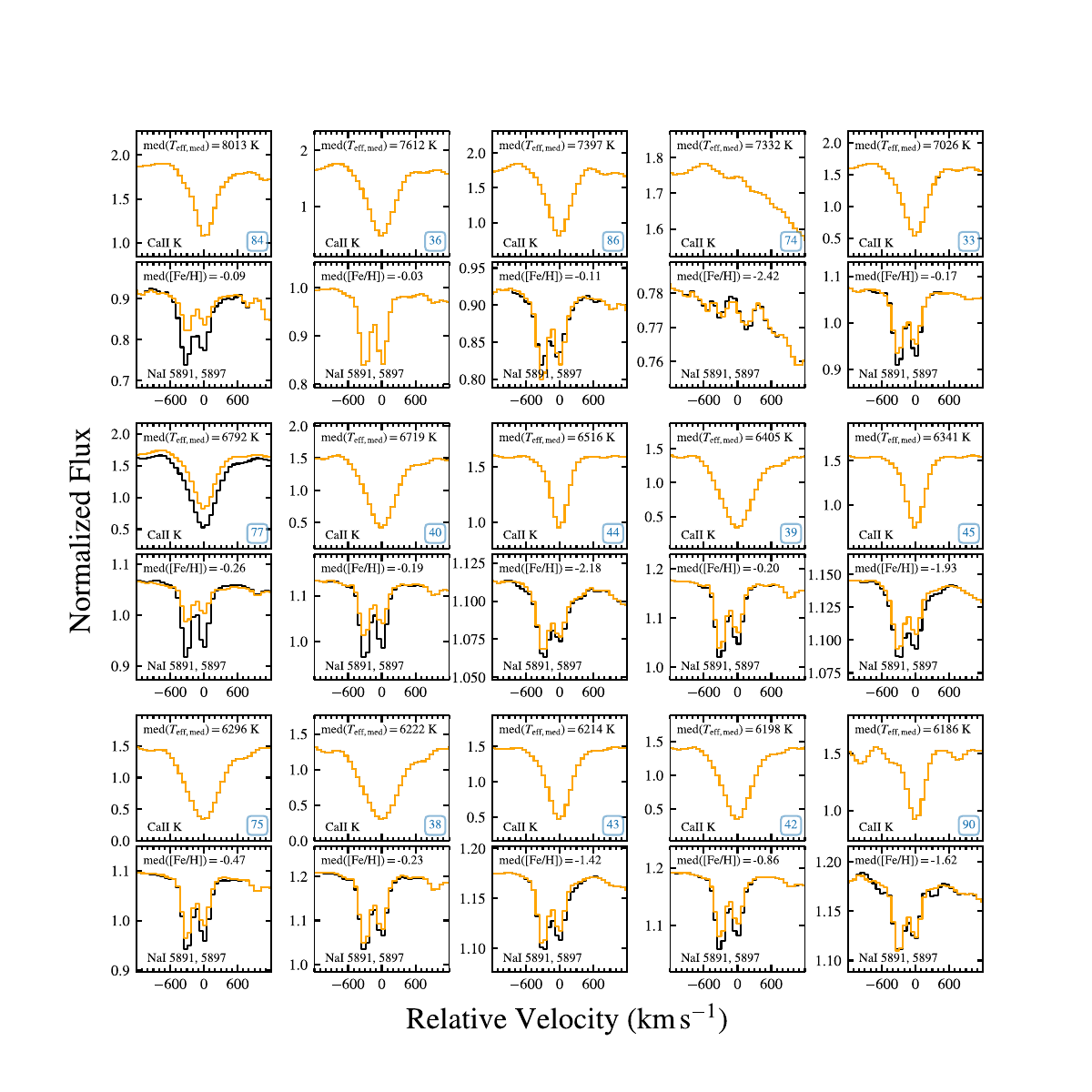}
\caption{Comparison of hierarchically-clustered MaStar spectral templates constructed without correction for ISM absorption (black) and from our ``cleaned'' spectral sample (orange).  This is a multipage continuation of Figure~\ref{fig:hc_velplots}. Each pair of stacked panels shows the same template at the locations of the \ion{Ca}{2} K and \ion{Na}{1} $\lambda \lambda 5891, 5897$ transitions on top and bottom, respectively.  Velocities are computed relative to the \ion{Ca}{2} K $\lambda3934$ and \ion{Na}{1} $\lambda 5897$ rest wavelengths. 
The templates have been ordered according to the median $T_{\rm eff,med}$ value of the stars used in each.  This value, along with the median $\rm [Fe/H]$ value, is noted in each panel pair.  The cluster ID of each template is indicated in blue. \label{fig:hc_velplots_2}}
\end{figure*}
\setcounter{figure}{24}
\begin{figure*}[th]
\includegraphics[width=1.0\textwidth]{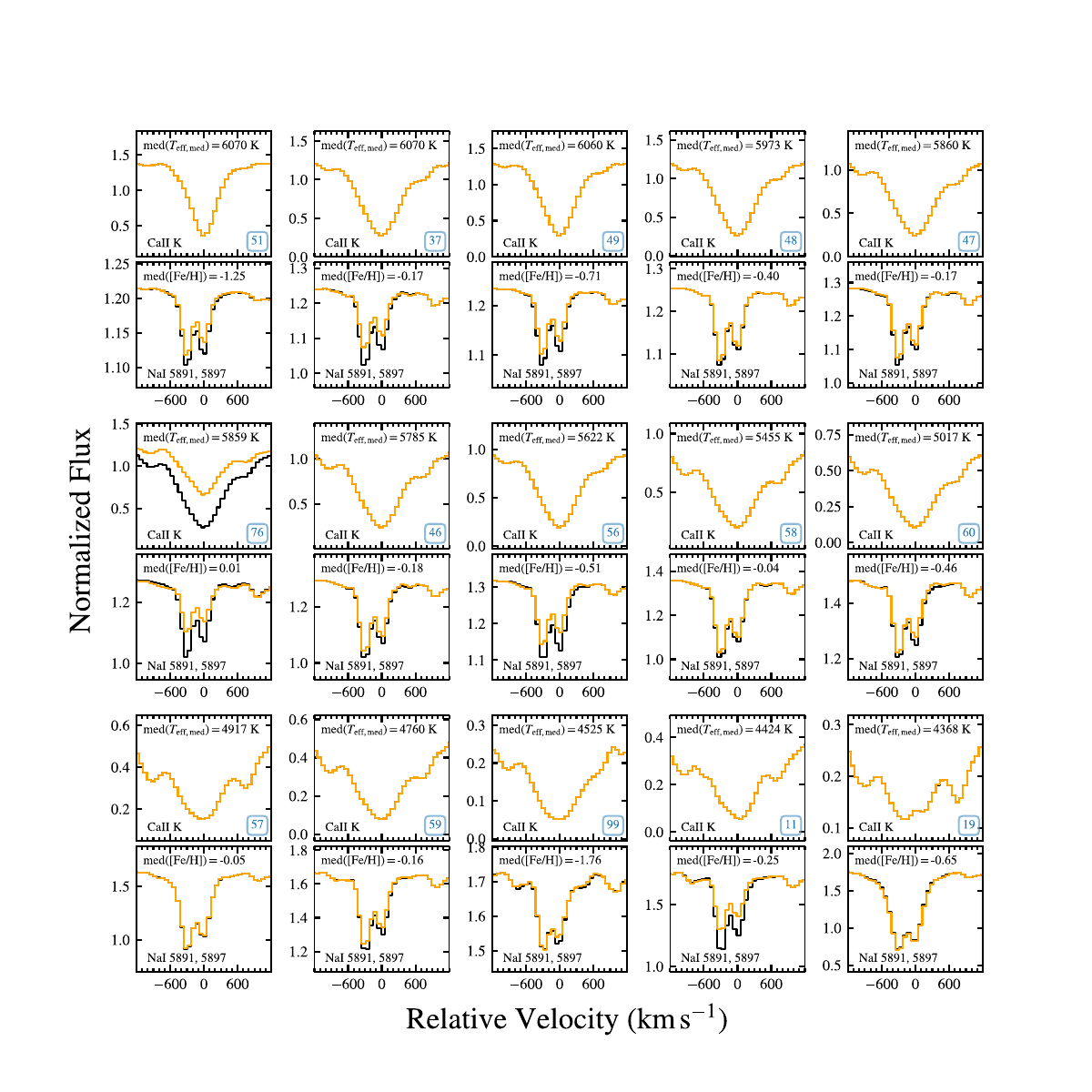}
\caption{-- continued \label{fig:velplot3}}
\end{figure*}
\setcounter{figure}{24}
\begin{figure*}[th]
\includegraphics[width=1.0\textwidth]{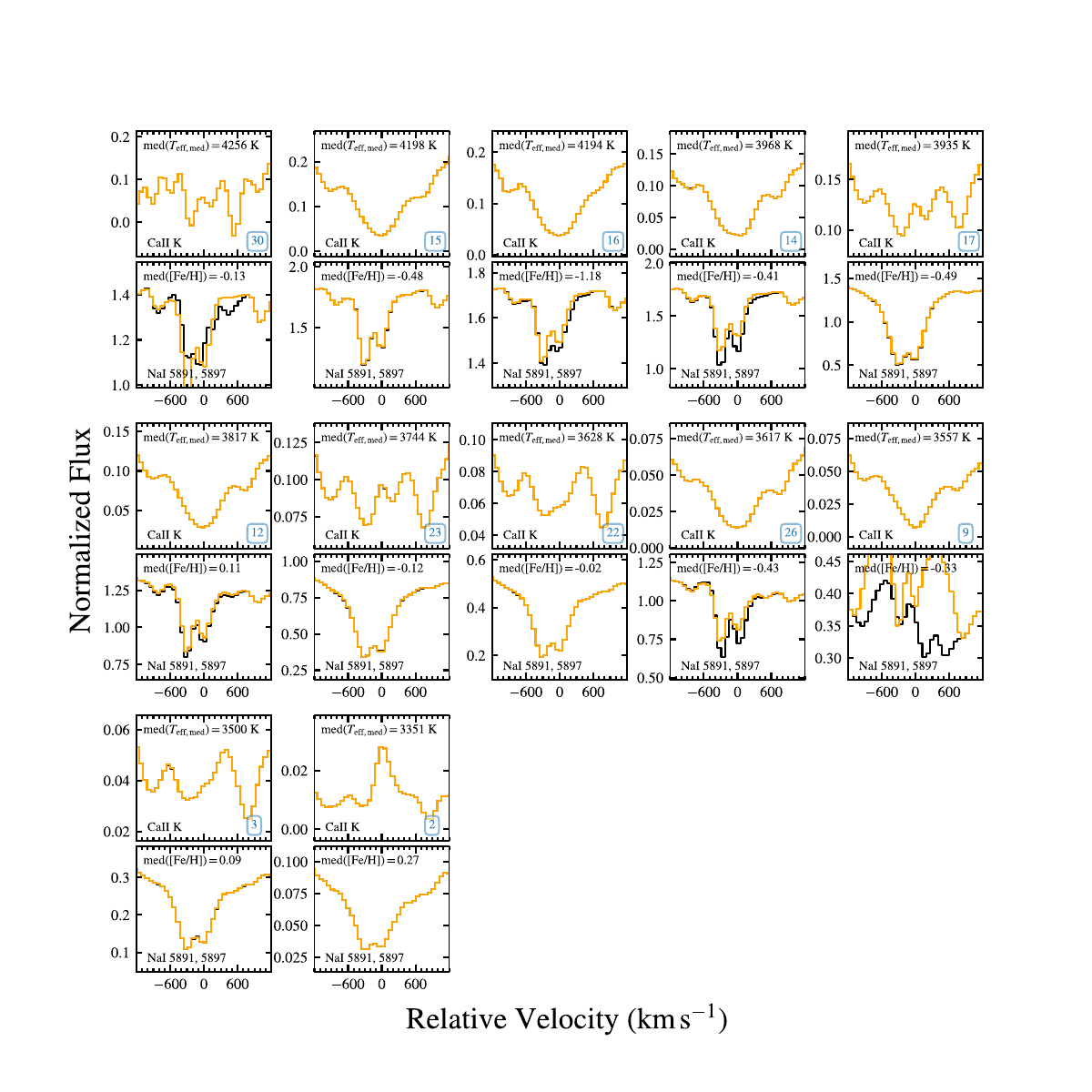}
\caption{-- continued \label{fig:velplot4}}
\end{figure*}

\clearpage 
\bibliography{mastar-ism-proofs}{}

\begin{thebibliography}{}
\expandafter\ifx\csname natexlab\endcsname\relax\def\natexlab#1{#1}\fi
\providecommand{\url}[1]{\href{#1}{#1}}
\providecommand{\dodoi}[1]{doi:~\href{http://doi.org/#1}{\nolinkurl{#1}}}
\providecommand{\doeprint}[1]{\href{http://ascl.net/#1}{\nolinkurl{http://ascl.net/#1}}}
\providecommand{\doarXiv}[1]{\href{https://arxiv.org/abs/#1}{\nolinkurl{https://arxiv.org/abs/#1}}}

\bibitem[{{Abdurro'uf} {et~al.}(2022){Abdurro'uf}, {Accetta}, {Aerts}, {Silva
  Aguirre}, {Ahumada}, {Ajgaonkar}, {Filiz Ak}, {Alam}, {Allende Prieto},
  {Almeida}, {Anders}, {Anderson}, {Andrews}, {Anguiano}, {Aquino-Ort{\'\i}z},
  {Arag{\'o}n-Salamanca}, {Argudo-Fern{\'a}ndez}, {Ata}, {Aubert},
  {Avila-Reese}, {Badenes}, {Barb{\'a}}, {Barger}, {Barrera-Ballesteros},
  {Beaton}, {Beers}, {Belfiore}, {Bender}, {Bernardi}, {Bershady}, {Beutler},
  {Bidin}, {Bird}, {Bizyaev}, {Blanc}, {Blanton}, {Boardman}, {Bolton},
  {Boquien}, {Borissova}, {Bovy}, {Brandt}, {Brown}, {Brownstein}, {Brusa},
  {Buchner}, {Bundy}, {Burchett}, {Bureau}, {Burgasser}, {Cabang}, {Campbell},
  {Cappellari}, {Carlberg}, {Wanderley}, {Carrera}, {Cash}, {Chen}, {Chen},
  {Cherinka}, {Chiappini}, {Choi}, {Chojnowski}, {Chung}, {Clerc}, {Cohen},
  {Comerford}, {Comparat}, {da Costa}, {Covey}, {Crane}, {Cruz-Gonzalez},
  {Culhane}, {Cunha}, {Dai}, {Damke}, {Darling}, {Davidson}, {Davies},
  {Dawson}, {De Lee}, {Diamond-Stanic}, {Cano-D{\'\i}az}, {S{\'a}nchez},
  {Donor}, {Duckworth}, {Dwelly}, {Eisenstein}, {Elsworth}, {Emsellem},
  {Eracleous}, {Escoffier}, {Fan}, {Farr}, {Feng}, {Fern{\'a}ndez-Trincado},
  {Feuillet}, {Filipp}, {Fillingham}, {Frinchaboy}, {Fromenteau}, {Galbany},
  {Garc{\'\i}a}, {Garc{\'\i}a-Hern{\'a}ndez}, {Ge}, {Geisler}, {Gelfand},
  {G{\'e}ron}, {Gibson}, {Goddy}, {Godoy-Rivera}, {Grabowski}, {Green},
  {Greener}, {Grier}, {Griffith}, {Guo}, {Guy}, {Hadjara}, {Harding},
  {Hasselquist}, {Hayes}, {Hearty}, {Hern{\'a}ndez}, {Hill}, {Hogg},
  {Holtzman}, {Horta}, {Hsieh}, {Hsu}, {Hsu}, {Huber}, {Huertas-Company},
  {Hutchinson}, {Hwang}, {Ibarra-Medel}, {Chitham}, {Ilha}, {Imig}, {Jaekle},
  {Jayasinghe}, {Ji}, {Johnson}, {Jones}, {J{\"o}nsson}, {Katkov}, {Khalatyan},
  {Kinemuchi}, {Kisku}, {Knapen}, {Kneib}, {Kollmeier}, {Kong}, {Kounkel},
  {Kreckel}, {Krishnarao}, {Lacerna}, {Lane}, {Langgin}, {Lavender}, {Law},
  {Lazarz}, {Leung}, {Leung}, {Lewis}, {Li}, {Li}, {Lian}, {Liang}, {Lin},
  {Lin}, {Lin}, {Lintott}, {Long}, {Longa-Pe{\~n}a}, {L{\'o}pez-Cob{\'a}},
  {Lu}, {Lundgren}, {Luo}, {Mackereth}, {de la Macorra}, {Mahadevan},
  {Majewski}, {Manchado}, {Mandeville}, {Maraston}, {Margalef-Bentabol},
  {Masseron}, {Masters}, {Mathur}, {McDermid}, {Mckay}, {Merloni},
  {Merrifield}, {Meszaros}, {Miglio}, {Di Mille}, {Minniti}, {Minsley},
  {Monachesi}, {Moon}, {Mosser}, {Mulchaey}, {Muna}, {Mu{\~n}oz}, {Myers},
  {Myers}, {Nadathur}, {Nair}, {Nandra}, {Neumann}, {Newman}, {Nidever},
  {Nikakhtar}, {Nitschelm}, {O'Connell}, {Garma-Oehmichen}, {Luan Souza de
  Oliveira}, {Olney}, {Oravetz}, {Ortigoza-Urdaneta}, {Osorio}, {Otter},
  {Pace}, {Padilla}, {Pan}, {Pan}, {Parikh}, {Parker}, {Peirani}, {Pe{\~n}a
  Ram{\'\i}rez}, {Penny}, {Percival}, {Perez-Fournon}, {Pinsonneault},
  {Poidevin}, {Poovelil}, {Price-Whelan}, {B{\'a}rbara de Andrade Queiroz},
  {Raddick}, {Ray}, {Rembold}, {Riddle}, {Riffel}, {Riffel}, {Rix}, {Robin},
  {Rodr{\'\i}guez-Puebla}, {Roman-Lopes}, {Rom{\'a}n-Z{\'u}{\~n}iga}, {Rose},
  {Ross}, {Rossi}, {Rubin}, {Salvato}, {S{\'a}nchez}, {S{\'a}nchez-Gallego},
  {Sanderson}, {Santana Rojas}, {Sarceno}, {Sarmiento}, {Sayres}, {Sazonova},
  {Schaefer}, {Schiavon}, {Schlegel}, {Schneider}, {Schultheis}, {Schwope},
  {Serenelli}, {Serna}, {Shao}, {Shapiro}, {Sharma}, {Shen}, {Shetrone}, {Shu},
  {Simon}, {Skrutskie}, {Smethurst}, {Smith}, {Sobeck}, {Spoo}, {Sprague},
  {Stark}, {Stassun}, {Steinmetz}, {Stello}, {Stone-Martinez},
  {Storchi-Bergmann}, {Stringfellow}, {Stutz}, {Su}, {Taghizadeh-Popp},
  {Talbot}, {Tayar}, {Telles}, {Teske}, {Thakar}, {Theissen}, {Tkachenko},
  {Thomas}, {Tojeiro}, {Hernandez Toledo}, {Troup}, {Trump}, {Trussler},
  {Turner}, {Tuttle}, {Unda-Sanzana}, {V{\'a}zquez-Mata}, {Valentini},
  {Valenzuela}, {Vargas-Gonz{\'a}lez}, {Vargas-Maga{\~n}a}, {Alfaro},
  {Villanova}, {Vincenzo}, {Wake}, {Warfield}, {Washington}, {Weaver},
  {Weijmans}, {Weinberg}, {Weiss}, {Westfall}, {Wild}, {Wilde}, {Wilson},
  {Wilson}, {Wilson}, {Wolf}, {Wood-Vasey}, {Yan}, {Zamora}, {Zasowski},
  {Zhang}, {Zhao}, {Zheng}, {Zheng}, \& {Zhu}}]{Abdurrouf2022}
{Abdurro'uf}, {Accetta}, K., {Aerts}, C., {et~al.} 2022, \apjs, 259, 35,
  \dodoi{10.3847/1538-4365/ac4414}

\bibitem[{{Alatalo} {et~al.}(2013){Alatalo}, {Davis}, {Bureau}, {Young},
  {Blitz}, {Crocker}, {Bayet}, {Bois}, {Bournaud}, {Cappellari}, {Davies}, {de
  Zeeuw}, {Duc}, {Emsellem}, {Khochfar}, {Krajnovi{\'c}}, {Kuntschner},
  {Lablanche}, {Morganti}, {McDermid}, {Naab}, {Oosterloo}, {Sarzi}, {Scott},
  {Serra}, \& {Weijmans}}]{Alatalo2013}
{Alatalo}, K., {Davis}, T.~A., {Bureau}, M., {et~al.} 2013, \mnras, 432, 1796,
  \dodoi{10.1093/mnras/sts299}

\bibitem[{{Alton} {et~al.}(2017){Alton}, {Smith}, \& {Lucey}}]{Alton2017}
{Alton}, P.~D., {Smith}, R.~J., \& {Lucey}, J.~R. 2017, \mnras, 468, 1594,
  \dodoi{10.1093/mnras/stx464}

\bibitem[{{Avery} {et~al.}(2022){Avery}, {Wuyts}, {F{\"o}rster Schreiber},
  {Villforth}, {Bertemes}, {Hamer}, {Sharma}, {Toshikawa}, \&
  {Zhang}}]{Avery2022}
{Avery}, C.~R., {Wuyts}, S., {F{\"o}rster Schreiber}, N.~M., {et~al.} 2022,
  \mnras, 511, 4223, \dodoi{10.1093/mnras/stac190}

\bibitem[{{Bailer-Jones} {et~al.}(2021){Bailer-Jones}, {Rybizki}, {Fouesneau},
  {Demleitner}, \& {Andrae}}]{Bailer-Jones2021}
{Bailer-Jones}, C.~A.~L., {Rybizki}, J., {Fouesneau}, M., {Demleitner}, M., \&
  {Andrae}, R. 2021, \aj, 161, 147, \dodoi{10.3847/1538-3881/abd806}

\bibitem[{{Baron} {et~al.}(2016){Baron}, {Stern}, {Poznanski}, \&
  {Netzer}}]{Baron2016}
{Baron}, D., {Stern}, J., {Poznanski}, D., \& {Netzer}, H. 2016, \apj, 832, 8,
  \dodoi{10.3847/0004-637X/832/1/8}

\bibitem[{{Belfiore} {et~al.}(2017){Belfiore}, {Maiolino}, {Tremonti},
  {S{\'a}nchez}, {Bundy}, {Bershady}, {Westfall}, {Lin}, {Drory}, {Boquien},
  {Thomas}, \& {Brinkmann}}]{Belfiore2017}
{Belfiore}, F., {Maiolino}, R., {Tremonti}, C., {et~al.} 2017, \mnras, 469,
  151, \dodoi{10.1093/mnras/stx789}

\bibitem[{{Belfiore} {et~al.}(2019){Belfiore}, {Westfall}, {Schaefer},
  {Cappellari}, {Ji}, {Bershady}, {Tremonti}, {Law}, {Yan}, {Bundy}, {Shetty},
  {Drory}, {Thomas}, {Emsellem}, \& {S{\'a}nchez}}]{Belfiore2019}
{Belfiore}, F., {Westfall}, K.~B., {Schaefer}, A., {et~al.} 2019, \aj, 158,
  160, \dodoi{10.3847/1538-3881/ab3e4e}

\bibitem[{{Ben Bekhti} {et~al.}(2008){Ben Bekhti}, {Richter}, {Westmeier}, \&
  {Murphy}}]{BenBekhti2008}
{Ben Bekhti}, N., {Richter}, P., {Westmeier}, T., \& {Murphy}, M.~T. 2008,
  \aap, 487, 583, \dodoi{10.1051/0004-6361:20079067}

\bibitem[{{Ben Bekhti} {et~al.}(2012){Ben Bekhti}, {Winkel}, {Richter}, {Kerp},
  {Klein}, \& {Murphy}}]{BenBekhti2012}
{Ben Bekhti}, N., {Winkel}, B., {Richter}, P., {et~al.} 2012, \aap, 542, A110,
  \dodoi{10.1051/0004-6361/201118673}

\bibitem[{{Bensby} {et~al.}(2014){Bensby}, {Feltzing}, \& {Oey}}]{Bensby2014}
{Bensby}, T., {Feltzing}, S., \& {Oey}, M.~S. 2014, \aap, 562, A71,
  \dodoi{10.1051/0004-6361/201322631}

\bibitem[{{Bensby} {et~al.}(2017){Bensby}, {Feltzing}, {Gould}, {Yee},
  {Johnson}, {Asplund}, {Mel{\'e}ndez}, {Lucatello}, {Howes}, {McWilliam},
  {Udalski}, {Szyma{\'n}ski}, {Soszy{\'n}ski}, {Poleski}, {Wyrzykowski},
  {Ulaczyk}, {Koz{\l}owski}, {Pietrukowicz}, {Skowron}, {Mr{\'o}z}, {Pawlak},
  {Abe}, {Asakura}, {Bhattacharya}, {Bond}, {Bennett}, {Hirao}, {Nagakane},
  {Koshimoto}, {Sumi}, {Suzuki}, \& {Tristram}}]{Bensby2017}
{Bensby}, T., {Feltzing}, S., {Gould}, A., {et~al.} 2017, \aap, 605, A89,
  \dodoi{10.1051/0004-6361/201730560}

\bibitem[{{Bernardi} {et~al.}(2023){Bernardi}, {Dom{\'\i}nguez S{\'a}nchez},
  {Sheth}, {Brownstein}, \& {Lane}}]{Bernardi2022}
{Bernardi}, M., {Dom{\'\i}nguez S{\'a}nchez}, H., {Sheth}, R.~K., {Brownstein},
  J.~R., \& {Lane}, R.~R. 2023, \mnras, 518, 4713,
  \dodoi{10.1093/mnras/stac3287}

\bibitem[{{Bernardi} {et~al.}(2006){Bernardi}, {Nichol}, {Sheth}, {Miller}, \&
  {Brinkmann}}]{Bernardi2006}
{Bernardi}, M., {Nichol}, R.~C., {Sheth}, R.~K., {Miller}, C.~J., \&
  {Brinkmann}, J. 2006, \aj, 131, 1288, \dodoi{10.1086/499522}

\bibitem[{{Bish} {et~al.}(2019){Bish}, {Werk}, {Prochaska}, {Rubin}, {Zheng},
  {O'Meara}, \& {Deason}}]{Bish2019}
{Bish}, H.~V., {Werk}, J.~K., {Prochaska}, J.~X., {et~al.} 2019, \apj, 882, 76,
  \dodoi{10.3847/1538-4357/ab3414}

\bibitem[{{Bloom} {et~al.}(2017){Bloom}, {Croom}, {Bryant}, {Callingham},
  {Schaefer}, {Cortese}, {Hopkins}, {D'Eugenio}, {Scott}, {Glazebrook},
  {Tonini}, {McElroy}, {Clark}, {Catinella}, {Allen}, {Bland-Hawthorn},
  {Goodwin}, {Green}, {Konstantopoulos}, {Lawrence}, {Lorente}, {Medling},
  {Owers}, {Richards}, \& {Sharp}}]{Bloom2017}
{Bloom}, J.~V., {Croom}, S.~M., {Bryant}, J.~J., {et~al.} 2017, \mnras, 472,
  1809, \dodoi{10.1093/mnras/stx1701}

\bibitem[{{Bohlin} {et~al.}(2017){Bohlin}, {M{\'e}sz{\'a}ros}, {Fleming},
  {Gordon}, {Koekemoer}, \& {Kov{\'a}cs}}]{BOSZ2017}
{Bohlin}, R.~C., {M{\'e}sz{\'a}ros}, S., {Fleming}, S.~W., {et~al.} 2017, \aj,
  153, 234, \dodoi{10.3847/1538-3881/aa6ba9}

\bibitem[{{Borisov} {et~al.}(2023){Borisov}, {Chilingarian}, {Rubtsov},
  {Ledoux}, {Melo}, {Grishin}, {Katkov}, {Goradzhanov}, {Afanasiev},
  {Kasparova}, \& {Saburova}}]{Borisov2023}
{Borisov}, S.~B., {Chilingarian}, I.~V., {Rubtsov}, E.~V., {et~al.} 2023,
  \apjs, 266, 11, \dodoi{10.3847/1538-4365/acc321}

\bibitem[{{Bruzual}(1983)}]{Bruzual1983}
{Bruzual}, A.~G. 1983, \apj, 273, 105, \dodoi{10.1086/161352}

\bibitem[{{Bruzual} \& {Charlot}(2003)}]{BC2003}
{Bruzual}, G., \& {Charlot}, S. 2003, \mnras, 344, 1000,
  \dodoi{10.1046/j.1365-8711.2003.06897.x}

\bibitem[{{Bryant} {et~al.}(2019){Bryant}, {Croom}, {van de Sande}, {Scott},
  {Fogarty}, {Bland-Hawthorn}, {Bloom}, {Taylor}, {Brough}, {Robotham},
  {Cortese}, {Couch}, {Owers}, {Medling}, {Federrath}, {Bekki}, {Richards},
  {Lawrence}, \& {Konstantopoulos}}]{Bryant2019}
{Bryant}, J.~J., {Croom}, S.~M., {van de Sande}, J., {et~al.} 2019, \mnras,
  483, 458, \dodoi{10.1093/mnras/sty3122}

\bibitem[{{Bundy} {et~al.}(2015){Bundy}, {Bershady}, {Law}, {Yan}, {Drory},
  {MacDonald}, {Wake}, {Cherinka}, {S{\'a}nchez-Gallego}, {Weijmans}, {Thomas},
  {Tremonti}, {Masters}, {Coccato}, {Diamond-Stanic}, {Arag{\'o}n-Salamanca},
  {Avila-Reese}, {Badenes}, {Falc{\'o}n-Barroso}, {Belfiore}, {Bizyaev},
  {Blanc}, {Bland-Hawthorn}, {Blanton}, {Brownstein}, {Byler}, {Cappellari},
  {Conroy}, {Dutton}, {Emsellem}, {Etherington}, {Frinchaboy}, {Fu}, {Gunn},
  {Harding}, {Johnston}, {Kauffmann}, {Kinemuchi}, {Klaene}, {Knapen},
  {Leauthaud}, {Li}, {Lin}, {Maiolino}, {Malanushenko}, {Malanushenko}, {Mao},
  {Maraston}, {McDermid}, {Merrifield}, {Nichol}, {Oravetz}, {Pan}, {Parejko},
  {Sanchez}, {Schlegel}, {Simmons}, {Steele}, {Steinmetz}, {Thanjavur},
  {Thompson}, {Tinker}, {van den Bosch}, {Westfall}, {Wilkinson}, {Wright},
  {Xiao}, \& {Zhang}}]{Bundy2015}
{Bundy}, K., {Bershady}, M.~A., {Law}, D.~R., {et~al.} 2015, \apj, 798, 7,
  \dodoi{10.1088/0004-637X/798/1/7}

\bibitem[{{Burstein} {et~al.}(1984){Burstein}, {Faber}, {Gaskell}, \&
  {Krumm}}]{Burstein1984}
{Burstein}, D., {Faber}, S.~M., {Gaskell}, C.~M., \& {Krumm}, N. 1984, \apj,
  287, 586, \dodoi{10.1086/162718}

\bibitem[{{Burstein} {et~al.}(1986){Burstein}, {Faber}, \&
  {Gonzalez}}]{Burstein1986}
{Burstein}, D., {Faber}, S.~M., \& {Gonzalez}, J.~J. 1986, \aj, 91, 1130,
  \dodoi{10.1086/114090}

\bibitem[{{Byrne} \& {Stanway}(2023)}]{Byrne2023}
{Byrne}, C.~M., \& {Stanway}, E.~R. 2023, \mnras, 521, 4995,
  \dodoi{10.1093/mnras/stad832}

\bibitem[{{Cappellari}(2016)}]{Cappellari2016}
{Cappellari}, M. 2016, \araa, 54, 597,
  \dodoi{10.1146/annurev-astro-082214-122432}

\bibitem[{{Cappellari}(2017)}]{Cappellari2017}
---. 2017, \mnras, 466, 798, \dodoi{10.1093/mnras/stw3020}

\bibitem[{{Cappellari} {et~al.}(2011){Cappellari}, {Emsellem}, {Krajnovi{\'c}},
  {McDermid}, {Scott}, {Verdoes Kleijn}, {Young}, {Alatalo}, {Bacon}, {Blitz},
  {Bois}, {Bournaud}, {Bureau}, {Davies}, {Davis}, {de Zeeuw}, {Duc},
  {Khochfar}, {Kuntschner}, {Lablanche}, {Morganti}, {Naab}, {Oosterloo},
  {Sarzi}, {Serra}, \& {Weijmans}}]{Cappellari2011}
{Cappellari}, M., {Emsellem}, E., {Krajnovi{\'c}}, D., {et~al.} 2011, \mnras,
  413, 813, \dodoi{10.1111/j.1365-2966.2010.18174.x}

\bibitem[{{Cappellari} {et~al.}(2012){Cappellari}, {McDermid}, {Alatalo},
  {Blitz}, {Bois}, {Bournaud}, {Bureau}, {Crocker}, {Davies}, {Davis}, {de
  Zeeuw}, {Duc}, {Emsellem}, {Khochfar}, {Krajnovi{\'c}}, {Kuntschner},
  {Lablanche}, {Morganti}, {Naab}, {Oosterloo}, {Sarzi}, {Scott}, {Serra},
  {Weijmans}, \& {Young}}]{Cappellari2012}
{Cappellari}, M., {McDermid}, R.~M., {Alatalo}, K., {et~al.} 2012, \nat, 484,
  485, \dodoi{10.1038/nature10972}

\bibitem[{{Cassisi} {et~al.}(1997{\natexlab{a}}){Cassisi}, {Castellani}, \&
  {Castellani}}]{Cassisi1997a}
{Cassisi}, S., {Castellani}, M., \& {Castellani}, V. 1997{\natexlab{a}}, \aap,
  317, 108, \dodoi{10.48550/arXiv.astro-ph/9603023}

\bibitem[{{Cassisi} {et~al.}(2000){Cassisi}, {Castellani}, {Ciarcelluti},
  {Piotto}, \& {Zoccali}}]{Cassisi2000}
{Cassisi}, S., {Castellani}, V., {Ciarcelluti}, P., {Piotto}, G., \& {Zoccali},
  M. 2000, \mnras, 315, 679, \dodoi{10.1046/j.1365-8711.2000.03457.x}

\bibitem[{{Cassisi} {et~al.}(1997{\natexlab{b}}){Cassisi}, {degl'Innocenti}, \&
  {Salaris}}]{Cassisi1997b}
{Cassisi}, S., {degl'Innocenti}, S., \& {Salaris}, M. 1997{\natexlab{b}},
  \mnras, 290, 515, \dodoi{10.1093/mnras/290.3.515}

\bibitem[{{Castelli} \& {Kurucz}(2003)}]{CastelliKurucz2003}
{Castelli}, F., \& {Kurucz}, R.~L. 2003, in Modelling of Stellar Atmospheres,
  ed. N.~{Piskunov}, W.~W. {Weiss}, \& D.~F. {Gray}, Vol. 210, A20.
\newblock \doarXiv{astro-ph/0405087}

\bibitem[{{Cazzoli} {et~al.}(2014){Cazzoli}, {Arribas}, {Colina},
  {Piqueras-L{\'o}pez}, {Bellocchi}, {Emonts}, \& {Maiolino}}]{Cazzoli2014}
{Cazzoli}, S., {Arribas}, S., {Colina}, L., {et~al.} 2014, \aap, 569, A14,
  \dodoi{10.1051/0004-6361/201323296}

\bibitem[{{Chen} {et~al.}(2010{\natexlab{a}}){Chen}, {Helsby}, {Gauthier},
  {Shectman}, {Thompson}, \& {Tinker}}]{Chen2010a}
{Chen}, H.-W., {Helsby}, J.~E., {Gauthier}, J.-R., {et~al.} 2010{\natexlab{a}},
  \apj, 714, 1521, \dodoi{10.1088/0004-637X/714/2/1521}

\bibitem[{{Chen} {et~al.}(2010{\natexlab{b}}){Chen}, {Tremonti}, {Heckman},
  {Kauffmann}, {Weiner}, {Brinchmann}, \& {Wang}}]{ChenTremonti2010}
{Chen}, Y.-M., {Tremonti}, C.~A., {Heckman}, T.~M., {et~al.}
  2010{\natexlab{b}}, \aj, 140, 445, \dodoi{10.1088/0004-6256/140/2/445}

\bibitem[{{Chen} {et~al.}(2014){Chen}, {Trager}, {Peletier}, {Lan{\c{c}}on},
  {Vazdekis}, {Prugniel}, {Silva}, \& {Gonneau}}]{Chen2014}
{Chen}, Y.-P., {Trager}, S.~C., {Peletier}, R.~F., {et~al.} 2014, \aap, 565,
  A117, \dodoi{10.1051/0004-6361/201322505}

\bibitem[{{Chen} {et~al.}(2020){Chen}, {Yan}, {Maraston}, {Thomas},
  {Stringfellow}, {Bizyaev}, {Gelfand}, {Beers}, {Fern{\'a}ndez-Trincado},
  {Lazarz}, {Hill}, {Drory}, \& {Stassun}}]{Chen2020}
{Chen}, Y.-P., {Yan}, R., {Maraston}, C., {et~al.} 2020, \apj, 899, 62,
  \dodoi{10.3847/1538-4357/ab9f35}

\bibitem[{{Cid Fernandes} {et~al.}(2005){Cid Fernandes}, {Mateus}, {Sodr{\'e}},
  {Stasi{\'n}ska}, \& {Gomes}}]{CidFernandes2005}
{Cid Fernandes}, R., {Mateus}, A., {Sodr{\'e}}, L., {Stasi{\'n}ska}, G., \&
  {Gomes}, J.~M. 2005, \mnras, 358, 363,
  \dodoi{10.1111/j.1365-2966.2005.08752.x}

\bibitem[{{Clemens} {et~al.}(2006){Clemens}, {Bressan}, {Nikolic}, {Alexander},
  {Annibali}, \& {Rampazzo}}]{Clemens2006}
{Clemens}, M.~S., {Bressan}, A., {Nikolic}, B., {et~al.} 2006, \mnras, 370,
  702, \dodoi{10.1111/j.1365-2966.2006.10530.x}

\bibitem[{{Clough} {et~al.}(1992){Clough}, {Iacono}, \& {Moncet}}]{Clough1992}
{Clough}, S.~A., {Iacono}, M.~J., \& {Moncet}, J.-L. 1992, \jgr, 97, 15,761,
  \dodoi{10.1029/92JD01419}

\bibitem[{{Clough} {et~al.}(2005){Clough}, {Shephard}, {Mlawer}, {Delamere},
  {Iacono}, {Cady-Pereira}, {Boukabara}, \& {Brown}}]{Clough2005}
{Clough}, S.~A., {Shephard}, M.~W., {Mlawer}, E.~J., {et~al.} 2005, \jqsrt, 91,
  233, \dodoi{10.1016/j.jqsrt.2004.05.058}

\bibitem[{{Coelho} {et~al.}(2005){Coelho}, {Barbuy}, {Mel{\'e}ndez},
  {Schiavon}, \& {Castilho}}]{Coelho2005}
{Coelho}, P., {Barbuy}, B., {Mel{\'e}ndez}, J., {Schiavon}, R.~P., \&
  {Castilho}, B.~V. 2005, \aap, 443, 735, \dodoi{10.1051/0004-6361:20053511}

\bibitem[{{Coelho}(2014)}]{Coelho2014}
{Coelho}, P.~R.~T. 2014, \mnras, 440, 1027, \dodoi{10.1093/mnras/stu365}

\bibitem[{{Concas} {et~al.}(2019){Concas}, {Popesso}, {Brusa}, {Mainieri}, \&
  {Thomas}}]{Concas2019}
{Concas}, A., {Popesso}, P., {Brusa}, M., {Mainieri}, V., \& {Thomas}, D. 2019,
  \aap, 622, A188, \dodoi{10.1051/0004-6361/201732152}

\bibitem[{{Conroy}(2013)}]{Conroy2013}
{Conroy}, C. 2013, \araa, 51, 393, \dodoi{10.1146/annurev-astro-082812-141017}

\bibitem[{{Conroy} {et~al.}(2014){Conroy}, {Graves}, \& {van
  Dokkum}}]{Conroy2014}
{Conroy}, C., {Graves}, G.~J., \& {van Dokkum}, P.~G. 2014, \apj, 780, 33,
  \dodoi{10.1088/0004-637X/780/1/33}

\bibitem[{{Conroy} \& {van Dokkum}(2012{\natexlab{a}})}]{ConroyvanDokkum2012a}
{Conroy}, C., \& {van Dokkum}, P. 2012{\natexlab{a}}, \apj, 747, 69,
  \dodoi{10.1088/0004-637X/747/1/69}

\bibitem[{{Conroy} \& {van Dokkum}(2012{\natexlab{b}})}]{ConroyvanDokkum2012b}
{Conroy}, C., \& {van Dokkum}, P.~G. 2012{\natexlab{b}}, \apj, 760, 71,
  \dodoi{10.1088/0004-637X/760/1/71}

\bibitem[{{Conroy} {et~al.}(2018){Conroy}, {Villaume}, {van Dokkum}, \&
  {Lind}}]{Conroy2018}
{Conroy}, C., {Villaume}, A., {van Dokkum}, P.~G., \& {Lind}, K. 2018, \apj,
  854, 139, \dodoi{10.3847/1538-4357/aaab49}

\bibitem[{{Couch} \& {Sharples}(1987)}]{CouchSharples1987}
{Couch}, W.~J., \& {Sharples}, R.~M. 1987, \mnras, 229, 423,
  \dodoi{10.1093/mnras/229.3.423}

\bibitem[{{Crawford}(1992)}]{Crawford1992}
{Crawford}, I.~A. 1992, \mnras, 259, 47, \dodoi{10.1093/mnras/259.1.47}

\bibitem[{{Crowther}(2022)}]{Crowther2022}
{Crowther}, P.~A. 2022, arXiv e-prints, arXiv:2207.08690.
\newblock \doarXiv{2207.08690}

\bibitem[{{Davis} {et~al.}(2019){Davis}, {Greene}, {Ma}, {Blakeslee}, {Dawson},
  {Pandya}, {Veale}, \& {Zabel}}]{Davis2019}
{Davis}, T.~A., {Greene}, J.~E., {Ma}, C.-P., {et~al.} 2019, \mnras, 486, 1404,
  \dodoi{10.1093/mnras/stz871}

\bibitem[{{Davis} {et~al.}(2013){Davis}, {Alatalo}, {Bureau}, {Cappellari},
  {Scott}, {Young}, {Blitz}, {Crocker}, {Bayet}, {Bois}, {Bournaud}, {Davies},
  {de Zeeuw}, {Duc}, {Emsellem}, {Khochfar}, {Krajnovi{\'c}}, {Kuntschner},
  {Lablanche}, {McDermid}, {Morganti}, {Naab}, {Oosterloo}, {Sarzi}, {Serra},
  \& {Weijmans}}]{Davis2013}
{Davis}, T.~A., {Alatalo}, K., {Bureau}, M., {et~al.} 2013, \mnras, 429, 534,
  \dodoi{10.1093/mnras/sts353}

\bibitem[{{den Brok} {et~al.}(2024){den Brok}, {Krajnovi{\'c}}, {Emsellem},
  {Mercier}, {Steinmetz}, \& {Weilbacher}}]{denBrok2024}
{den Brok}, M., {Krajnovi{\'c}}, D., {Emsellem}, E., {et~al.} 2024, \mnras,
  530, 3278, \dodoi{10.1093/mnras/stae912}

\bibitem[{{Dressler} \& {Gunn}(1983)}]{DresslerGunn1983}
{Dressler}, A., \& {Gunn}, J.~E. 1983, \apj, 270, 7, \dodoi{10.1086/161093}

\bibitem[{{Eldridge} {et~al.}(2017){Eldridge}, {Stanway}, {Xiao}, {McClelland},
  {Taylor}, {Ng}, {Greis}, \& {Bray}}]{Eldridge2017}
{Eldridge}, J.~J., {Stanway}, E.~R., {Xiao}, L., {et~al.} 2017, \pasa, 34,
  e058, \dodoi{10.1017/pasa.2017.51}

\bibitem[{{Faber}(1972)}]{Faber1972}
{Faber}, S.~M. 1972, \aap, 20, 361

\bibitem[{{Faber} \& {French}(1980)}]{FaberFrench1980}
{Faber}, S.~M., \& {French}, H.~B. 1980, \apj, 235, 405, \dodoi{10.1086/157644}

\bibitem[{{Feldmeier-Krause} {et~al.}(2021){Feldmeier-Krause}, {Lonoce}, \&
  {Freedman}}]{Feldmeier-Krause2021}
{Feldmeier-Krause}, A., {Lonoce}, I., \& {Freedman}, W.~L. 2021, \apj, 923, 65,
  \dodoi{10.3847/1538-4357/ac281e}

\bibitem[{{Fitzpatrick} {et~al.}(2019){Fitzpatrick}, {Massa}, {Gordon},
  {Bohlin}, \& {Clayton}}]{Fitzpatrick2019}
{Fitzpatrick}, E.~L., {Massa}, D., {Gordon}, K.~D., {Bohlin}, R., \& {Clayton},
  G.~C. 2019, \apj, 886, 108, \dodoi{10.3847/1538-4357/ab4c3a}

\bibitem[{{Foreman-Mackey} {et~al.}(2013){Foreman-Mackey}, {Hogg}, {Lang}, \&
  {Goodman}}]{Foreman-Mackey2013}
{Foreman-Mackey}, D., {Hogg}, D.~W., {Lang}, D., \& {Goodman}, J. 2013, \pasp,
  125, 306, \dodoi{10.1086/670067}

\bibitem[{{Franx} \& {Illingworth}(1990)}]{FranxIllingworth1990}
{Franx}, M., \& {Illingworth}, G. 1990, \apjl, 359, L41, \dodoi{10.1086/185791}

\bibitem[{{Gaia Collaboration} {et~al.}(2023{\natexlab{a}}){Gaia
  Collaboration}, {Schultheis}, {Zhao}, {Zwitter}, {Bailer-Jones}, {Carballo},
  {Sordo}, {Drimmel}, {Ordenovic}, {Pailler}, {Fouesneau}, {Creevey}, {Heiter},
  {Recio-Blanco}, {Kordopatis}, {de Laverny}, {Marshall}, {Dharmawardena},
  {Brown}, {Vallenari}, {Prusti}, {de Bruijne}, {Arenou}, {Babusiaux},
  {Barbier}, {Biermann}, {Ducourant}, {Evans}, {Eyer}, {Guerra}, {Hutton},
  {Jordi}, {Klioner}, {Lammers}, {Lindegren}, {Luri}, {Mignard}, {Randich},
  {Sartoretti}, {Smiljanic}, {Tanga}, {Walton}, {Bastian}, {Cropper}, {Katz},
  {Soubiran}, {van Leeuwen}, {Andrae}, {Audard}, {Bakker}, {Blomme},
  {Casta{\~n}eda}, {De Angeli}, {Fabricius}, {Fr{\'e}mat}, {Galluccio},
  {Guerrier}, {Masana}, {Messineo}, {Nicolas}, {Nienartowicz}, {Panuzzo},
  {Riclet}, {Roux}, {Seabroke}, {Th{\'e}venin}, {Gracia-Abril}, {Portell},
  {Teyssier}, {Altmann}, {Benson}, {Berthier}, {Burgess}, {Busonero}, {Busso},
  {C{\'a}novas}, {Carry}, {Cheek}, {Clementini}, {Damerdji}, {Davidson}, {de
  Teodoro}, {Delchambre}, {Dell'Oro}, {Fraile Garcia}, {Garabato},
  {Garc{\'\i}a-Lario}, {Garralda Torres}, {Gavras}, {Haigron}, {Hambly},
  {Harrison}, {Hatzidimitriou}, {Hern{\'a}ndez}, {Hodgkin}, {Holl}, {Jamal},
  {Jordan}, {Krone-Martins}, {Lanzafame}, {L{\"o}ffler}, {Lorca}, {Marchal},
  {Marrese}, {Moitinho}, {Muinonen}, {Nu{\~n}ez Campos}, {Oreshina-Slezak},
  {Osborne}, {Pancino}, {Pauwels}, {Riello}, {Rimoldini}, {Robin}, {Roegiers},
  {Sarro}, {Siopis}, {Smith}, {Sozzetti}, {Utrilla}, {van Leeuwen},
  {Weingrill}, {Abbas}, {{\'A}brah{\'a}m}, {Abreu Aramburu}, {Aerts},
  {Altavilla}, {{\'A}lvarez}, {Alves}, {Anders}, {Anderson}, {Antoja},
  {Baines}, {Baker}, {Balog}, {Barache}, {Barbato}, {Barros}, {Barstow},
  {Bartolom{\'e}}, {Bashi}, {Bauchet}, {Baudeau}, {Becciani}, {Bedin},
  {Bellas-Velidis}, {Bellazzini}, {Beordo}, {Berihuete}, {Bernet},
  {Bertolotto}, {Bertone}, {Bianchi}, {Binnenfeld}, {Blazere}, {Boch},
  {Bombrun}, {Bouquillon}, {Bragaglia}, {Braine}, {Bramante}, {Breedt},
  {Bressan}, {Brouillet}, {Brugaletta}, {Bucciarelli}, {Butkevich}, {Buzzi},
  {Caffau}, {Cancelliere}, {Cannizzo}, {Carlucci}, {Carnerero}, {Carrasco},
  {Carretero}, {Carton}, {Casamiquela}, {Castellani}, {Castro-Ginard},
  {Cesare}, {Charlot}, {Chemin}, {Chiaramida}, {Chiavassa}, {Chornay},
  {Collins}, {Contursi}, {Cooper}, {Cornez}, {Crosta}, {Crowley}, {Dafonte},
  {De Luise}, {De March}, {de Souza}, {de Torres}, {del Peloso}, {Delbo},
  {Delgado}, {Diakite}, {Diener}, {Distefano}, {Dolding}, {Dsilva},
  {Dur{\'a}n}, {Enke}, {Esquej}, {Fabre}, {Fabrizio}, {Faigler}, {Fatovi{\'c}},
  {Fedorets}, {Fern{\'a}ndez-Hern{\'a}ndez}, {Fernique}, {Figueras},
  {Fournier}, {Fouron}, {Gai}, {Galinier}, {Garcia-Gutierrez},
  {Garc{\'\i}a-Torres}, {Garofalo}, {Gerlach}, {Geyer}, {Giacobbe}, {Gilmore},
  {Girona}, {Giuffrida}, {Gomel}, {Gomez}, {Gonz{\'a}lez-N{\'u}{\~n}ez},
  {Gonz{\'a}lez-Santamar{\'\i}a}, {Gosset}, {Granvik}, {Gregori Barrera},
  {Guti{\'e}rrez-S{\'a}nchez}, {Haywood}, {Helmer}, {Helmi}, {Henares},
  {Hidalgo}, {Hilger}, {Hobbs}, {Hottier}, {Huckle}, {Jab{\l}o{\'n}ska},
  {Jansen}, {Jim{\'e}nez-Arranz}, {Juaristi Campillo}, {Khanna}, {Korn},
  {K{\'o}sp{\'a}l}, {Kostrzewa-Rutkowska}, {Kun}, {Lambert}, {Lanza}, {Le
  Campion}, {Lebreton}, {Lebzelter}, {Leccia}, {Lecoeur-Taibi}, {Lecoutre},
  {Liao}, {Liberato}, {Licata}, {Lindstr{\o}m}, {Lister}, {Livanou}, {Lobel},
  {Loup}, {Mahy}, {Mann}, {Manteiga}, {Marchant}, {Marconi}, {Mar{\'\i}n Pina},
  {Marinoni}, {Mart{\'\i}n Lozano}, {Mart{\'\i}n-Fleitas}, {Marton}, {Mary},
  {Masip}, {Massari}, {Mastrobuono-Battisti}, {Mazeh}, {McMillan}, {Meichsner},
  {Messina}, {Michalik}, {Millar}, {Mints}, {Molina}, {Molinaro}, {Moln{\'a}r},
  {Monari}, {Mongui{\'o}}, {Montegriffo}, {Montero}, {Mor}, {Mora},
  {Morbidelli}, {Morel}, {Morris}, {Mowlavi}, {Munoz}, {Muraveva}, {Murphy},
  {Musella}, {Nagy}, {Nieto}, {Noval}, {Ogden}, {Pagani}, {Pagano},
  {Palaversa}, {Palicio}, {Pallas-Quintela}, {Panahi}, {Panem},
  {Payne-Wardenaar}, {Pegoraro}, {Penttil{\"a}}, {Pesciullesi}, {Piersimoni},
  {Pinamonti}, {Pineau}, {Plachy}, {Plum}, {Poggio}, {Pourbaix}, {Pr{\v{s}}a},
  {Pulone}, {Racero}, {Rainer}, {Raiteri}, {Ramos}, {Ramos-Lerate},
  {Ratajczak}, {Re Fiorentin}, {Regibo}, {Reyl{\'e}}, {Ripepi}, {Riva}, {Rix},
  {Rixon}, {Robichon}, {Robin}, {Romero-G{\'o}mez}, {Rowell}, {Royer}, {Ruz
  Mieres}, {Rybicki}, {Sadowski}, {S{\'a}ez N{\'u}{\~n}ez}, {Sagrist{\`a}
  Sell{\'e}s}, {Sahlmann}, {Sanchez Gimenez}, {Sanna}, {Santove{\~n}a},
  {Sarasso}, {Sarrate Riera}, {Sciacca}, {Segovia}, {S{\'e}gransan}, {Shahaf},
  {Siebert}, {Siltala}, {Slezak}, {Smart}, {Snaith}, {Solano}, {Solitro},
  {Souami}, {Souchay}, {Spina}, {Spitoni}, {Spoto}, {Squillante}, {Steele},
  {Steidelm{\"u}ller}, {Surdej}, {Szabados}, {Taris}, {Taylor}, {Teixeira},
  {Tisani{\'c}}, {Tolomei}, {Torra}, {Torralba Elipe}, {Trabucchi}, {Tsantaki},
  {Ulla}, {Unger}, {Vanel}, {Vecchiato}, {Vicente}, {Voutsinas}, {Weiler},
  {Wyrzykowski}, {Zorec}, {Balaguer-N{\'u}{\~n}ez}, {Leclerc}, {Morgenthaler},
  {Robert}, \& {Zucker}}]{Gaia2023}
{Gaia Collaboration}, {Schultheis}, M., {Zhao}, H., {et~al.}
  2023{\natexlab{a}}, \aap, 680, A38, \dodoi{10.1051/0004-6361/202347103}

\bibitem[{{Gaia Collaboration} {et~al.}(2023{\natexlab{b}}){Gaia
  Collaboration}, {Vallenari}, {Brown}, {Prusti}, {de Bruijne}, {Arenou},
  {Babusiaux}, {Biermann}, {Creevey}, {Ducourant}, {Evans}, {Eyer}, {Guerra},
  {Hutton}, {Jordi}, {Klioner}, {Lammers}, {Lindegren}, {Luri}, {Mignard},
  {Panem}, {Pourbaix}, {Randich}, {Sartoretti}, {Soubiran}, {Tanga}, {Walton},
  {Bailer-Jones}, {Bastian}, {Drimmel}, {Jansen}, {Katz}, {Lattanzi}, {van
  Leeuwen}, {Bakker}, {Cacciari}, {Casta{\~n}eda}, {De Angeli}, {Fabricius},
  {Fouesneau}, {Fr{\'e}mat}, {Galluccio}, {Guerrier}, {Heiter}, {Masana},
  {Messineo}, {Mowlavi}, {Nicolas}, {Nienartowicz}, {Pailler}, {Panuzzo},
  {Riclet}, {Roux}, {Seabroke}, {Sordo}, {Th{\'e}venin}, {Gracia-Abril},
  {Portell}, {Teyssier}, {Altmann}, {Andrae}, {Audard}, {Bellas-Velidis},
  {Benson}, {Berthier}, {Blomme}, {Burgess}, {Busonero}, {Busso},
  {C{\'a}novas}, {Carry}, {Cellino}, {Cheek}, {Clementini}, {Damerdji},
  {Davidson}, {de Teodoro}, {Nu{\~n}ez Campos}, {Delchambre}, {Dell'Oro},
  {Esquej}, {Fern{\'a}ndez-Hern{\'a}ndez}, {Fraile}, {Garabato},
  {Garc{\'\i}a-Lario}, {Gosset}, {Haigron}, {Halbwachs}, {Hambly}, {Harrison},
  {Hern{\'a}ndez}, {Hestroffer}, {Hodgkin}, {Holl}, {Jan{\ss}en}, {Jevardat de
  Fombelle}, {Jordan}, {Krone-Martins}, {Lanzafame}, {L{\"o}ffler}, {Marchal},
  {Marrese}, {Moitinho}, {Muinonen}, {Osborne}, {Pancino}, {Pauwels},
  {Recio-Blanco}, {Reyl{\'e}}, {Riello}, {Rimoldini}, {Roegiers}, {Rybizki},
  {Sarro}, {Siopis}, {Smith}, {Sozzetti}, {Utrilla}, {van Leeuwen}, {Abbas},
  {{\'A}brah{\'a}m}, {Abreu Aramburu}, {Aerts}, {Aguado}, {Ajaj},
  {Aldea-Montero}, {Altavilla}, {{\'A}lvarez}, {Alves}, {Anders}, {Anderson},
  {Anglada Varela}, {Antoja}, {Baines}, {Baker}, {Balaguer-N{\'u}{\~n}ez},
  {Balbinot}, {Balog}, {Barache}, {Barbato}, {Barros}, {Barstow},
  {Bartolom{\'e}}, {Bassilana}, {Bauchet}, {Becciani}, {Bellazzini},
  {Berihuete}, {Bernet}, {Bertone}, {Bianchi}, {Binnenfeld}, {Blanco-Cuaresma},
  {Blazere}, {Boch}, {Bombrun}, {Bossini}, {Bouquillon}, {Bragaglia},
  {Bramante}, {Breedt}, {Bressan}, {Brouillet}, {Brugaletta}, {Bucciarelli},
  {Burlacu}, {Butkevich}, {Buzzi}, {Caffau}, {Cancelliere}, {Cantat-Gaudin},
  {Carballo}, {Carlucci}, {Carnerero}, {Carrasco}, {Casamiquela}, {Castellani},
  {Castro-Ginard}, {Chaoul}, {Charlot}, {Chemin}, {Chiaramida}, {Chiavassa},
  {Chornay}, {Comoretto}, {Contursi}, {Cooper}, {Cornez}, {Cowell}, {Crifo},
  {Cropper}, {Crosta}, {Crowley}, {Dafonte}, {Dapergolas}, {David}, {David},
  {de Laverny}, {De Luise}, \& {De March}}]{GaiaDR3}
{Gaia Collaboration}, {Vallenari}, A., {Brown}, A.~G.~A., {et~al.}
  2023{\natexlab{b}}, \aap, 674, A1, \dodoi{10.1051/0004-6361/202243940}

\bibitem[{{Girardi} {et~al.}(2000){Girardi}, {Bressan}, {Bertelli}, \&
  {Chiosi}}]{Girardi2000}
{Girardi}, L., {Bressan}, A., {Bertelli}, G., \& {Chiosi}, C. 2000, \aaps, 141,
  371, \dodoi{10.1051/aas:2000126}

\bibitem[{{Goddard} {et~al.}(2017){Goddard}, {Thomas}, {Maraston}, {Westfall},
  {Etherington}, {Riffel}, {Mallmann}, {Zheng}, {Argudo-Fern{\'a}ndez}, {Lian},
  {Bershady}, {Bundy}, {Drory}, {Law}, {Yan}, {Wake}, {Weijmans}, {Bizyaev},
  {Brownstein}, {Lane}, {Maiolino}, {Masters}, {Merrifield}, {Nitschelm},
  {Pan}, {Roman-Lopes}, {Storchi-Bergmann}, \& {Schneider}}]{Goddard2017}
{Goddard}, D., {Thomas}, D., {Maraston}, C., {et~al.} 2017, \mnras, 466, 4731,
  \dodoi{10.1093/mnras/stw3371}

\bibitem[{{Gonz{\'a}lez Delgado} {et~al.}(2014){Gonz{\'a}lez Delgado},
  {P{\'e}rez}, {Cid Fernandes}, {Garc{\'\i}a-Benito}, {de Amorim},
  {S{\'a}nchez}, {Husemann}, {Cortijo-Ferrero}, {L{\'o}pez Fern{\'a}ndez},
  {S{\'a}nchez-Bl{\'a}zquez}, {Bekeraite}, {Walcher}, {Falc{\'o}n-Barroso},
  {Gallazzi}, {van de Ven}, {Alves}, {Bland-Hawthorn}, {Kennicutt}, {Kupko},
  {Lyubenova}, {Mast}, {Moll{\'a}}, {Marino}, {Quirrenbach}, {V{\'\i}lchez}, \&
  {Wisotzki}}]{Gonzalez-Delgado2014}
{Gonz{\'a}lez Delgado}, R.~M., {P{\'e}rez}, E., {Cid Fernandes}, R., {et~al.}
  2014, \aap, 562, A47, \dodoi{10.1051/0004-6361/201322011}

\bibitem[{{Gonz{\'a}lez Delgado} {et~al.}(2015){Gonz{\'a}lez Delgado},
  {Garc{\'\i}a-Benito}, {P{\'e}rez}, {Cid Fernandes}, {de Amorim},
  {Cortijo-Ferrero}, {Lacerda}, {L{\'o}pez Fern{\'a}ndez}, {Vale-Asari},
  {S{\'a}nchez}, {Moll{\'a}}, {Ruiz-Lara}, {S{\'a}nchez-Bl{\'a}zquez},
  {Walcher}, {Alves}, {Aguerri}, {Bekerait{\'e}}, {Bland-Hawthorn}, {Galbany},
  {Gallazzi}, {Husemann}, {Iglesias-P{\'a}ramo}, {Kalinova},
  {L{\'o}pez-S{\'a}nchez}, {Marino}, {M{\'a}rquez}, {Masegosa}, {Mast},
  {M{\'e}ndez-Abreu}, {Mendoza}, {del Olmo}, {P{\'e}rez}, {Quirrenbach}, \&
  {Zibetti}}]{Gonzalez-Delgado2015}
{Gonz{\'a}lez Delgado}, R.~M., {Garc{\'\i}a-Benito}, R., {P{\'e}rez}, E.,
  {et~al.} 2015, \aap, 581, A103, \dodoi{10.1051/0004-6361/201525938}

\bibitem[{Gordon {et~al.}(2022)Gordon, Rothman, Hargreaves, Hashemi, Karlovets,
  Skinner, Conway, Hill, Kochanov, Tan, Wcisło, Finenko, Nelson, Bernath,
  Birk, Boudon, Campargue, Chance, Coustenis, Drouin, Flaud, Gamache, Hodges,
  Jacquemart, Mlawer, Nikitin, Perevalov, Rotger, Tennyson, Toon, Tran,
  Tyuterev, Adkins, Baker, Barbe, Canè, Császár, Dudaryonok, Egorov,
  Fleisher, Fleurbaey, Foltynowicz, Furtenbacher, Harrison, Hartmann, Horneman,
  Huang, Karman, Karns, Kassi, Kleiner, Kofman, Kwabia–Tchana, Lavrentieva,
  Lee, Long, Lukashevskaya, Lyulin, Makhnev, Matt, Massie, Melosso,
  Mikhailenko, Mondelain, Müller, Naumenko, Perrin, Polyansky, Raddaoui,
  Raston, Reed, Rey, Richard, Tóbiás, Sadiek, Schwenke, Starikova, Sung,
  Tamassia, Tashkun, {Vander Auwera}, Vasilenko, Vigasin, Villanueva, Vispoel,
  Wagner, Yachmenev, \& Yurchenko}]{HITRAN2020}
Gordon, I., Rothman, L., Hargreaves, R., {et~al.} 2022, Journal of Quantitative
  Spectroscopy and Radiative Transfer, 277, 107949,
  \dodoi{https://doi.org/10.1016/j.jqsrt.2021.107949}

\bibitem[{{Green}(2018)}]{Green2018}
{Green}, G. 2018, The Journal of Open Source Software, 3, 695,
  \dodoi{10.21105/joss.00695}

\bibitem[{{Green} {et~al.}(2019){Green}, {Schlafly}, {Zucker}, {Speagle}, \&
  {Finkbeiner}}]{Green2019}
{Green}, G.~M., {Schlafly}, E., {Zucker}, C., {Speagle}, J.~S., \&
  {Finkbeiner}, D. 2019, \apj, 887, 93, \dodoi{10.3847/1538-4357/ab5362}

\bibitem[{{Greene} {et~al.}(2015){Greene}, {Janish}, {Ma}, {McConnell},
  {Blakeslee}, {Thomas}, \& {Murphy}}]{Greene2015}
{Greene}, J.~E., {Janish}, R., {Ma}, C.-P., {et~al.} 2015, \apj, 807, 11,
  \dodoi{10.1088/0004-637X/807/1/11}

\bibitem[{{Gu} {et~al.}(2022){Gu}, {Greene}, {Newman}, {Kreisch},
  {Quenneville}, {Ma}, \& {Blakeslee}}]{Gu2022}
{Gu}, M., {Greene}, J.~E., {Newman}, A.~B., {et~al.} 2022, \apj, 932, 103,
  \dodoi{10.3847/1538-4357/ac69ea}

\bibitem[{{Guiderdoni} \& {Rocca-Volmerange}(1987)}]{Guiderdoni1987}
{Guiderdoni}, B., \& {Rocca-Volmerange}, B. 1987, \aap, 186, 1

\bibitem[{{Habets} \& {Heintze}(1981)}]{HabetsHeintze1981}
{Habets}, G.~M.~H.~J., \& {Heintze}, J.~R.~W. 1981, \aaps, 46, 193

\bibitem[{{Hartmann}(1904)}]{Hartmann1904}
{Hartmann}, J. 1904, \apj, 19, 268, \dodoi{10.1086/141112}

\bibitem[{{Heckman} {et~al.}(2000){Heckman}, {Lehnert}, {Strickland}, \&
  {Armus}}]{Heckman2000}
{Heckman}, T.~M., {Lehnert}, M.~D., {Strickland}, D.~K., \& {Armus}, L. 2000,
  \apjs, 129, 493, \dodoi{10.1086/313421}

\bibitem[{{Hill} {et~al.}(2022){Hill}, {Thomas}, {Maraston}, {Yan}, {Neumann},
  {Lundgren}, {Lazarz}, {Chen}, {Cappellari}, {Holtzman}, {Imig}, {Cunha},
  {Stringfellow}, {Bizyaev}, {Law}, {Stassun}, {Drory}, {Merrifield}, \&
  {Beers}}]{Hill2022}
{Hill}, L., {Thomas}, D., {Maraston}, C., {et~al.} 2022, \mnras, 509, 4308,
  \dodoi{10.1093/mnras/stab3263}

\bibitem[{{Ho} {et~al.}(2015){Ho}, {Kudritzki}, {Kewley}, {Zahid}, {Dopita},
  {Bresolin}, \& {Rupke}}]{Ho2015}
{Ho}, I.~T., {Kudritzki}, R.-P., {Kewley}, L.~J., {et~al.} 2015, \mnras, 448,
  2030, \dodoi{10.1093/mnras/stv067}

\bibitem[{{Hobbs}(1969)}]{Hobbs1969}
{Hobbs}, L.~M. 1969, \apj, 158, 461, \dodoi{10.1086/150210}

\bibitem[{{Hobbs}(1974)}]{Hobbs1974}
---. 1974, \apj, 191, 381, \dodoi{10.1086/152976}

\bibitem[{{Howk} {et~al.}(2003){Howk}, {Sembach}, \& {Savage}}]{Howk2003}
{Howk}, J.~C., {Sembach}, K.~R., \& {Savage}, B.~D. 2003, \apj, 586, 249,
  \dodoi{10.1086/346262}

\bibitem[{{Imig} {et~al.}(2022){Imig}, {Holtzman}, {Yan}, {Lazarz}, {Chen},
  {Hill}, {Thomas}, {Maraston}, {Prescott}, {Stringfellow}, {Bizyaev},
  {Beaton}, \& {Drory}}]{Imig2022}
{Imig}, J., {Holtzman}, J.~A., {Yan}, R., {et~al.} 2022, \aj, 163, 56,
  \dodoi{10.3847/1538-3881/ac3ca7}

\bibitem[{{Johansson} {et~al.}(2010){Johansson}, {Thomas}, \&
  {Maraston}}]{Johansson2010}
{Johansson}, J., {Thomas}, D., \& {Maraston}, C. 2010, \mnras, 406, 165,
  \dodoi{10.1111/j.1365-2966.2010.16683.x}

\bibitem[{{Johansson} {et~al.}(2012){Johansson}, {Thomas}, \&
  {Maraston}}]{Johansson2012}
---. 2012, \mnras, 421, 1908, \dodoi{10.1111/j.1365-2966.2011.20316.x}

\bibitem[{{Kauffmann} {et~al.}(2003){Kauffmann}, {Heckman}, {White}, {Charlot},
  {Tremonti}, {Brinchmann}, {Bruzual}, {Peng}, {Seibert}, {Bernardi},
  {Blanton}, {Brinkmann}, {Castander}, {Cs{\'a}bai}, {Fukugita}, {Ivezic},
  {Munn}, {Nichol}, {Padmanabhan}, {Thakar}, {Weinberg}, \&
  {York}}]{Kauffmann2003}
{Kauffmann}, G., {Heckman}, T.~M., {White}, S. D.~M., {et~al.} 2003, \mnras,
  341, 33, \dodoi{10.1046/j.1365-8711.2003.06291.x}

\bibitem[{{Kaviraj} {et~al.}(2007){Kaviraj}, {Kirkby}, {Silk}, \&
  {Sarzi}}]{Kaviraj2007}
{Kaviraj}, S., {Kirkby}, L.~A., {Silk}, J., \& {Sarzi}, M. 2007, \mnras, 382,
  960, \dodoi{10.1111/j.1365-2966.2007.12475.x}

\bibitem[{{Kobayashi} {et~al.}(2020){Kobayashi}, {Karakas}, \&
  {Lugaro}}]{Kobayashi2020}
{Kobayashi}, C., {Karakas}, A.~I., \& {Lugaro}, M. 2020, \apj, 900, 179,
  \dodoi{10.3847/1538-4357/abae65}

\bibitem[{{Kobayashi} {et~al.}(2006){Kobayashi}, {Umeda}, {Nomoto}, {Tominaga},
  \& {Ohkubo}}]{Kobayashi2006}
{Kobayashi}, C., {Umeda}, H., {Nomoto}, K., {Tominaga}, N., \& {Ohkubo}, T.
  2006, \apj, 653, 1145, \dodoi{10.1086/508914}

\bibitem[{{Kos} {et~al.}(2013){Kos}, {Zwitter}, {Grebel}, {Bienayme}, {Binney},
  {Bland-Hawthorn}, {Freeman}, {Gibson}, {Gilmore}, {Kordopatis}, {Navarro},
  {Parker}, {Reid}, {Seabroke}, {Siebert}, {Siviero}, {Steinmetz}, {Watson}, \&
  {Wyse}}]{Kos2013}
{Kos}, J., {Zwitter}, T., {Grebel}, E.~K., {et~al.} 2013, \apj, 778, 86,
  \dodoi{10.1088/0004-637X/778/2/86}

\bibitem[{{Kuntschner}(2000)}]{Kuntschner2000}
{Kuntschner}, H. 2000, \mnras, 315, 184,
  \dodoi{10.1046/j.1365-8711.2000.03377.x}

\bibitem[{{Kurucz}(1993)}]{Kurucz1993}
{Kurucz}, R.~L. 1993, {SYNTHE spectrum synthesis programs and line data}

\bibitem[{{Kurucz}(2011)}]{Kurucz2011}
---. 2011, Canadian Journal of Physics, 89, 417, \dodoi{10.1139/p10-104}

\bibitem[{{La Barbera} {et~al.}(2013){La Barbera}, {Ferreras}, {Vazdekis}, {de
  la Rosa}, {de Carvalho}, {Trevisan}, {Falc{\'o}n-Barroso}, \&
  {Ricciardelli}}]{LaBarbera2013}
{La Barbera}, F., {Ferreras}, I., {Vazdekis}, A., {et~al.} 2013, \mnras, 433,
  3017, \dodoi{10.1093/mnras/stt943}

\bibitem[{{La Barbera} {et~al.}(2019){La Barbera}, {Vazdekis}, {Ferreras},
  {Pasquali}, {Allende Prieto}, {Mart{\'\i}n-Navarro}, {Aguado}, {de Carvalho},
  {Rembold}, {Falc{\'o}n-Barroso}, \& {van de Ven}}]{LaBarbera2019}
{La Barbera}, F., {Vazdekis}, A., {Ferreras}, I., {et~al.} 2019, \mnras, 489,
  4090, \dodoi{10.1093/mnras/stz2192}

\bibitem[{{Lallement} {et~al.}(1993){Lallement}, {Bertin}, {Chassefiere}, \&
  {Scott}}]{Lallement1993}
{Lallement}, R., {Bertin}, P., {Chassefiere}, E., \& {Scott}, N. 1993, \aap,
  271, 734

\bibitem[{{Law} {et~al.}(2022){Law}, {Belfiore}, {Bershady}, {Cappellari},
  {Drory}, {Masters}, {Westfall}, {Bizyaev}, {Bundy}, {Pan}, \&
  {Yan}}]{Law2022}
{Law}, D.~R., {Belfiore}, F., {Bershady}, M.~A., {et~al.} 2022, \apj, 928, 58,
  \dodoi{10.3847/1538-4357/ac5620}

\bibitem[{{Lazarz} {et~al.}(2022){Lazarz}, {Yan}, {Wilhelm}, {Chen}, {Hill},
  {Holtzman}, {Imig}, {Maraston}, {M{\'e}sz{\'a}ros}, {Stringfellow}, {Thomas},
  {Beers}, {Bizyaev}, {Drory}, {Lane}, \& {Nitschelm}}]{Lazarz2022}
{Lazarz}, D., {Yan}, R., {Wilhelm}, R., {et~al.} 2022, \aap, 668, A21,
  \dodoi{10.1051/0004-6361/202243701}

\bibitem[{{Le Borgne} {et~al.}(2003){Le Borgne}, {Bruzual}, {Pell{\'o}},
  {Lan{\c{c}}on}, {Rocca-Volmerange}, {Sanahuja}, {Schaerer}, {Soubiran}, \&
  {V{\'\i}lchez-G{\'o}mez}}]{LeBorgne2003}
{Le Borgne}, J.~F., {Bruzual}, G., {Pell{\'o}}, R., {et~al.} 2003, \aap, 402,
  433, \dodoi{10.1051/0004-6361:20030243}

\bibitem[{{Lehner} \& {Howk}(2011)}]{LehnerHowk2011}
{Lehner}, N., \& {Howk}, J.~C. 2011, Science, 334, 955,
  \dodoi{10.1126/science.1209069}

\bibitem[{{Leitherer} \& {Heckman}(1995)}]{LeithererHeckman1995}
{Leitherer}, C., \& {Heckman}, T.~M. 1995, \apjs, 96, 9, \dodoi{10.1086/192112}

\bibitem[{{Lejeune} {et~al.}(1997){Lejeune}, {Cuisinier}, \&
  {Buser}}]{Lejeune1997}
{Lejeune}, T., {Cuisinier}, F., \& {Buser}, R. 1997, \aaps, 125, 229,
  \dodoi{10.1051/aas:1997373}

\bibitem[{{Leonardi} \& {Rose}(1996)}]{Leonardi1996}
{Leonardi}, A.~J., \& {Rose}, J.~A. 1996, \aj, 111, 182, \dodoi{10.1086/117772}

\bibitem[{{Leonardi} \& {Rose}(2003)}]{Leonardi2003}
---. 2003, \aj, 126, 1811, \dodoi{10.1086/377617}

\bibitem[{{Le{\'s}niewska} {et~al.}(2023){Le{\'s}niewska}, {Micha{\l}owski},
  {Gall}, {Hjorth}, {Nadolny}, {Ryzhov}, \& {Solar}}]{Lesniewska2023}
{Le{\'s}niewska}, A., {Micha{\l}owski}, M.~J., {Gall}, C., {et~al.} 2023, \apj,
  953, 27, \dodoi{10.3847/1538-4357/acdcfc}

\bibitem[{{Lonoce} {et~al.}(2023){Lonoce}, {Freedman}, \&
  {Feldmeier-Krause}}]{Lonoce2023}
{Lonoce}, I., {Freedman}, W.~L., \& {Feldmeier-Krause}, A. 2023, \apj, 948, 65,
  \dodoi{10.3847/1538-4357/acc025}

\bibitem[{{Lyubenova} {et~al.}(2016){Lyubenova}, {Mart{\'\i}n-Navarro}, {van de
  Ven}, {Falc{\'o}n-Barroso}, {Galbany}, {Gallazzi}, {Garc{\'\i}a-Benito},
  {Gonz{\'a}lez Delgado}, {Husemann}, {La Barbera}, {Marino}, {Mast},
  {Mendez-Abreu}, {Peletier}, {S{\'a}nchez-Bl{\'a}zquez}, {S{\'a}nchez},
  {Trager}, {van den Bosch}, {Vazdekis}, {Walcher}, {Zhu}, {Zibetti},
  {Ziegler}, {Bland-Hawthorn}, \& {CALIFA Collaboration}}]{Lyubenova2016}
{Lyubenova}, M., {Mart{\'\i}n-Navarro}, I., {van de Ven}, G., {et~al.} 2016,
  \mnras, 463, 3220, \dodoi{10.1093/mnras/stw2434}

\bibitem[{{Maksymowicz-Maciata} {et~al.}(2024){Maksymowicz-Maciata},
  {Spiniello}, {Mart{\'\i}n-Navarro}, {Ferr{\'e}-Mateu}, {Bevacqua},
  {Cappellari}, {D'Ago}, {Tortora}, {Arnaboldi}, {Hartke}, {Napolitano},
  {Saracco}, \& {Scognamiglio}}]{Maksymowicz-Maciata2024}
{Maksymowicz-Maciata}, M., {Spiniello}, C., {Mart{\'\i}n-Navarro}, I., {et~al.}
  2024, \mnras, 531, 2864, \dodoi{10.1093/mnras/stae1318}

\bibitem[{{Maraston}(1998)}]{Maraston1998}
{Maraston}, C. 1998, \mnras, 300, 872, \dodoi{10.1046/j.1365-8711.1998.01947.x}

\bibitem[{{Maraston}(2005)}]{Maraston2005}
---. 2005, \mnras, 362, 799, \dodoi{10.1111/j.1365-2966.2005.09270.x}

\bibitem[{{Maraston} {et~al.}(2009){Maraston}, {Nieves Colmen{\'a}rez},
  {Bender}, \& {Thomas}}]{Maraston2009}
{Maraston}, C., {Nieves Colmen{\'a}rez}, L., {Bender}, R., \& {Thomas}, D.
  2009, \aap, 493, 425, \dodoi{10.1051/0004-6361:20066907}

\bibitem[{{Maraston} \& {Str{\"o}mb{\"a}ck}(2011)}]{Maraston2011}
{Maraston}, C., \& {Str{\"o}mb{\"a}ck}, G. 2011, \mnras, 418, 2785,
  \dodoi{10.1111/j.1365-2966.2011.19738.x}

\bibitem[{{Maraston} {et~al.}(2020){Maraston}, {Hill}, {Thomas}, {Yan}, {Chen},
  {Lian}, {Parikh}, {Neumann}, {Meneses-Goytia}, {Bershady}, {Drory},
  {Bizyaev}, {Concas}, {Brownstein}, {Lazarz}, {Stringfellow}, \&
  {Stassun}}]{Maraston2020}
{Maraston}, C., {Hill}, L., {Thomas}, D., {et~al.} 2020, \mnras, 496, 2962,
  \dodoi{10.1093/mnras/staa1489}

\bibitem[{{Mart{\'\i}n-Navarro} {et~al.}(2015){Mart{\'\i}n-Navarro}, {La
  Barbera}, {Vazdekis}, {Falc{\'o}n-Barroso}, \&
  {Ferreras}}]{Martin-Navarro2015}
{Mart{\'\i}n-Navarro}, I., {La Barbera}, F., {Vazdekis}, A.,
  {Falc{\'o}n-Barroso}, J., \& {Ferreras}, I. 2015, \mnras, 447, 1033,
  \dodoi{10.1093/mnras/stu2480}

\bibitem[{{Mart{\'\i}n-Navarro} {et~al.}(2023){Mart{\'\i}n-Navarro},
  {Spiniello}, {Tortora}, {Coccato}, {D'Ago}, {Ferr{\'e}-Mateu}, {Pulsoni},
  {Hartke}, {Arnaboldi}, {Hunt}, {Napolitano}, {Scognamiglio}, \&
  {Spavone}}]{Martin-Navarro2023}
{Mart{\'\i}n-Navarro}, I., {Spiniello}, C., {Tortora}, C., {et~al.} 2023,
  \mnras, 521, 1408, \dodoi{10.1093/mnras/stad503}

\bibitem[{{McConnell} {et~al.}(2016){McConnell}, {Lu}, \&
  {Mann}}]{McConnell2016}
{McConnell}, N.~J., {Lu}, J.~R., \& {Mann}, A.~W. 2016, \apj, 821, 39,
  \dodoi{10.3847/0004-637X/821/1/39}

\bibitem[{{Mehlert} {et~al.}(2003){Mehlert}, {Thomas}, {Saglia}, {Bender}, \&
  {Wegner}}]{Mehlert2003}
{Mehlert}, D., {Thomas}, D., {Saglia}, R.~P., {Bender}, R., \& {Wegner}, G.
  2003, \aap, 407, 423, \dodoi{10.1051/0004-6361:20030886}

\bibitem[{{M{\'e}sz{\'a}ros} {et~al.}(2012){M{\'e}sz{\'a}ros}, {Allende
  Prieto}, {Edvardsson}, {Castelli}, {Garc{\'\i}a P{\'e}rez}, {Gustafsson},
  {Majewski}, {Plez}, {Schiavon}, {Shetrone}, \& {de Vicente}}]{Meszaros2012}
{M{\'e}sz{\'a}ros}, S., {Allende Prieto}, C., {Edvardsson}, B., {et~al.} 2012,
  \aj, 144, 120, \dodoi{10.1088/0004-6256/144/4/120}

\bibitem[{{Munari} \& {Zwitter}(1997)}]{MunariZwitter1997}
{Munari}, U., \& {Zwitter}, T. 1997, \aap, 318, 269

\bibitem[{{M{\"u}nch} \& {Zirin}(1961)}]{MunchZirin1961}
{M{\"u}nch}, G., \& {Zirin}, H. 1961, \apj, 133, 11, \dodoi{10.1086/146999}

\bibitem[{{Murga} {et~al.}(2015){Murga}, {Zhu}, {M{\'e}nard}, \&
  {Lan}}]{Murga2015}
{Murga}, M., {Zhu}, G., {M{\'e}nard}, B., \& {Lan}, T.-W. 2015, \mnras, 452,
  511, \dodoi{10.1093/mnras/stv1277}

\bibitem[{{Neumann} {et~al.}(2021){Neumann}, {Thomas}, {Maraston}, {Goddard},
  {Lian}, {Hill}, {Dom{\'\i}nguez S{\'a}nchez}, {Bernardi},
  {Margalef-Bentabol}, {Barrera-Ballesteros}, {Bizyaev}, {Boardman}, {Drory},
  {Fern{\'a}ndez-Trincado}, \& {Lane}}]{Neumann2021}
{Neumann}, J., {Thomas}, D., {Maraston}, C., {et~al.} 2021, \mnras, 508, 4844,
  \dodoi{10.1093/mnras/stab286810.48550/arXiv.2109.11564}

\bibitem[{{O'Connell}(1976)}]{OConnell1976}
{O'Connell}, R.~W. 1976, \apj, 206, 370, \dodoi{10.1086/154392}

\bibitem[{{Ocvirk} {et~al.}(2006){Ocvirk}, {Pichon}, {Lan{\c{c}}on}, \&
  {Thi{\'e}baut}}]{Ocvirk2006}
{Ocvirk}, P., {Pichon}, C., {Lan{\c{c}}on}, A., \& {Thi{\'e}baut}, E. 2006,
  \mnras, 365, 46, \dodoi{10.1111/j.1365-2966.2005.09182.x}

\bibitem[{{Oosterloo} {et~al.}(2010){Oosterloo}, {Morganti}, {Crocker},
  {J{\"u}tte}, {Cappellari}, {de Zeeuw}, {Krajnovi{\'c}}, {McDermid},
  {Kuntschner}, {Sarzi}, \& {Weijmans}}]{Oosterloo2010}
{Oosterloo}, T., {Morganti}, R., {Crocker}, A., {et~al.} 2010, \mnras, 409,
  500, \dodoi{10.1111/j.1365-2966.2010.17351.x}

\bibitem[{{Parikh} {et~al.}(2024){Parikh}, {Saglia}, {Thomas}, {Mehrgan},
  {Bender}, \& {Maraston}}]{Parikh2024}
{Parikh}, T., {Saglia}, R., {Thomas}, J., {et~al.} 2024, \mnras, 528, 7338,
  \dodoi{10.1093/mnras/stae448}

\bibitem[{{Parikh} {et~al.}(2021){Parikh}, {Thomas}, {Maraston}, {Westfall},
  {Andrews}, {Boardman}, {Drory}, \& {Oyarzun}}]{Parikh2021}
{Parikh}, T., {Thomas}, D., {Maraston}, C., {et~al.} 2021, \mnras, 502, 5508,
  \dodoi{10.1093/mnras/stab449}

\bibitem[{{Parikh} {et~al.}(2018){Parikh}, {Thomas}, {Maraston}, {Westfall},
  {Goddard}, {Lian}, {Meneses-Goytia}, {Jones}, {Vaughan}, {Andrews},
  {Bershady}, {Bizyaev}, {Brinkmann}, {Brownstein}, {Bundy}, {Drory},
  {Emsellem}, {Law}, {Newman}, {Roman-Lopes}, {Wake}, {Yan}, \&
  {Zheng}}]{Parikh2018}
---. 2018, \mnras, 477, 3954, \dodoi{10.1093/mnras/sty785}

\bibitem[{{Parikh} {et~al.}(2019){Parikh}, {Thomas}, {Maraston}, {Westfall},
  {Lian}, {Fraser-McKelvie}, {Andrews}, {Drory}, \&
  {Meneses-Goytia}}]{Parikh2019}
---. 2019, \mnras, 483, 3420, \dodoi{10.1093/mnras/sty3339}

\bibitem[{{Pellerin} {et~al.}(2002){Pellerin}, {Fullerton}, {Robert}, {Howk},
  {Hutchings}, {Walborn}, {Bianchi}, {Crowther}, \& {Sonneborn}}]{Pellerin2002}
{Pellerin}, A., {Fullerton}, A.~W., {Robert}, C., {et~al.} 2002, \apjs, 143,
  159, \dodoi{10.1086/342268}

\bibitem[{{P{\'e}rez} {et~al.}(2013){P{\'e}rez}, {Cid Fernandes}, {Gonz{\'a}lez
  Delgado}, {Garc{\'\i}a-Benito}, {S{\'a}nchez}, {Husemann}, {Mast},
  {Rod{\'o}n}, {Kupko}, {Backsmann}, {de Amorim}, {van de Ven}, {Walcher},
  {Wisotzki}, {Cortijo-Ferrero}, \& {CALIFA Collaboration}}]{Perez2013}
{P{\'e}rez}, E., {Cid Fernandes}, R., {Gonz{\'a}lez Delgado}, R.~M., {et~al.}
  2013, \apjl, 764, L1, \dodoi{10.1088/2041-8205/764/1/L1}

\bibitem[{{Perna} {et~al.}(2020){Perna}, {Arribas}, {Catal{\'a}n-Torrecilla},
  {Colina}, {Bellocchi}, {Fluetsch}, {Maiolino}, {Cazzoli}, {Hern{\'a}n
  Caballero}, {Pereira Santaella}, {Piqueras L{\'o}pez}, \& {Rodr{\'\i}guez del
  Pino}}]{Perna2020}
{Perna}, M., {Arribas}, S., {Catal{\'a}n-Torrecilla}, C., {et~al.} 2020, \aap,
  643, A139, \dodoi{10.1051/0004-6361/202038328}

\bibitem[{{Perna} {et~al.}(2021){Perna}, {Arribas}, {Pereira Santaella},
  {Colina}, {Bellocchi}, {Catal{\'a}n-Torrecilla}, {Cazzoli}, {Crespo
  G{\'o}mez}, {Maiolino}, {Piqueras L{\'o}pez}, \& {Rodr{\'\i}guez del
  Pino}}]{Perna2021}
{Perna}, M., {Arribas}, S., {Pereira Santaella}, M., {et~al.} 2021, \aap, 646,
  A101, \dodoi{10.1051/0004-6361/202039702}

\bibitem[{{Phillips} {et~al.}(1984){Phillips}, {Pettini}, \&
  {Gondhalekar}}]{Phillips1984}
{Phillips}, A.~P., {Pettini}, M., \& {Gondhalekar}, P.~M. 1984, \mnras, 206,
  337, \dodoi{10.1093/mnras/206.2.337}

\bibitem[{{Phillips} {et~al.}(2013){Phillips}, {Simon}, {Morrell}, {Burns},
  {Cox}, {Foley}, {Karakas}, {Patat}, {Sternberg}, {Williams}, {Gal-Yam},
  {Hsiao}, {Leonard}, {Persson}, {Stritzinger}, {Thompson}, {Campillay},
  {Contreras}, {Folatelli}, {Freedman}, {Hamuy}, {Roth}, {Shields}, {Suntzeff},
  {Chomiuk}, {Ivans}, {Madore}, {Penprase}, {Perley}, {Pignata}, {Preston}, \&
  {Soderberg}}]{Phillips2013}
{Phillips}, M.~M., {Simon}, J.~D., {Morrell}, N., {et~al.} 2013, \apj, 779, 38,
  \dodoi{10.1088/0004-637X/779/1/38}

\bibitem[{{Pickles}(1985)}]{Pickles1985}
{Pickles}, A.~J. 1985, \apj, 296, 340, \dodoi{10.1086/163454}

\bibitem[{{Poznanski} {et~al.}(2012){Poznanski}, {Prochaska}, \&
  {Bloom}}]{Poznanski2012}
{Poznanski}, D., {Prochaska}, J.~X., \& {Bloom}, J.~S. 2012, \mnras, 426, 1465,
  \dodoi{10.1111/j.1365-2966.2012.21796.x}

\bibitem[{{Puspitarini} \& {Lallement}(2012)}]{Puspitarini2012}
{Puspitarini}, L., \& {Lallement}, R. 2012, \aap, 545, A21,
  \dodoi{10.1051/0004-6361/201219284}

\bibitem[{{Renzini} \& {Buzzoni}(1986)}]{RenziniBuzzoni1986}
{Renzini}, A., \& {Buzzoni}, A. 1986, in Astrophysics and Space Science
  Library, Vol. 122, Spectral Evolution of Galaxies, ed. C.~{Chiosi} \&
  A.~{Renzini}, 195--231, \dodoi{10.1007/978-94-009-4598-2_19}

\bibitem[{{Richter} {et~al.}(2011){Richter}, {Krause}, {Fechner}, {Charlton},
  \& {Murphy}}]{Richter2011}
{Richter}, P., {Krause}, F., {Fechner}, C., {Charlton}, J.~C., \& {Murphy},
  M.~T. 2011, \aap, 528, A12, \dodoi{10.1051/0004-6361/201015566}

\bibitem[{{Richter} {et~al.}(2001{\natexlab{a}}){Richter}, {Savage}, {Wakker},
  {Sembach}, \& {Kalberla}}]{Richter2001b}
{Richter}, P., {Savage}, B.~D., {Wakker}, B.~P., {Sembach}, K.~R., \&
  {Kalberla}, P. M.~W. 2001{\natexlab{a}}, \apj, 549, 281,
  \dodoi{10.1086/319070}

\bibitem[{{Richter} {et~al.}(2001{\natexlab{b}}){Richter}, {Sembach}, {Wakker},
  {Savage}, {Tripp}, {Murphy}, {Kalberla}, \& {Jenkins}}]{Richter2001a}
{Richter}, P., {Sembach}, K.~R., {Wakker}, B.~P., {et~al.} 2001{\natexlab{b}},
  \apj, 559, 318, \dodoi{10.1086/322401}

\bibitem[{{Robert} {et~al.}(2003){Robert}, {Pellerin}, {Aloisi}, {Leitherer},
  {Hoopes}, \& {Heckman}}]{Robert2003}
{Robert}, C., {Pellerin}, A., {Aloisi}, A., {et~al.} 2003, \apjs, 144, 21,
  \dodoi{10.1086/344478}

\bibitem[{Robert {et~al.}(2009)Robert, Chopin, \& Rousseau}]{Robert2009}
Robert, C.~P., Chopin, N., \& Rousseau, J. 2009, Statistical Science, 24, 141 ,
  \dodoi{10.1214/09-STS284}

\bibitem[{{Roberts-Borsani} \& {Saintonge}(2019)}]{RobertsBorsani2019}
{Roberts-Borsani}, G.~W., \& {Saintonge}, A. 2019, \mnras, 482, 4111,
  \dodoi{10.1093/mnras/sty2824}

\bibitem[{{Roberts-Borsani} {et~al.}(2020){Roberts-Borsani}, {Saintonge},
  {Masters}, \& {Stark}}]{RobertsBorsani2020}
{Roberts-Borsani}, G.~W., {Saintonge}, A., {Masters}, K.~L., \& {Stark}, D.~V.
  2020, \mnras, 493, 3081, \dodoi{10.1093/mnras/staa464}

\bibitem[{{Rodr{\'\i}guez-Merino} {et~al.}(2005){Rodr{\'\i}guez-Merino},
  {Chavez}, {Bertone}, \& {Buzzoni}}]{Rodriguez-Merino2005}
{Rodr{\'\i}guez-Merino}, L.~H., {Chavez}, M., {Bertone}, E., \& {Buzzoni}, A.
  2005, \apj, 626, 411, \dodoi{10.1086/429858}

\bibitem[{{Roig} {et~al.}(2015){Roig}, {Blanton}, \& {Yan}}]{Roig2015}
{Roig}, B., {Blanton}, M.~R., \& {Yan}, R. 2015, \apj, 808, 26,
  \dodoi{10.1088/0004-637X/808/1/26}

\bibitem[{{Rose}(1985)}]{Rose1985}
{Rose}, J.~A. 1985, \aj, 90, 1927, \dodoi{10.1086/113898}

\bibitem[{{Rubin} {et~al.}(2018){Rubin}, {Diamond-Stanic}, {Coil}, {Crighton},
  \& {Moustakas}}]{Rubin2018a}
{Rubin}, K. H.~R., {Diamond-Stanic}, A.~M., {Coil}, A.~L., {Crighton}, N.
  H.~M., \& {Moustakas}, J. 2018, \apj, 853, 95,
  \dodoi{10.3847/1538-4357/aa9792}

\bibitem[{{Rubin} {et~al.}(2022){Rubin}, {Juarez}, {Cooksey}, {Werk},
  {Prochaska}, {O'Meara}, {Burchett}, {Rickards Vaught}, {Kulkarni}, \&
  {Straka}}]{Rubin2022}
{Rubin}, K. H.~R., {Juarez}, C., {Cooksey}, K.~L., {et~al.} 2022, \apj, 936,
  171, \dodoi{10.3847/1538-4357/ac7b88}

\bibitem[{{Ruffa} {et~al.}(2019){Ruffa}, {Prandoni}, {Laing}, {Paladino},
  {Parma}, {de Ruiter}, {Mignano}, {Davis}, {Bureau}, \& {Warren}}]{Ruffa2019}
{Ruffa}, I., {Prandoni}, I., {Laing}, R.~A., {et~al.} 2019, \mnras, 484, 4239,
  \dodoi{10.1093/mnras/stz255}

\bibitem[{{Rupke} {et~al.}(2005){Rupke}, {Veilleux}, \& {Sanders}}]{Rupke2005a}
{Rupke}, D.~S., {Veilleux}, S., \& {Sanders}, D.~B. 2005, \apjs, 160, 87,
  \dodoi{10.1086/432886}

\bibitem[{{Rupke} {et~al.}(2021){Rupke}, {Thomas}, \& {Dopita}}]{Rupke2021}
{Rupke}, D. S.~N., {Thomas}, A.~D., \& {Dopita}, M.~A. 2021, \mnras, 503, 4748,
  \dodoi{10.1093/mnras/stab743}

\bibitem[{{S{\'a}nchez-Bl{\'a}zquez} {et~al.}(2006){S{\'a}nchez-Bl{\'a}zquez},
  {Peletier}, {Jim{\'e}nez-Vicente}, {Cardiel}, {Cenarro},
  {Falc{\'o}n-Barroso}, {Gorgas}, {Selam}, \&
  {Vazdekis}}]{Sanchez-Blazquez2006}
{S{\'a}nchez-Bl{\'a}zquez}, P., {Peletier}, R.~F., {Jim{\'e}nez-Vicente}, J.,
  {et~al.} 2006, \mnras, 371, 703, \dodoi{10.1111/j.1365-2966.2006.10699.x}

\bibitem[{{S{\'a}nchez-Bl{\'a}zquez} {et~al.}(2014){S{\'a}nchez-Bl{\'a}zquez},
  {Rosales-Ortega}, {M{\'e}ndez-Abreu}, {P{\'e}rez}, {S{\'a}nchez}, {Zibetti},
  {Aguerri}, {Bland-Hawthorn}, {Catal{\'a}n-Torrecilla}, {Cid Fernandes}, {de
  Amorim}, {de Lorenzo-Caceres}, {Falc{\'o}n-Barroso}, {Galazzi}, {Garc{\'\i}a
  Benito}, {Gil de Paz}, {Gonz{\'a}lez Delgado}, {Husemann},
  {Iglesias-P{\'a}ramo}, {Jungwiert}, {Marino}, {M{\'a}rquez}, {Mast},
  {Mendoza}, {Moll{\'a}}, {Papaderos}, {Ruiz-Lara}, {van de Ven}, {Walcher}, \&
  {Wisotzki}}]{Sanchez-Blazquez2014}
{S{\'a}nchez-Bl{\'a}zquez}, P., {Rosales-Ortega}, F.~F., {M{\'e}ndez-Abreu},
  J., {et~al.} 2014, \aap, 570, A6, \dodoi{10.1051/0004-6361/201423635}

\bibitem[{{Sandford} {et~al.}(2023){Sandford}, {Weisz}, \&
  {Ting}}]{Sandford2023}
{Sandford}, N.~R., {Weisz}, D.~R., \& {Ting}, Y.-S. 2023, \apjs, 267, 18,
  \dodoi{10.3847/1538-4365/acd37b}

\bibitem[{{Schaller} {et~al.}(1992){Schaller}, {Schaerer}, {Meynet}, \&
  {Maeder}}]{Schaller1992}
{Schaller}, G., {Schaerer}, D., {Meynet}, G., \& {Maeder}, A. 1992, \aaps, 96,
  269

\bibitem[{{Schlegel} {et~al.}(1998){Schlegel}, {Finkbeiner}, \&
  {Davis}}]{SFD98}
{Schlegel}, D.~J., {Finkbeiner}, D.~P., \& {Davis}, M. 1998, \apj, 500, 525,
  \dodoi{10.1086/305772}

\bibitem[{Schmidt-Kaler(1982)}]{Schmidt-Kaler1982}
Schmidt-Kaler. 1982, Numerical Data and Functional Relationships in Science and
  Technology, Landolt/Bornstein No. Group IV, Vol. 2(b) (Springer,Berlin)

\bibitem[{{Sembach} \& {Danks}(1994)}]{Sembach1994}
{Sembach}, K.~R., \& {Danks}, A.~C. 1994, \aap, 289, 539

\bibitem[{{Sembach} {et~al.}(1993){Sembach}, {Danks}, \&
  {Savage}}]{Sembach1993}
{Sembach}, K.~R., {Danks}, A.~C., \& {Savage}, B.~D. 1993, \aaps, 100, 107

\bibitem[{{Sembach} {et~al.}(1999){Sembach}, {Savage}, \&
  {Hurwitz}}]{Sembach1999}
{Sembach}, K.~R., {Savage}, B.~D., \& {Hurwitz}, M. 1999, \apj, 524, 98,
  \dodoi{10.1086/307811}

\bibitem[{{Serven} {et~al.}(2005){Serven}, {Worthey}, \& {Briley}}]{Serven2005}
{Serven}, J., {Worthey}, G., \& {Briley}, M.~M. 2005, \apj, 627, 754,
  \dodoi{10.1086/430400}

\bibitem[{{Smith} {et~al.}(2012){Smith}, {Lucey}, \& {Carter}}]{Smith2012}
{Smith}, R.~J., {Lucey}, J.~R., \& {Carter}, D. 2012, \mnras, 426, 2994,
  \dodoi{10.1111/j.1365-2966.2012.21922.x}

\bibitem[{{Spiniello} {et~al.}(2012){Spiniello}, {Trager}, {Koopmans}, \&
  {Chen}}]{Spiniello2012}
{Spiniello}, C., {Trager}, S.~C., {Koopmans}, L.~V.~E., \& {Chen}, Y.~P. 2012,
  \apjl, 753, L32, \dodoi{10.1088/2041-8205/753/2/L32}

\bibitem[{{Spinrad} \& {Taylor}(1971)}]{SpinradTaylor1971}
{Spinrad}, H., \& {Taylor}, B.~J. 1971, \apjs, 22, 445, \dodoi{10.1086/190232}

\bibitem[{{Spolaor} {et~al.}(2010){Spolaor}, {Kobayashi}, {Forbes}, {Couch}, \&
  {Hau}}]{Spolaor2010}
{Spolaor}, M., {Kobayashi}, C., {Forbes}, D.~A., {Couch}, W.~J., \& {Hau}, G.
  K.~T. 2010, \mnras, 408, 272, \dodoi{10.1111/j.1365-2966.2010.17080.x}

\bibitem[{{Starkenburg} {et~al.}(2017){Starkenburg}, {Martin}, {Youakim},
  {Aguado}, {Allende Prieto}, {Arentsen}, {Bernard}, {Bonifacio}, {Caffau},
  {Carlberg}, {C{\^o}t{\'e}}, {Fouesneau}, {Fran{\c{c}}ois}, {Franke},
  {Gonz{\'a}lez Hern{\'a}ndez}, {Gwyn}, {Hill}, {Ibata}, {Jablonka},
  {Longeard}, {McConnachie}, {Navarro}, {S{\'a}nchez-Janssen}, {Tolstoy}, \&
  {Venn}}]{Starkenburg2017}
{Starkenburg}, E., {Martin}, N., {Youakim}, K., {et~al.} 2017, \mnras, 471,
  2587, \dodoi{10.1093/mnras/stx1068}

\bibitem[{{Straka} {et~al.}(2015){Straka}, {Noterdaeme}, {Srianand},
  {Nutalaya}, {Kulkarni}, {Khare}, {Bowen}, {Bishof}, \& {York}}]{Straka2015}
{Straka}, L.~A., {Noterdaeme}, P., {Srianand}, R., {et~al.} 2015, \mnras, 447,
  3856, \dodoi{10.1093/mnras/stu2739}

\bibitem[{Strand(1963)}]{Johnson1963}
Strand, K. 1963, Basic Astronomical Data, Basic Astronomical Data No. v. 3
  (University of Chicago Press).
\newblock \url{https://books.google.com/books?id=qInvAAAAMAAJ}

\bibitem[{{S{\={u}}d{\v{z}}ius} \& {Bobinas}(1994)}]{SudziusBobinas1994}
{S{\={u}}d{\v{z}}ius}, J., \& {Bobinas}, V. 1994, Baltic Astronomy, 3, 158,
  \dodoi{10.1515/astro-1994-1-221}

\bibitem[{{Taresch} {et~al.}(1997){Taresch}, {Kudritzki}, {Hurwitz}, {Bowyer},
  {Pauldrach}, {Puls}, {Butler}, {Lennon}, \& {Haser}}]{Taresch1997}
{Taresch}, G., {Kudritzki}, R.~P., {Hurwitz}, M., {et~al.} 1997, \aap, 321, 531

\bibitem[{{Thomas} {et~al.}(2011{\natexlab{a}}){Thomas}, {Johansson}, \&
  {Maraston}}]{TJM2011}
{Thomas}, D., {Johansson}, J., \& {Maraston}, C. 2011{\natexlab{a}}, \mnras,
  412, 2199, \dodoi{10.1111/j.1365-2966.2010.18108.x}

\bibitem[{{Thomas} {et~al.}(2003){Thomas}, {Maraston}, \&
  {Bender}}]{Thomas2003}
{Thomas}, D., {Maraston}, C., \& {Bender}, R. 2003, \mnras, 339, 897,
  \dodoi{10.1046/j.1365-8711.2003.06248.x}

\bibitem[{{Thomas} {et~al.}(2005){Thomas}, {Maraston}, {Bender}, \& {Mendes de
  Oliveira}}]{Thomas2005}
{Thomas}, D., {Maraston}, C., {Bender}, R., \& {Mendes de Oliveira}, C. 2005,
  \apj, 621, 673, \dodoi{10.1086/426932}

\bibitem[{{Thomas} {et~al.}(2011{\natexlab{b}}){Thomas}, {Maraston}, \&
  {Johansson}}]{Thomas2011}
{Thomas}, D., {Maraston}, C., \& {Johansson}, J. 2011{\natexlab{b}}, \mnras,
  412, 2183, \dodoi{10.1111/j.1365-2966.2010.18049.x}

\bibitem[{{Thomas} {et~al.}(2011{\natexlab{c}}){Thomas}, {Saglia}, {Bender},
  {Thomas}, {Gebhardt}, {Magorrian}, {Corsini}, {Wegner}, \&
  {Seitz}}]{ThomasJ2011}
{Thomas}, J., {Saglia}, R.~P., {Bender}, R., {et~al.} 2011{\natexlab{c}},
  \mnras, 415, 545, \dodoi{10.1111/j.1365-2966.2011.18725.x}

\bibitem[{{Tinsley}(1978)}]{Tinsley1978}
{Tinsley}, B.~M. 1978, \apj, 222, 14, \dodoi{10.1086/156116}

\bibitem[{{Trager} {et~al.}(1998){Trager}, {Worthey}, {Faber}, {Burstein}, \&
  {Gonz{\'a}lez}}]{Trager1998}
{Trager}, S.~C., {Worthey}, G., {Faber}, S.~M., {Burstein}, D., \&
  {Gonz{\'a}lez}, J.~J. 1998, \apjs, 116, 1, \dodoi{10.1086/313099}

\bibitem[{{Tremonti} {et~al.}(2004){Tremonti}, {Heckman}, {Kauffmann},
  {Brinchmann}, {Charlot}, {White}, {Seibert}, {Peng}, {Schlegel}, {Uomoto},
  {Fukugita}, \& {Brinkmann}}]{Tremonti2004}
{Tremonti}, C.~A., {Heckman}, T.~M., {Kauffmann}, G., {et~al.} 2004, \apj, 613,
  898, \dodoi{10.1086/423264}

\bibitem[{{Treu} {et~al.}(2010){Treu}, {Auger}, {Koopmans}, {Gavazzi},
  {Marshall}, \& {Bolton}}]{Treu2010}
{Treu}, T., {Auger}, M.~W., {Koopmans}, L. V.~E., {et~al.} 2010, \apj, 709,
  1195, \dodoi{10.1088/0004-637X/709/2/1195}

\bibitem[{{Tumlinson} {et~al.}(2002){Tumlinson}, {Shull}, {Rachford},
  {Browning}, {Snow}, {Fullerton}, {Jenkins}, {Savage}, {Crowther}, {Moos},
  {Sembach}, {Sonneborn}, \& {York}}]{Tumlinson2002}
{Tumlinson}, J., {Shull}, J.~M., {Rachford}, B.~L., {et~al.} 2002, \apj, 566,
  857, \dodoi{10.1086/338112}

\bibitem[{{Turnrose}(1976)}]{Turnrose1976}
{Turnrose}, B.~E. 1976, \apj, 210, 33, \dodoi{10.1086/154801}

\bibitem[{{Valdes} {et~al.}(2004){Valdes}, {Gupta}, {Rose}, {Singh}, \&
  {Bell}}]{Valdes2004}
{Valdes}, F., {Gupta}, R., {Rose}, J.~A., {Singh}, H.~P., \& {Bell}, D.~J.
  2004, \apjs, 152, 251, \dodoi{10.1086/386343}

\bibitem[{{van Dokkum} {et~al.}(2017){van Dokkum}, {Conroy}, {Villaume},
  {Brodie}, \& {Romanowsky}}]{vanDokkum2017}
{van Dokkum}, P., {Conroy}, C., {Villaume}, A., {Brodie}, J., \& {Romanowsky},
  A.~J. 2017, \apj, 841, 68, \dodoi{10.3847/1538-4357/aa7135}

\bibitem[{{van Dokkum} \& {Conroy}(2010)}]{vanDokkumConroy2010}
{van Dokkum}, P.~G., \& {Conroy}, C. 2010, \nat, 468, 940,
  \dodoi{10.1038/nature09578}

\bibitem[{{van Dokkum} \& {Conroy}(2012)}]{vanDokkumConroy2012}
---. 2012, \apj, 760, 70, \dodoi{10.1088/0004-637X/760/1/70}

\bibitem[{{VanderPlas}(2014)}]{VanderPlas2014}
{VanderPlas}, J. 2014, arXiv e-prints, arXiv:1411.5018,
  \dodoi{10.48550/arXiv.1411.5018}

\bibitem[{{Vaughan} {et~al.}(2018){Vaughan}, {Davies}, {Zieleniewski}, \&
  {Houghton}}]{Vaughan2018}
{Vaughan}, S.~P., {Davies}, R.~L., {Zieleniewski}, S., \& {Houghton}, R. C.~W.
  2018, \mnras, 479, 2443, \dodoi{10.1093/mnras/sty1434}

\bibitem[{{Vazdekis}(1999)}]{Vazdekis1999}
{Vazdekis}, A. 1999, \apj, 513, 224, \dodoi{10.1086/306843}

\bibitem[{{Vazdekis} {et~al.}(2010){Vazdekis}, {S{\'a}nchez-Bl{\'a}zquez},
  {Falc{\'o}n-Barroso}, {Cenarro}, {Beasley}, {Cardiel}, {Gorgas}, \&
  {Peletier}}]{Vazdekis2010}
{Vazdekis}, A., {S{\'a}nchez-Bl{\'a}zquez}, P., {Falc{\'o}n-Barroso}, J.,
  {et~al.} 2010, \mnras, 404, 1639, \dodoi{10.1111/j.1365-2966.2010.16407.x}

\bibitem[{{Veilleux} {et~al.}(2020){Veilleux}, {Maiolino}, {Bolatto}, \&
  {Aalto}}]{Veilleux2020}
{Veilleux}, S., {Maiolino}, R., {Bolatto}, A.~D., \& {Aalto}, S. 2020, \aapr,
  28, 2, \dodoi{10.1007/s00159-019-0121-9}

\bibitem[{{Vogrin{\v{c}}i{\v{c}}} {et~al.}(2023){Vogrin{\v{c}}i{\v{c}}}, {Kos},
  {Zwitter}, {Traven}, {Beeson}, {{\v{C}}otar}, {Munari}, {Buder}, {Martell},
  {Lewis}, {De Silva}, {Hayden}, {Bland-Hawthorn}, \&
  {D'Orazi}}]{Vogrincic2023}
{Vogrin{\v{c}}i{\v{c}}}, R., {Kos}, J., {Zwitter}, T., {et~al.} 2023, \mnras,
  521, 3727, \dodoi{10.1093/mnras/stad678}

\bibitem[{{Wakker}(2001)}]{Wakker2001}
{Wakker}, B.~P. 2001, \apjs, 136, 463, \dodoi{10.1086/321783}

\bibitem[{{Walborn}(1972)}]{Walborn1972}
{Walborn}, N.~R. 1972, \aj, 77, 312, \dodoi{10.1086/111285}

\bibitem[{{Walborn}(1973)}]{Walborn1973}
---. 1973, \aj, 78, 1067, \dodoi{10.1086/111509}

\bibitem[{{Welsh} {et~al.}(2010){Welsh}, {Lallement}, {Vergely}, \&
  {Raimond}}]{Welsh2010}
{Welsh}, B.~Y., {Lallement}, R., {Vergely}, J.~L., \& {Raimond}, S. 2010, \aap,
  510, A54, \dodoi{10.1051/0004-6361/200913202}

\bibitem[{{Welty} {et~al.}(2006){Welty}, {Federman}, {Gredel}, {Thorburn}, \&
  {Lambert}}]{Welty2006}
{Welty}, D.~E., {Federman}, S.~R., {Gredel}, R., {Thorburn}, J.~A., \&
  {Lambert}, D.~L. 2006, \apjs, 165, 138, \dodoi{10.1086/504153}

\bibitem[{{Welty} {et~al.}(1996){Welty}, {Morton}, \& {Hobbs}}]{Welty1996}
{Welty}, D.~E., {Morton}, D.~C., \& {Hobbs}, L.~M. 1996, \apjs, 106, 533,
  \dodoi{10.1086/192347}

\bibitem[{{Welty} {et~al.}(2012){Welty}, {Xue}, \& {Wong}}]{Welty2012}
{Welty}, D.~E., {Xue}, R., \& {Wong}, T. 2012, \apj, 745, 173,
  \dodoi{10.1088/0004-637X/745/2/173}

\bibitem[{{Westfall} {et~al.}(2019){Westfall}, {Cappellari}, {Bershady},
  {Bundy}, {Belfiore}, {Ji}, {Law}, {Schaefer}, {Shetty}, {Tremonti}, {Yan},
  {Andrews}, {Brownstein}, {Cherinka}, {Coccato}, {Drory}, {Maraston},
  {Parikh}, {S{\'a}nchez-Gallego}, {Thomas}, {Weijmans}, {Barrera-Ballesteros},
  {Du}, {Goddard}, {Li}, {Masters}, {Ibarra Medel}, {S{\'a}nchez}, {Yang},
  {Zheng}, \& {Zhou}}]{Westfall2019}
{Westfall}, K.~B., {Cappellari}, M., {Bershady}, M.~A., {et~al.} 2019, \aj,
  158, 231, \dodoi{10.3847/1538-3881/ab44a2}

\bibitem[{{Whitford}(1977)}]{Whitford1977}
{Whitford}, A.~E. 1977, \apj, 211, 527, \dodoi{10.1086/154959}

\bibitem[{{Wild} \& {Hewett}(2005)}]{Wild2005}
{Wild}, V., \& {Hewett}, P.~C. 2005, \mnras, 361, L30,
  \dodoi{10.1111/j.1745-3933.2005.00058.x}

\bibitem[{{Wild} {et~al.}(2007){Wild}, {Kauffmann}, {Heckman}, {Charlot},
  {Lemson}, {Brinchmann}, {Reichard}, \& {Pasquali}}]{WildCaHK2007}
{Wild}, V., {Kauffmann}, G., {Heckman}, T., {et~al.} 2007, \mnras, 381, 543,
  \dodoi{10.1111/j.1365-2966.2007.12256.x}

\bibitem[{{Wilkinson} {et~al.}(2017){Wilkinson}, {Maraston}, {Goddard},
  {Thomas}, \& {Parikh}}]{Wilkinson2017}
{Wilkinson}, D.~M., {Maraston}, C., {Goddard}, D., {Thomas}, D., \& {Parikh},
  T. 2017, \mnras, 472, 4297, \dodoi{10.1093/mnras/stx2215}

\bibitem[{{Wing} \& {Ford}(1969)}]{WingFord1969}
{Wing}, R.~F., \& {Ford}, W.~Kent, J. 1969, \pasp, 81, 527,
  \dodoi{10.1086/128814}

\bibitem[{{Woosley} \& {Weaver}(1995)}]{WoosleyWeaver1995}
{Woosley}, S.~E., \& {Weaver}, T.~A. 1995, \apjs, 101, 181,
  \dodoi{10.1086/192237}

\bibitem[{{Worthey} {et~al.}(1992){Worthey}, {Faber}, \&
  {Gonzalez}}]{Worthey1992}
{Worthey}, G., {Faber}, S.~M., \& {Gonzalez}, J.~J. 1992, \apj, 398, 69,
  \dodoi{10.1086/171836}

\bibitem[{{Worthey} {et~al.}(1994){Worthey}, {Faber}, {Gonzalez}, \&
  {Burstein}}]{Worthey1994}
{Worthey}, G., {Faber}, S.~M., {Gonzalez}, J.~J., \& {Burstein}, D. 1994,
  \apjs, 94, 687, \dodoi{10.1086/192087}

\bibitem[{{Worthey} \& {Ottaviani}(1997)}]{WortheyOttaviani1997}
{Worthey}, G., \& {Ottaviani}, D.~L. 1997, \apjs, 111, 377,
  \dodoi{10.1086/313021}

\bibitem[{{Worthey} {et~al.}(2014){Worthey}, {Tang}, \& {Serven}}]{Worthey2014}
{Worthey}, G., {Tang}, B., \& {Serven}, J. 2014, \apj, 783, 20,
  \dodoi{10.1088/0004-637X/783/1/20}

\bibitem[{{Yan} {et~al.}(2016){Yan}, {Bundy}, {Law}, {Bershady}, {Andrews},
  {Cherinka}, {Diamond-Stanic}, {Drory}, {MacDonald}, {S{\'a}nchez-Gallego},
  {Thomas}, {Wake}, {Weijmans}, {Westfall}, {Zhang}, {Arag{\'o}n-Salamanca},
  {Belfiore}, {Bizyaev}, {Blanc}, {Blanton}, {Brownstein}, {Cappellari},
  {D'Souza}, {Emsellem}, {Fu}, {Gaulme}, {Graham}, {Goddard}, {Gunn},
  {Harding}, {Jones}, {Kinemuchi}, {Li}, {Li}, {Maiolino}, {Mao}, {Maraston},
  {Masters}, {Merrifield}, {Oravetz}, {Pan}, {Parejko}, {Sanchez}, {Schlegel},
  {Simmons}, {Thanjavur}, {Tinker}, {Tremonti}, {van den Bosch}, \&
  {Zheng}}]{Yan2016}
{Yan}, R., {Bundy}, K., {Law}, D.~R., {et~al.} 2016, \aj, 152, 197,
  \dodoi{10.3847/0004-6256/152/6/197}

\bibitem[{{Yan} {et~al.}(2019){Yan}, {Chen}, {Lazarz}, {Bizyaev}, {Maraston},
  {Stringfellow}, {McCarthy}, {Meneses-Goytia}, {Law}, {Thomas}, {Falcon
  Barroso}, {S{\'a}nchez-Gallego}, {Schlafly}, {Zheng}, {Argudo-Fern{\'a}ndez},
  {Beaton}, {Beers}, {Bershady}, {Blanton}, {Brownstein}, {Bundy}, {Chambers},
  {Cherinka}, {De Lee}, {Drory}, {Galbany}, {Holtzman}, {Imig}, {Kaiser},
  {Kinemuchi}, {Liu}, {Luo}, {Magnier}, {Majewski}, {Nair}, {Oravetz},
  {Oravetz}, {Pan}, {Sobeck}, {Stassun}, {Talbot}, {Tremonti}, {Waters},
  {Weijmans}, {Wilhelm}, {Zasowski}, {Zhao}, \& {Zhao}}]{Yan2019}
{Yan}, R., {Chen}, Y., {Lazarz}, D., {et~al.} 2019, \apj, 883, 175,
  \dodoi{10.3847/1538-4357/ab3ebc}

\bibitem[{{Yao} {et~al.}(2009){Yao}, {Tripp}, {Wang}, {Danforth}, {Canizares},
  {Shull}, {Marshall}, \& {Song}}]{Yao2009}
{Yao}, Y., {Tripp}, T.~M., {Wang}, Q.~D., {et~al.} 2009, \apj, 697, 1784,
  \dodoi{10.1088/0004-637X/697/2/1784}

\bibitem[{{Youakim} {et~al.}(2017){Youakim}, {Starkenburg}, {Aguado}, {Martin},
  {Fouesneau}, {Gonz{\'a}lez Hern{\'a}ndez}, {Allende Prieto}, {Bonifacio},
  {Gentile}, {Kielty}, {C{\^o}t{\'e}}, {Jablonka}, {McConnachie}, {S{\'a}nchez
  Janssen}, {Tolstoy}, \& {Venn}}]{Youakim2017}
{Youakim}, K., {Starkenburg}, E., {Aguado}, D.~S., {et~al.} 2017, \mnras, 472,
  2963, \dodoi{10.1093/mnras/stx2005}

\bibitem[{{Young} {et~al.}(2011){Young}, {Bureau}, {Davis}, {Combes},
  {McDermid}, {Alatalo}, {Blitz}, {Bois}, {Bournaud}, {Cappellari}, {Davies},
  {de Zeeuw}, {Emsellem}, {Khochfar}, {Krajnovi{\'c}}, {Kuntschner},
  {Lablanche}, {Morganti}, {Naab}, {Oosterloo}, {Sarzi}, {Scott}, {Serra}, \&
  {Weijmans}}]{Young2011}
{Young}, L.~M., {Bureau}, M., {Davis}, T.~A., {et~al.} 2011, \mnras, 414, 940,
  \dodoi{10.1111/j.1365-2966.2011.18561.x}

\bibitem[{{Zabludoff} {et~al.}(1996){Zabludoff}, {Zaritsky}, {Lin}, {Tucker},
  {Hashimoto}, {Shectman}, {Oemler}, \& {Kirshner}}]{Zabludoff1996}
{Zabludoff}, A.~I., {Zaritsky}, D., {Lin}, H., {et~al.} 1996, \apj, 466, 104,
  \dodoi{10.1086/177495}

\bibitem[{{Zych} {et~al.}(2009){Zych}, {Murphy}, {Hewett}, \&
  {Prochaska}}]{Zych2009}
{Zych}, B.~J., {Murphy}, M.~T., {Hewett}, P.~C., \& {Prochaska}, J.~X. 2009,
  \mnras, 392, 1429, \dodoi{10.1111/j.1365-2966.2008.14157.x}

\bibitem[{{Zych} {et~al.}(2007){Zych}, {Murphy}, {Pettini}, {Hewett},
  {Ryan-Weber}, \& {Ellison}}]{Zych2007}
{Zych}, B.~J., {Murphy}, M.~T., {Pettini}, M., {et~al.} 2007, \mnras, 379,
  1409, \dodoi{10.1111/j.1365-2966.2007.12015.x}

\end{thebibliography}
\bibliographystyle{aasjournal}

\end{document}